\documentclass[twocolumn]{aastex631}

\newcommand{\Msun}{ \mbox{M$_\odot$}}
\newcommand{\Mjup}{\mbox{M$_{\rm Jup}$}}

\newcommand\changed[1]{\textcolor{black}{#1}}

\usepackage{amssymb,gensymb,times,graphicx,morefloats,color}
\usepackage{amsmath}
\usepackage{graphics,graphicx,url,verbatim,xspace}

\usepackage{changepage}
\usepackage{mathtools} 
\usepackage{comment}
\usepackage{gensymb}
\usepackage{hyperref}

\shorttitle{MagAO/Clio Binary Differential Imaging Survey}
\shortauthors{Pearce et al.}

\graphicspath{{./}{figures/}}

\begin{document}
\nolinenumbers
\defcitealias{rodigas_direct_2015}{R15}
\defcitealias{Baraffe2015BHAC}{BHAC15}

\title{Companion Mass Limits for 17 Binary Systems Obtained with Binary Differential Imaging and MagAO/Clio}

\author[0000-0003-3904-7378]{Logan A. Pearce}
\affiliation{Steward Observatory, University of Arizona, Tucson, AZ 85721, USA}
\affiliation{NSF Graduate Research Fellow}

\author[0000-0002-2346-3441]{Jared R. Males}
\affiliation{Steward Observatory, University of Arizona, Tucson, AZ 85721, USA}

\author[0000-0001-6654-7859]{Alycia J. Weinberger}
\affiliation{Earth and Planets Laboratory, Carnegie Institution for Science, 5241 Broad Branch Road NW, Washington, DC 20015-1305}

\author[0000-0003-1905-9443]{Joseph D. Long}
\affiliation{Steward Observatory, University of Arizona, Tucson, AZ 85721, USA}

\author[0000-0002-1384-0063]{Katie M. Morzinski}
\affiliation{Steward Observatory, University of Arizona, Tucson, AZ 85721, USA}

\author[0000-0002-2167-8246]{Laird M. Close}
\affiliation{Steward Observatory, University of Arizona, Tucson, AZ 85721, USA}

\author[0000-0002-1954-4564]{Philip M. Hinz}
\affiliation{UC Santa Cruz, 1156 High St, Santa Cruz CA 95064, USA}

\begin{abstract}
Improving direct detection capability close to the star through improved star-subtraction and post-processing techniques is vital for discovering new low-mass companions and characterizing known ones at longer wavelengths.  We present results of 17 binary star systems observed with the Magellan Adaptive Optics system (MagAO) and the Clio infrared camera on the Magellan Clay Telescope using Binary Differential Imaging (BDI).  BDI is an application of Reference Differential Imaging (RDI) and Angular Differential Imaging (ADI) applied to wide binary star systems (2\arcsec~$<\Delta \rho<$~10\arcsec) within the isoplanatic patch in the infrared. Each star serves as the point-spread-function (PSF) reference for the other, and we performed PSF estimation and subtraction using Principal Component Analysis.  \changed{We report contrast and mass limits for the 35 stars in our initial survey using BDI with MagAO/Clio in L$^\prime$ and 3.95$\mu$m bands.  Our achieved contrasts varied between systems, and spanned a range of contrasts from 3.0-7.5 magnitudes and a range of separations from 0.2\arcsec to $\sim$2\arcsec.  Stars in our survey span a range of masses, and our achieved contrasts correspond to late-type M dwarf masses down to $\sim$10 \Mjup.  We also report detection of a candidate companion signal at 0.2\arcsec (18~AU) around  HIP~67506~A (SpT G5V, mass $\sim$1.2\Msun), which we estimate to be $\sim60-90~\Mjup$.  We found that the effectiveness of BDI is highest for approximately equal brightness binaries in high-Strehl conditions.
}

\end{abstract}


\section{Introduction} \label{sec:intro}
Giant planets on wide enough orbits to be accessible by direct imaging are rare (\changed{occurrence rate $9^{+5}_{-4}$\% for 5-13~\Mjup\ companions within 10-100 AU in the recent results from the Gemini Planet Imager Exoplanet Survey (GPIES); \citealt{nielsen_gemini_2019})}.  
Brown dwarf companions appear to be even more rare, with an occurrence rate of $\sim0.8^{+0.8}_{-0.5}$\% for 13-80~\Mjup\ \changed{from GPIES.}  Yet radial velocity, transit, and microlensing surveys have found that giant planets close to their stars are common in regions promising for future direct imaging. \changed{\citealt{Bryan2019JupiterAnalogs} found an occurrence rate of 39\%$\pm$7\% for masses 0.5-20~\Mjup\ and separations 1-20 AU from radial velocity surveys; \citealt{Herman2019LongPeriodTransit} found 0.7$^{+0.40}_{-0.20}$ planets per solar-type star for radius 0.3-1~R$_{\rm{Jup}}$ and 2-10 yr periods from \textsl{Kepler} \changed{\citep{Borucki2010Kepler}}; \citealt{Poleski2021MicrolensingWidePlanetsCommon} observed 1.4$^{+0.9}_{-0.6}$ ice giants per microlensing star with separations $\approx$5-15~AU from 20 years of the OGLE microlensing survey.  }
Improving direct detection capability close to the star is of paramount importance for increasing the directly-imaged companion sample size and inferring population property statistics.

In addition to building larger telescopes and better instruments for ground- and space-based direct imaging, improving on observational and data analysis techniques can push detection limits closer and deeper.  Point-spread function (PSF) subtraction via Reference Differential Imaging (RDI; commonly used with space telescopes) images the science target and a PSF reference star, but is hindered by time-varying PSFs, and requires observing two stars to reduce one.  An improvement on RDI utilizes a library of PSF reference images \citep[e.g.][]{Sanghi2021RDIwithWFCArchive} and a Locally Optimized Combination of Images \citep[LOCI;][]{lafreniere_new_2007} to optimally reconstruct the PSF, but is still susceptible to time variation between reference and science images. Spectral Differential Imaging (SDI; \citealt{Racine1999SDI,Marois2000SDI}), in which the science target is imaged simultaneously in multiple filter bands, does not obtain photon-noise limited PSF subtraction due to the chromatic variation in speckles that doesn't scale with wavelength \citep{Rameau2015SDIDetLimits}, suffers from a difference in Strehl ratios between images in different bands, and depends on spectral features like Methane in the companion's atmosphere.  With Angular Differential Imaging (ADI; \citealt{marois_angular_2006}), the star serves as its own PSF reference through sky rotation; however, it is susceptible to self-subtraction of candidate companion signals and requires significant sky rotation to avoid flux attenuation, especially close to the star.  A way to avoid these various drawbacks is to simultaneously image a science and reference star in the same filter band.

\cite{Kasper2007LBandImaging} simultaneously imaged 22 young stars in the Tucana and $\beta$ Pictoris moving groups in L$^\prime$ ($\sim$~4$\mu$m) band on NACO/VLT with adaptive optics, including two medium separation binaries (HIP 116748, $\rho$~=~5.8\arcsec\ and GJ 799, $\rho$~=~2.8\arcsec). For these systems, both target and reference star fit on the detector simultaneously, yet were separated such that their PSFs did not overlap. They used each star in the binary to subtract the starlight from the other, termed this Binary Differential Imaging (BDI; in Figure 5 of their paper), and saw improved contrast at close separations compared to contemporaneously imaged single stars.  Similarly, \cite{Heinze2010LongPeriodPlanets} applied ``binary star subtraction" to binaries in their L$^\prime$ and M$^\prime$ nearby star survey with MMT AO with the Clio instrument, in which the PSF of the secondary was scaled and subtracted from the primary and vice versa.  They also saw improvement in achievable contrast with binary star subtraction compared to single stars in their survey.  

\cite{rodigas_direct_2015} \citepalias[hereafter][]{rodigas_direct_2015} expanded on the BDI technique by combining the advantages of simultaneous imaging with advanced data analysis algorithms like Karhunen-Lo\'eve Image Processing (KLIP; \citealt{soummer_detection_2012}) --- an application of Principle Component Analysis (PCA) to image data ---
to better remove the speckle structure in the PSF.  They compared the expected signal-to-noise ratio (S/N) for BDI to ADI and determined that BDI is advantageous close to the star, achieving $\sim$~0.5 mag better contrast within $\sim$~1\arcsec, which they estimate translates to $\sim$1\Mjup\ improvement in sensitivity.  They also note that observing binaries near 4~$\mu$m takes advantage of the large isoplanatic patch ($\sim$10\arcsec--30\arcsec), in addition to being where young substellar companions will be bright \cite[][hereafter BHAC15]{Baraffe2015BHAC}.  They note that a limitation of BDI is the potential for companion flux around one star to be attenuated by flux from a companion around the other star, but that there is low probability of this ($\sim$2\% at 0.15\arcsec, and even smaller farther out).  Additionally, while coronagraphs employing a focal plane occulter cannot easily be used with BDI, pupil-plane only coronagraphs (such as apodizing phase plates, \citealt{Kenworthy2007APP,Otten2014APP}) that affect the PSFs of both stars could further increase sensitivity.  

\citetalias{rodigas_direct_2015} identified a target list of $\sim$140 binary systems optimized for effective BDI.  Targets are young ($\lesssim$ 200 Myr) so that brown dwarf and planetary companions will be bright at near-infrared (NIR) wavelengths.  Their binary separations are between 2\arcsec--10\arcsec\ so that their PSFs do not overlap and are within the isoplanatic patch at L$^\prime$, and their apparent magnitudes in L$^\prime$ are similar to $\lesssim$2~mag, making their PSF features have similar signal-to-noise. 

In this paper, we describe the results of our MagAO/Clio NIR survey of 17 of the \citetalias{rodigas_direct_2015} binary star target list.  \changed{In Section \ref{sec:motivation} we review current relevant studies of substellar companions in binaries.  In Section \ref{sec:systems} we describe our survey sample, with detailed descriptions of each system in Appendix \ref{appendixA}.}  In Section \ref{sec:methods} we describe our BDI observations and KLIP data reduction application.  In Section \ref{sec:results} we show contrast limits for each binary system, discuss limitations on our achievable contrast, and discuss our detection of a candidate companion to HIP~67506~A.  

\section{Motivation: Substellar Companions in Wide Binaries}\label{sec:motivation}
The occurrence rate of planet and brown dwarf companions in binaries, and the influence the binary has on the formation and evolution of the planetary environment, is not well understood, and is hampered by small numbers of observed systems.  

Circumstellar planets in wide binaries (S-type, in which the companions orbits one component of the binary) have been shown to be fully suppressed for close binaries (semi-major axis (sma)~$\lesssim$~1 AU, \citealt{moe_impact_2019}), an occurrence rate of $\sim$15\% at sma~$\sim$10~AU, and increasing with binary separation out to sma~$\sim$100~AU \citep{kraus_impact_2016,moe_impact_2019,ziegler_soar_2020}.  Several recent observational studies have found a higher fraction of close-in S-type companions in multiple systems \changed{compared to single stars.} \citep[][]{Knutson2014FOHJ1, Ngo2015FOHJ2, Piskorz2015FOHJ3}.  \cite{Ngo2016FOHJ4} found a $\sim$3$\times$ inflation in occurrence rate of hot Jupiters in multiple systems over single stars, and infer that stellar companions beyond 50 AU might actually facilitate giant planet formation. \cite{Fontanive2019HighBinarity} found an inflated binary fraction of 80\% with separations from 20-10,000 AU for stars hosting close in higher-mass planetary and brown dwarf companions (7-60 \Mjup).  \cite{Cadman2022BinariesGravitationalInstability} showed that the binary companions can trigger instability and fragmentation in gravitationally unstable disks, leading to formation of these giant planet and brown dwarf companions in outer regions of the disk, which somehow move to the close-in orbits currently observed. However other studies have concluded that the frequency of planets in binaries is not statistically different from that of single stars \citep[e.g.][]{Bonavita2007PlanetsInMultipleSystems,Harris2012TaurusMultipleStarSys}.  \cite{deacon_pan-starrs_2016} found no evidence that binaries with $\rho$~$>$~3000 AU affected occurrence rate of \textit{Kepler} planets with P~$<$~300~days around FGK stars. \cite{moe_impact_2019} found that for sma~$\gtrsim$~100~AU the binary does not suppress planet occurrence, and the apparent inflated occurrence is due to selection effects.

Although it is unclear if wide ($\gtrsim$~100~AU) stellar companions are consequential to the formation of planetary systems, it is likely to impact the evolution of a planetary system through gravitational scattering and migration.  While the wide binary may be too wide to induce binary von Zeipel-Kozai-Lidov oscillations \citep{VonZeipel1910,kozai_secular_1962,lidov_evolution_1962} on an S-type planet directly \citep{Ngo2016FOHJ4}, it could still induce chaos in the system. Mean motion resonance overlap from the companion star leads to regions of chaotic diffusion and eventual planet ejection even in circular binary orbits \citep{Holman1999PlanetsInBinaries, Mudryk2006ResonanceOverlap, Kratter2012StarHoppers}.  Simulations by \cite{kaib_planetary_2013} and \cite{correa-otto_galactic_2017} showed the influence of the galactic gravitational potential and stellar flybys perturbs the wide companion's orbit over time, causing S-type companion orbits to be disrupted, pushed into high-eccentricity orbits, and potentially scattered (see also \citealt{Hamers2017HJinGC}).  The presence of an additional giant planet can further induce secular resonances \citep{bazso_fear_2020} or high-e migration \citep{Hamers2017HintsofHiddenCompanions, Hamers2017HJinGC} interior to the giant planet, planet-planet scattering, and push the surviving planet into high-eccentricity orbits which could be further boosted by Kozai-Lidov cycles from the stellar companion \citep{Mustill2021HR5183bKozaiLidoz}. Wide-orbit stellar companion(s) can also be sufficient to explain stellar spin-planetary orbit misalignment even if the companion is orders of magnitude more distant, \changed{including} inducing retrograde obliquities \citep{Best2022RetrogradeMultiplanetChaos}.

More population members at various stages of evolution are needed to better develop the picture observationally.  In addition to the surveys of \citealt{Kasper2007LBandImaging} and \citealt{Heinze2010LongPeriodPlanets} discussed in the introduction, several other recent surveys have looked for companions in binaries using other starlight subtraction techniques.  \cite{Hagelberg2020VIBES} targeted 26 visible binary and multiple star young moving group members with SPHERE on VLT \citep{Beuzit2008SPHERE} in dual-H band filters, with a Lyot coronagraph masking the brighter star, and used ADI to subtract the starlight.  The SPOTS survey \citep{Thalmann2014SPOTS,Bonavita2016SPOTS,Asensio-Torres2018SPOTS} searched for wide circumbinary planets in the 30-300 AU range.  While neither survey detected new substellar companions, they placed upper limits on occurrence rates.  \cite{Dupuy2022OrbitalArchII} found evidence for mutual alignment between S-type planet and binary orbits $\lesssim$30$^\circ$ for \textit{Kepler}  planet hosts in visual binaries. Additionally, the precise astrometry of Gaia \changed{\citep{gaia_collaboration_gaia_2016}} enabled identification of 1.3 million spatially resolved binaries \citep{el-badry_million_2021}, and several recent studies utilized the Gaia astrometric information to examine the orbit of the wide stellar companion to transiting planet host stars \citep[e.g.][]{newton_tess_2019, venner_true_2021, Newton2021ThreeSmallTESSPlanets}, search for new unresolved companions \citep[e.g.][]{kervella_stellar_2019,Currie2021SurveyofAcceleratingStars}, refine the masses of known companions \citep[e.g.][]{brandt_precise_2019,Brandt2021DynamicalMassBetaPicBBetaPicC}, and observe an overabundance of alignments between planet and wide binary orbits for binaries with semi-major axis $<$~700~AU\citep{Christian2022OrbitalAlignments}.  
Orbital obliquity alignment studies such as \citealt{bryan_obliquity_2020} and \citealt{Xuan2020MutualInclinations} are important probes of the angular momentum evolution of planetary systems and the influence of scattering and/or Kozai-Lidov mechanisms \citep{Mustill2021HR5183bKozaiLidoz}, especially in the presence of a wide stellar companion \citep{Hjorth2021BackwardSpinning}.  Future Gaia data releases will contain improved astrometry and acceleration information for hundreds of millions of sources\footnote{\url{https://www.cosmos.esa.int/web/gaia/dr3}}, making new companion identification through astrometry common place, and promising to deliver numerous planets and brown dwarf companions in wide binaries

Multiple star systems should be prioritized as prime direct imaging targets for probing planetary system formation and evolution, population statistics, and planet characterization studies.

\movetabledown=0.6in
\begin{longrotatetable}
\begin{deluxetable*}{ccccccccccc}
\tablecaption{Summary of Binary Systems\label{tab:system-summary}}
\tablewidth{\textwidth}
\tablehead{
\colhead{HD Name} & \colhead{Alt Name} & \colhead{Separation$^{\rm{a,*}}$} &  \colhead{Distance$^{\rm{a},**}$} & \colhead{Age} &
\colhead{SpT} & \colhead{Group} & \colhead{RUWE$^{\rm{a}}$} &\colhead{G Mag$^{\rm{a}}$} & \colhead{WISE W1 Mag} &
\colhead{WISE W2 Mag}\\
\colhead{} & \colhead{} & \colhead{(arcsec)} & \colhead{(pc)} & \colhead{(Myr)} &
\colhead{} & \colhead{Membership\tablenotemark{$\S$}} & \colhead{A / B} &  \colhead{A / B} & \colhead{(3.35$\mu$m)} &
\colhead{(4.6$\mu$m)}
}
\startdata
HD 36705 & AB Dor & 8.8609 $\pm$ 50 & 14.93 $\pm$ 0.02 & 100$^{\rm{b}}$ & K0V + M5-6$^{\rm{c}}$ & AB Dor & 25.13 / 3.52 &  6.69 / 11.35 & 4.598 $\pm$ 0.121\tablenotemark{$\dagger$} & 4.189 $\pm$ 0.057\tablenotemark{$\dagger$}\\
HD 37551 & WX Col & 4.00175 $\pm$ 1 & 80.45 $\pm$ 0.07 & \changed{130$\pm$20$^{\rm{d}}$} & G7V + K1V$^{\rm{c}}$ & AB Dor$^{\rm{e}}$ &0.96 / 0.97 & 9.45 / 10.35 & 7.284 $\pm$ 0.027\tablenotemark{$\dagger$} & 7.385 $\pm$ 0.019\tablenotemark{$\dagger$} \\
HD 47787 & HIP 31821 & 2.15685 $\pm$ 2 & 47.83 $\pm$ 0.04 & 16.5 $\pm$ 6.5$^{\rm{f}}$ & K1IV + K1IV$^{\rm{c}}$ & Field$^{\rm{j}}$ & 1.11 / 1.11 & 8.91 / 9.01 & 6.348 $\pm$ 0.042\tablenotemark{$\dagger$} & 6.457 $\pm$ 0.042\tablenotemark{$\dagger$}\\
HD 76534 & OU Vel & 2.06874 $\pm$ 2 & 869 $\pm$ 14 & 0.27$^{\rm{h}}$ & B2Vn$^{\rm{i}}$ & Field$^{\rm{j}}$ & 1.53 / 0.89 & 8.25 / 9.42 & 7.271 $\pm$ 0.029\tablenotemark{$\dagger$} & 7.066 $\pm$ 0.020\tablenotemark{$\dagger$} \\
HD 82984 & HIP 46914 & 2.0041$ \pm$ 30 & 274 $\pm$ 7 & 53.4 $\pm$ 15.1$^{\rm{f}}$ & B4IV$^{\rm{f}}$ & Field$^{\rm{j}}$ & 1.09 / 0.89 & 5.53 / 6.26 & 5.346 $\pm$ 0.064\tablenotemark{$\dagger$} & 5.202 $\pm$ 0.030\tablenotemark{$\dagger$} \\
HD 104231 & HIP 58528 & 4.45718 $\pm$ 5 & 102.7 $\pm$ 0.5 & 21$^{\rm{k}}$ & F5V$^{\rm{l}}$ & LCC$^{\rm{m}}$ & 0.83 / 2.29 & 8.45 / 13.43 & A: 7.198 $\pm$ 0.028 & 7.248 $\pm$ 0.020 \\
 &  &  &  &  &  &  &  &  & B: 9.499 $\pm$ 0.228 & 9.338 $\pm$ 0.119 \\
HD 118072 & HIP 66273 & 2.27647 $\pm$ 7 & 79.5 $\pm$ 0.4 & 40-50$^{\rm{n}}$ & G3V$^{\rm{c}}$ & 90\% ARG$^{\rm{j}}$ & 1.20 / 1.20 & 9.02 / 9.14 & 6.875 $\pm$ 0.034\tablenotemark{$\dagger$} & 6.941 $\pm$ 0.020\tablenotemark{$\dagger$}\\
HD 118991 & Q Cen & 5.56444 $\pm$ 6 & 88.3 $\pm$ 0.3 & 130-140$^{\rm{p}}$ & B8.5 + A2.5$^{\rm{q}}$ & Sco-Cen$^{\rm{j}}$ & 1.11 / 1.07 & 5.24 / 6.60 & 4.975 $\pm$ 0.070\tablenotemark{$\dagger$} & 4.629 $\pm$ 0.036\tablenotemark{$\dagger$} \\
HD 137727 & HIP 75769 & 2.20358 $\pm$ 3 & 111.7 $\pm$ 0.3 & 8.2 $\pm$ 0.6$^{\rm{f}}$ & G9III + G6IV$^{\rm{c}}$ & Field$^{\rm{j}}$ & 1.42 / 0.88 & 9.16 / 9.66 & 6.739 $\pm$ 0.038\tablenotemark{$\dagger$} & 6.815 $\pm$ 0.020\tablenotemark{$\dagger$}\\
HD 147553 & HIP 80324 & 6.23216 $\pm$ 7 & 138.2 $\pm$ 1.3 & 16 $\pm$ 1$^{\rm{k}}$ & B9.5V + A1V$^{\rm{s}}$ & UCL$^{\rm{j}}$ & 0.93 / 0.89 & 7.00 / 7.46 & A: 7.039 $\pm$ 0.116 & 7.055 $\pm$ 0.026 \\
 &  &  &  &  &  &  &  &  & B: 7.219 $\pm$ 0.112 & 7.283 $\pm$ 0.037 \\
HD 151771  & HIP 82453 & 6.8957 $\pm$ 3 & 270 $\pm$ 2 & 200-300$^{\rm{t}}$ & B8III + B9.5$^{\rm{u}}$ & Field$^{\rm{j}}$ & 1.22 / 0.80 & 6.19 / 8.46 & A: 5.802 $\pm$ 0.069 & 5.696 $\pm$ 0.033\\
 &  &  &  &  &  &  &  &  & B: 7.412 $\pm$ 0.302 & 7.536 $\pm$ 0.157 \\
HD 164249 & HIP 88399 & 6.49406 $\pm$ 2 & 49.30 $\pm$ 0.06 & 25 $\pm$ 3$^{\rm{v}}$ & F6V + M2V$^{\rm{c}}$ & Beta Pic$^{\rm{w,x}}$ & 1.09 / 1.23 & 6.91 / 12.31 & 5.882 $\pm$ 0.057\tablenotemark{$\dagger$} & 5.841 $\pm$ 0.021\tablenotemark{$\dagger$}\\
HD 201247 & HIP 104526 & 4.17040 $\pm$ 3 & 33.20 $\pm$ 0.04 & 200-300$^{\rm{y}}$ & G5V + G7V$^{\rm{z}}$ & Field$^{\rm{j}}$ & 0.96 / 0.89 & 7.53 / 7.71 & 5.211 $\pm$ 0.0657\tablenotemark{$\dagger$} & 5.055 $\pm$ 0.041\tablenotemark{$\dagger$} \\
HD 222259 & DS Tuc & 5.36461 $\pm$ 3 & 44.12 $\pm$ 0.07 & 45 $\pm$ 4$^{\rm{\alpha}}$ & G6V + K3V$^{\rm{c}}$ & Tuc-Hor$^{\rm{g}}$ & 0.91 / 0.95 & 8.34 / 9.41 & A: 7.062 $\pm$ 0.068 & 7.072 $\pm$ 0.030 \\
 &  &  &  &  &  &  &  &  & B: 7.089 $\pm$ 0.179 & 7.140 $\pm$ 0.056 \\
-- & HIP 67506\tablenotemark{$\ddagger$} & 9.38117 $\pm$ 9 & 220 $\pm$ 2 & 210 $\pm$ 5$^{\rm{t}}$ & G5$^{\rm{\beta}}$ & Field$^{\rm{j}}$ & 2.01 & 10.67 & 9.189 $\pm$ 0.021 & 9.242 $\pm$ 0.023 \\
-- &\changed{TYC 7797-34-2}\tablenotemark{$\ddagger$} & & \changed{1700 $\pm$ 100} & -- & -- & Field$^{j}$ & 1.73 & 11.99 & 9.475 $\pm$ 0.023 & 9.561 $\pm$ 0.021 \\
-- & TWA 13 & 5.06925 $\pm$ 3 & 59.9 $\pm$ 0.1 & 10$^{+10; \rm{\gamma}}_{-7}$ & M1Ve + M1Ve$^{\rm{c}}$ & TW Hydra$^{\rm{\delta}}$ & 1.25 / 1.27 & 10.89 / 10.91 & A: 7.635 $\pm$ 0.052 & 7.545 $\pm$ 0.030 \\
 &  &  &  &  &  &  &  &  & B: 7.408 $\pm$ 0.087 & 7.470 $\pm$ 0.030 \\
-- & 2MASS J01535076-& 2.8543 $\pm$ 10 & 33.85 $\pm$ 0.09 & 25 $\pm$ 3$^{\rm{v}}$ & M3$^{\rm{\epsilon}}$ & Beta Pic$^{\rm{w}}$ & 1.36 / 1.38 & 11.49 / 11.52 & 6.810 $\pm$ 0.028\tablenotemark{$\dagger$} & 6.729 $\pm$ 0.014\tablenotemark{$\dagger$}\\
 & 1459503 & & & & & & & & \\
\enddata
\tablenotetext{*}{Uncertainties in units of 10$^{-5}$ arcsec}
\tablenotetext{**}{\changed{Distances and uncertainties are were computed using the method of \citealt{bailer-jones_estimating_2018} and Gaia EDR3 parallaxes.}}
\tablenotetext{$\S$}{Sco-Cen: Scorpius–Centaurus Association, UCL: Upper Centaurs Lupis association, Tuc-Hor: Tucana-Horologium Young Moving Group, ARG: Argus Association, Beta Pic: Beta Pictoris Moving Group, AB Dor: AB Doradus Moving Group, LCC: Lower Centaurus-Crux}
\tablenotetext{\dagger}{Binary is unresolved in WISE photometry}
\tablenotetext{\ddagger}{\changed{HIP~67506 and TYC 7797-34-2 (WDS J13500-4303~AB) were believed to be a 9\arcsec\ binary at the time of the survey, but we show in this work that they are an unassociated pair.  This has no impact on the BDI reduction of both stars. See Appendix \ref{appendixA}.  There are no age or spectral type estimates in literature for TYC 7797-34-2.
}}
\tablecomments{(a) Gaia EDR3, \citealt{gaiaEDR3}, (b) \citealt{MamajekAges2008}, (c) \citealt{torres_search_2006}, (d) \citealt{Binks2020YoungStars,Barrado2004LiDepl}, (e) \citealt{McCarthy2012SizesOfNearestYoungStars}, (f) \citealt{tetzlaff_catalogue_2011}, (g) \citealt{kraus_stellar_2014}, (h) \citealt{Arun2019MassIRExcessHerbigABStars}, (i) \citealt{Houk1978HDSpectralTypes}, (j) \citealt{gagne_banyan_2018}, (k) \citealt{pecaut_revised_2012}, (l) \citealt{Houk1975SpTofHDStarsSouthernDecl}, (m) \citealt{Hoogerwerf2000OBAssocMembers},  (n) \citealt{Zuckerman2019ArgusAssoc}, (p) \citealt{david_ages_2015}, (q) \citealt{GrayGarrison1987EarlyATypeStars}, (s) \citealt{Corbally1984BinarySpT}, (t) This work, Sec \ref{sec:systems}, (u) \citealt{Corbally1984BinarySpT}, (v) \citealt{messina_rotation-lithium_2016}, (w) \citealt{messina__2017}, (x) \citealt{deacon_wide_2020}, (y) \citealt{zuckerman_young_2013}, (z) \citealt{gray_contributions_2006},  ($\alpha$) \citealt{bell_self-consistent_2015}, ($\beta$) \citealt{SpencerJones1939SpT}, ($\gamma$) \citealt{barrado_y_navascues_age_2006}, ($\delta$) \citealt{Schneider2012TWAMembership}, ($\epsilon$) \citealt{Riaz2006NewMDwarfs},
}
\end{deluxetable*}
\end{longrotatetable}

\section{Binary Systems in our Survey} \label{sec:systems}
We observed 17 binary star systems between 2014 - 2017, chosen for their utility for BDI data reduction, to span a range of spectral types, and their availability between other observing programs.  Table \ref{tab:system-summary} summarizes the properties of each young binary system observed.  Binary separation, distance, and the primary's G-band magnitude were taken from Gaia EDR3 \citep{gaiaEDR3}; age and spectral type were taken from literature values; group membership is from literature and/or Banyan $\Sigma$ membership probabilities \citep{gagne_banyan_2018}.  Our observations were conducted in MKO L$^\prime$ and the narrowband 3.95$\mu$m ($\Delta\lambda_{\rm{eff}} = 0.08\mu$m, $\lambda_{\rm{0}} = 3.95\mu$m; hereafter [3.95]\footnote{Previous papers have called it the [3.9] filter, we here refer to it as [3.95] for clarity}) filters, so we have included the primary's WISE W1 and W2 magnitudes for reference \citep{Cutri2012WISE}.  A subset of systems were unresolved in WISE, so the photometry includes flux from both members, and are indicated with a dagger in Table \ref{tab:system-summary}. 

\changed{We have made use of literature ages for the estimation of mass limits in Section \ref{sec:results}.  Age estimates we adopted were derived using a variety of methods; specifics for each binary system are noted in Table \ref{tab:system-summary} and described in Appendix \ref{appendixA}.  We used the most-recent and lowest-uncertainty age estimate for an individual star where available; most were derived using isochrone model fitting to photometry, lithium equivalent widths, or chromospheric and coronal activity.  Where individual age estimates were not available we adopted the average age and uncertainty for the associated moving group. 
Two systems in our survey did not have literature ages or moving group membership (HIP~67506/TYC~7797-34-2 and HD~151771), and we estimated age using isochrone fitting (see Appendix \ref{appendixA} for details).  In Section \ref{sec:results} we discuss the impact the estimated age of the star has on our results.
}

\section{Methods}\label{sec:methods}

\subsection{Observations}
Observations for this survey were carried out between 2014 to 2017 with Magellan Adaptive Optics system (MagAO) \citep{close_diffraction-limited_2013} and Clio science camera on the 6.5~m Magellan Clay telescope at Las Campanas Observatory, Chile.  All images were obtained in [3.95] or MKO L$^\prime$ observing bands with the narrow camera (plate scale = 15.9 mas pixel$^{-1}$, field of view = 16$\times$18\arcsec;  \citealt{morzinski_magellan_2015}) in full frame mode (512$\times$1024 pixels), and with the telescope rotator off. 
Observation parameters varied between datasets and are documented in Table \ref{tab:obs-summary}.  There were two observing modes: ABBA Nod mode, in which two nod positions (A and B) with both stars on the detector, 10 frames each, were alternated in an ABBA pattern during the observations; and ``Sky" mode, where science frames were observed in a single nod and the telescope was offset to get starless ``sky frames".

\subsection{Data Reduction} 
Due to the difficulty of flat fielding Clio images \citep[see][Appendix B.3]{morzinski_magellan_2015}, we performed sky subtraction using Karhunen-Lo\'eve Image Processing (KLIP; \citealt{soummer_detection_2012}), an implementation of principle component analysis (PCA) applied to image data.  To sky subtract a science image from Nod A (in ABBA observing mode) with KLIP, we:
\begin{adjustwidth}{0.5cm}{}
1. masked the stars in every Nod B image in the dataset to a radius of 8 $\lambda$/D to capture variation in the sky alone,\\
2. constructed a PCA eigenimage basis set from the Nod B images in the dataset, following the prescription of \cite{soummer_detection_2012} Section 2.2 step 2,\\
3. projected the Nod A target image onto the eigenbasis constructed from Nod B up to a desired number of basis modes K$_{\rm{klip}}$ \citep[][Section 2.2 step 4]{soummer_detection_2012}, to create a sky estimator,\\
4. subtracted the sky estimator from the Nod A image.
\end{adjustwidth}
We repeated this process for Nod B images using a basis constructed from all Nod A images in the dataset.  For datasets observed in ``Sky" mode, we constructed the basis set from the sky frames.  All datasets were sky subtracted with K$_{\rm{klip}}$~$\leq$~5.  We then corrected bad pixels using the bad pixel maps of \cite{morzinski_magellan_2015}; we also used a high-pass filter and flagged pixels with excessive variation during the course of the dataset to identify and correct additional bad pixels.  Bad pixels within star PSFs were identified by eye and corrected.  Finally, images were inspected for quality by eye, and sharpest images were kept for use in starlight subtraction.  None of the images in our survey fell outside the linear regime and did not require linearity correction.

\subsection{KLIP PSF Subtraction}

\begin{figure}
\centering
\includegraphics[width=0.46\textwidth]{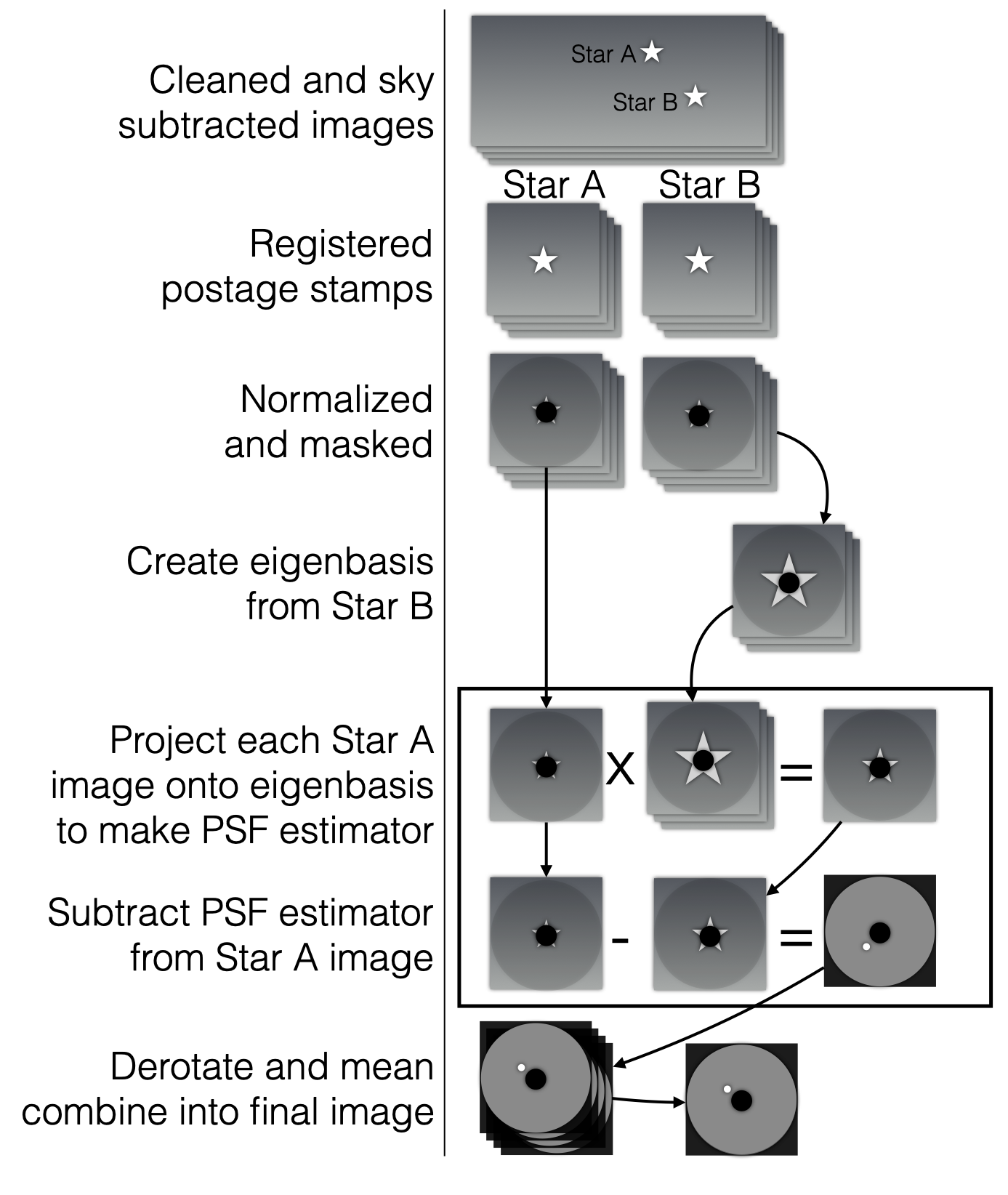}
\caption{\small{Illustration of our BDI implementation of the KLIP methodology.  This process is repeated for Star B using Star A as the eigenbasis.
}}
\label{fig:diagram}
\end{figure}

As with sky subtraction, we subtracted the star's PSF using a custom implementation of KLIP PSF subtraction.  Our algorithm, illustrated in Figure \ref{fig:diagram}, proceeds in the following way: 
\begin{adjustwidth}{0.5cm}{}
1. Each star is cut out of each cleaned and sky-subtracted full frame image into a ``postage stamp" and assembled into a cube of all images of Star A and another cube of Star B.  \\
2. Each image in each cube is registered (PSF core centered in frame), normalized (entire frame is divided by the sum of all pixels in the frame so that the pixel values all now sum to one), and the inner core of the PSF is masked to avoid fitting the PSF core and prioritize fitting the PSF wings.  We determined a radius of 1$\lambda$/D for the inner core mask was optimal for our data by inspection.  We did not have any saturated stars in our datasets.\\
3. For the Star A cube, a PCA eigenbasis set is constructed from the Star B cube, following the prescription of \cite{soummer_detection_2012} Section 2.2 as before.  \\
4. Each image in the Star A cube is projected onto the Star B basis set up to specified number of modes K$_{\rm{klip}}$ to create a PSF estimator, then the PSF estimator is subtracted from the Star A image. \\
5. Each image is rotated to North up/East left, then a sigma-clipped mean image of the cube is created as the final reduced image.  PSF estimation via ADI was not employed in our analysis.  \\
6. Repeat 3-5 for the Star B cube using Star A to create eigenbasis.\\
\end{adjustwidth}

Postage stamp size varied by dataset due to the binary separation (star PSFs must be able to be isolated), proximity to glints and detector defects, and proximity to the edge of the frame.  One system in our initial survey, 53 Aquarii, is a 1.2\arcsec\ binary, which ended up being too close to effectively separate the PSFs to serve as references for KLIP and was excluded.  Another system, WDS J00304-6236, is a triple system, with WDS J00304-6236 Aa,Ab separated by 0.1\arcsec, enough to cause elongation of Star A's PSF and disqualifying it from serving as a PSF reference to WDS J00304-6236 B and was excluded.

\subsection{Contrast and mass limits}\label{sec:cont limits}

\begin{figure}
\centering
\includegraphics[width=0.46\textwidth]{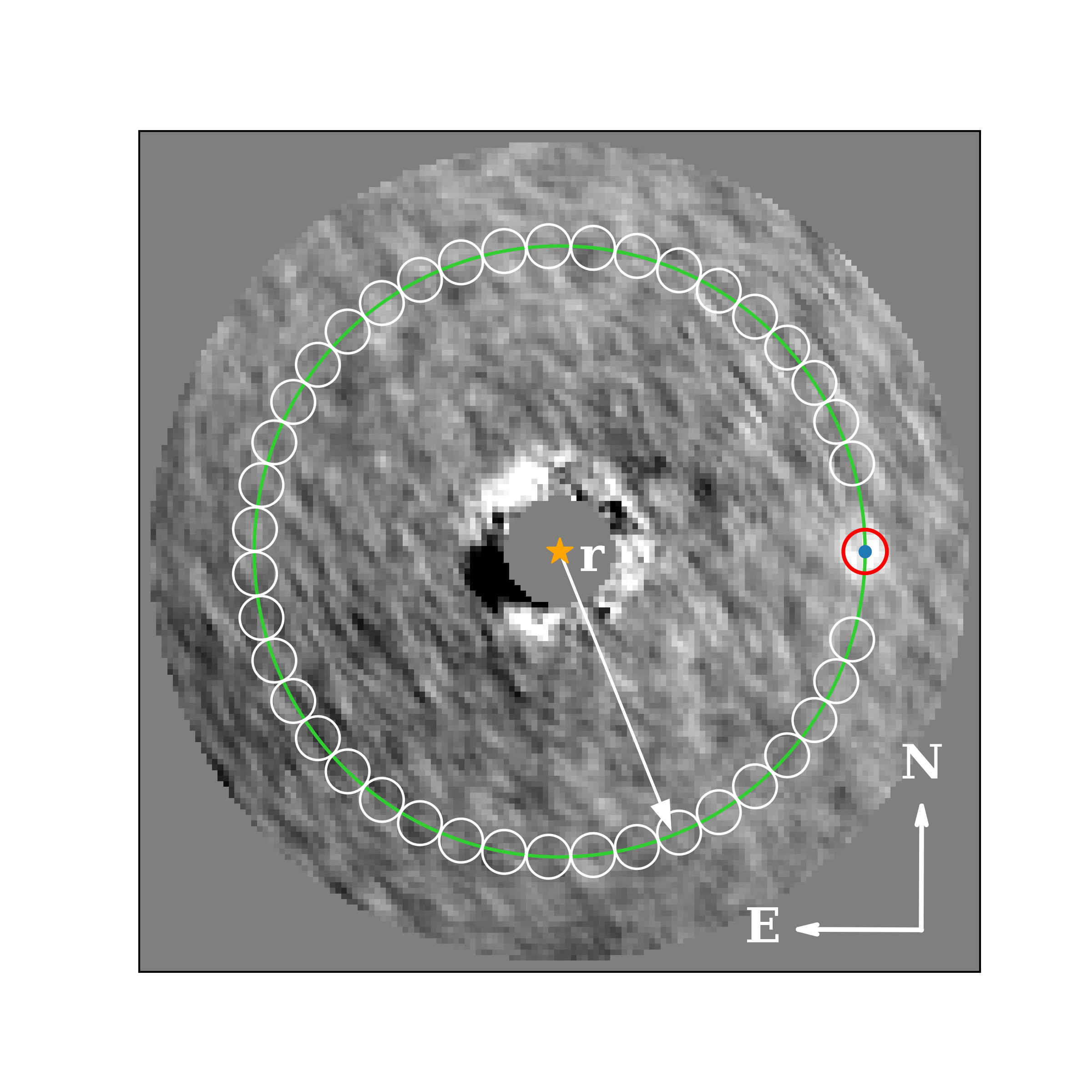}
\caption{\small{Illustration of our method for determining signal to noise ratio (S/N) based on \cite{mawet_fundamental_2014}.  The image is a post-BDI reduction of HD~82984~A with a fake signal injected right at the 5-$\sigma$ S/N limit at separation $r~=~7\lambda$/D and position angle 270$^{\circ}$ East of North. HD~82984~A is behind the mask and marked with an orange star. The sum of the pixels within the red aperture, with diameter~=~1$\lambda$/D, is $\bar x_1$ in Eqn \ref{eq:snr}; the mean and standard deviation of the sum of the pixels in the white apertures are $\bar x_2$ and $s_2$ respectively; $n_2$ is the number of white apertures.  This computation was repeated for all N = 2$\pi$r apertures along the ring of radius $r$, and for rings of radius $r~=~n \lambda$/D, where $n$ is an integer.
}}
\label{fig:SNR}
\end{figure}

To quantify achievable contrast limits for each system, we performed injection-recovery of synthetic ``planet" signals and determined the contrast at which injected signals can be recovered at the 5-$\sigma$ level. We produced the synthetic signal by scaling the star's image to a specified contrast,
injected the synthetic signals into Star A's postage stamp cube, then performed KLIP reduction using Star B as above, and measured the signal-to-noise ratio (S/N) of the resulting signal (repeating for Star B using Star A as basis).  To measure the S/N of recovered signals, we implemented the methodology of \cite{mawet_fundamental_2014} for small number statistics induced by close separations.  To summarize briefly, we injected a synthetic planet signal of a known contrast at a specific position angle and a separation $r = n\, \lambda/D$, where $n$ is an integer.  Figure \ref{fig:SNR} illustrates the S/N calculation for a synthetic signal injected at S/N~=~5 to the HD~82984~A dataset.  At separation $r$ (green circle) there are N = 2$\pi r$ resolution elements of size $\lambda$/D, the characteristic scale of speckle noise.  We defined a resolution element centered at the injected signal (Figure \ref{fig:SNR} red aperture) and in N-3 resolution elements at that radius (Figure \ref{fig:SNR} white apertures), neglecting those immediately to either side to avoid the wings of the injected PSF.  Then, using Eqn (9) of \cite{mawet_fundamental_2014}, which is simply the Student's two-sample t-test, we have
\begin{equation}\label{eq:snr}
    p(x,n2) = \frac{\bar x_1 - \bar x_2}{s_2 \sqrt{1 + \frac{1}{n_2}}}
\end{equation}
where $\bar x_1 = \Sigma$(pixels in red aperture), $\bar x_2 =$ mean[$\Sigma$(pixels in white apertures)], and $s_2 = $ stdev[$\Sigma$(pixels in white apertures)], n$_2$ = N-3, and S/N = p.  This calculation was repeated for signals injected in all N resolution elements in the ring at radius $r$, and we took the mean S/N value as the S/N for that specified radius and contrast.  We computed S/N for all $n\, \lambda/D$ radii from $r = 1.7\, \lambda/D$ (0.2\arcsec) to the outer extent of the postage stamp (indicated in Table \ref{tab:obs-summary}) and for various contrast values and interpolated the 5-$\sigma$ contrast limit.

For each system we determined an apparent L$^{\prime}$ or [3.95], as appropriate to the observation, magnitude for the primary star by retrieving the WISE W1 ($\lambda_{\rm{central}}$ = 3.35$\mu$m) and W2 ($\lambda_{\rm{central}}$ = 4.6$\mu$m) and interpolating an apparent magnitude at L$^{\prime}$ or [3.95] using spectral type models from CALSPEC (HST flux standard spectra, \citealt{Bohlin2014CALSPECRef}).  We converted the apparent 5-$\sigma$ contrast limits to absolute magnitudes using the distances in Table \ref{tab:system-summary}.  We determined an age for each system from literature, and used the age and contrast limit absolute magnitude as constraints to interpolate a mass from evolutionary models.  For mass limits in the stellar regime, we used isochrones from the \citetalias{Baraffe2015BHAC} evolutionary models; for substellar regime, we used the \cite{Marley2021SonoraBobcat} evolutionary models.  For observations in [3.95] filter, we re-interpreted for the [3.95] filter in Clio by computing synthetic photometry for each isochrone point under the assumption of a 2.3 mm PWV atmospheric transmission model \citep[ATRAN,][]{atran} and airmass of 1.0.  As noted in Section \ref{sec:systems}, we were unable to determine a literature age for two systems, HIP~67506 and HD~151771, and used BHAC15 and SYCLIST isochrones respectively to interpolate an estimated age, which we then used with BHAC15 to convert contrast limits to mass estimates in the same manner.

\begin{figure*}[!t]
\centering
\includegraphics[width=0.8\textwidth]{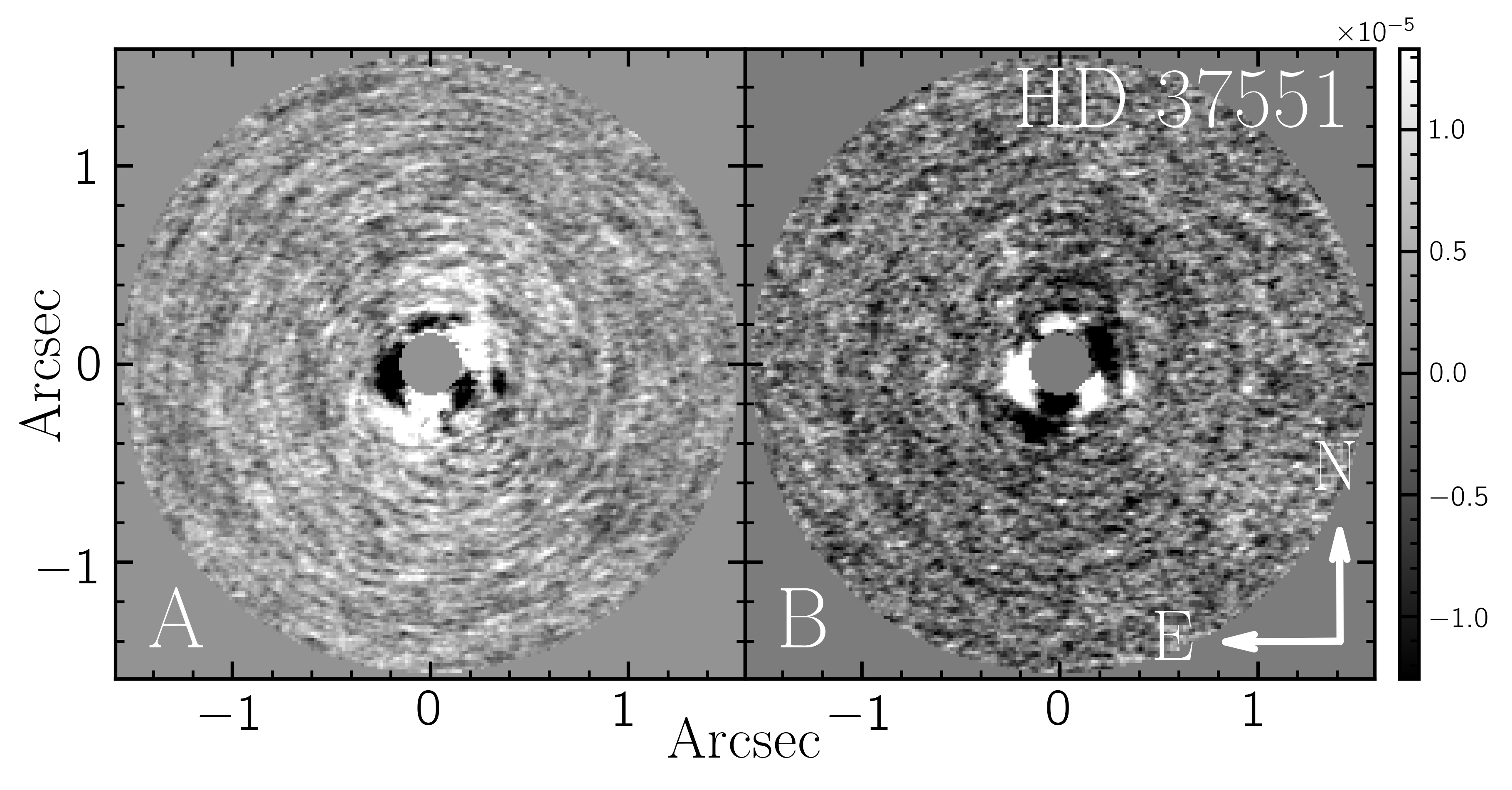}\\
\includegraphics[width=0.85\textwidth]{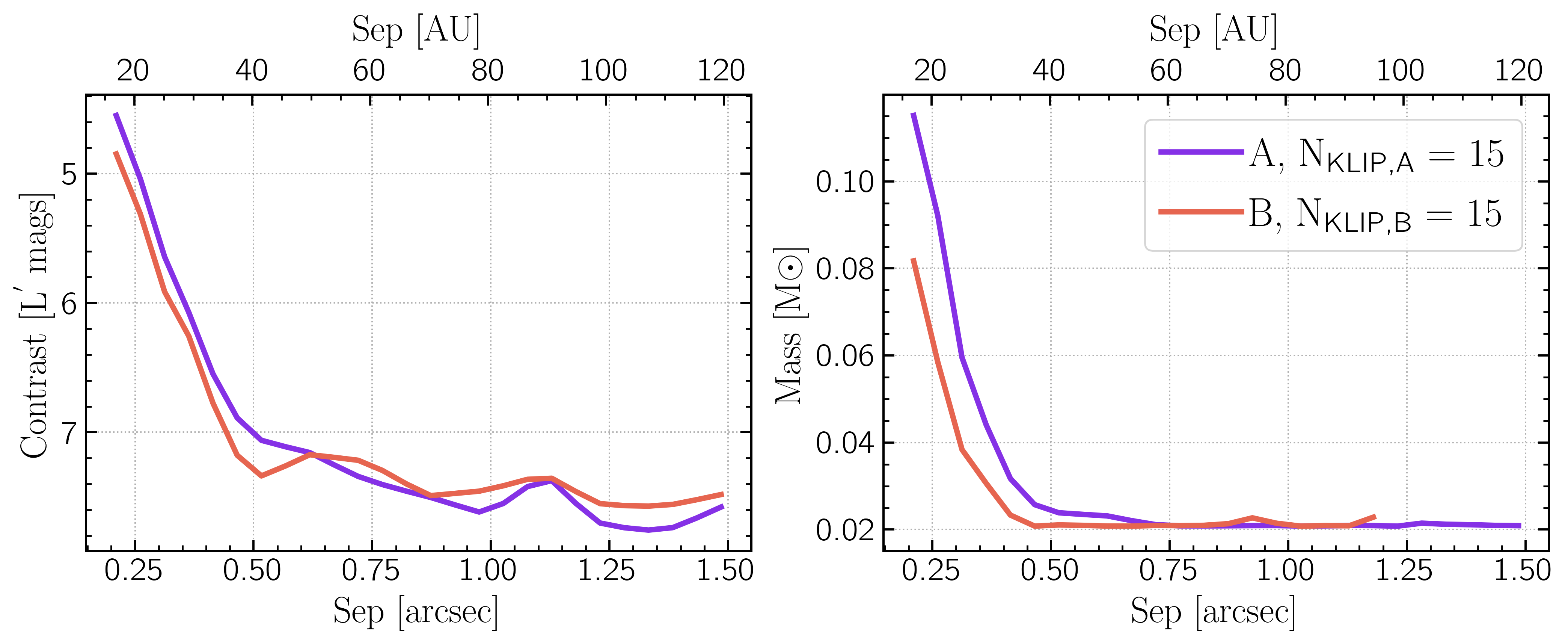}\\
\includegraphics[width=0.85\textwidth]{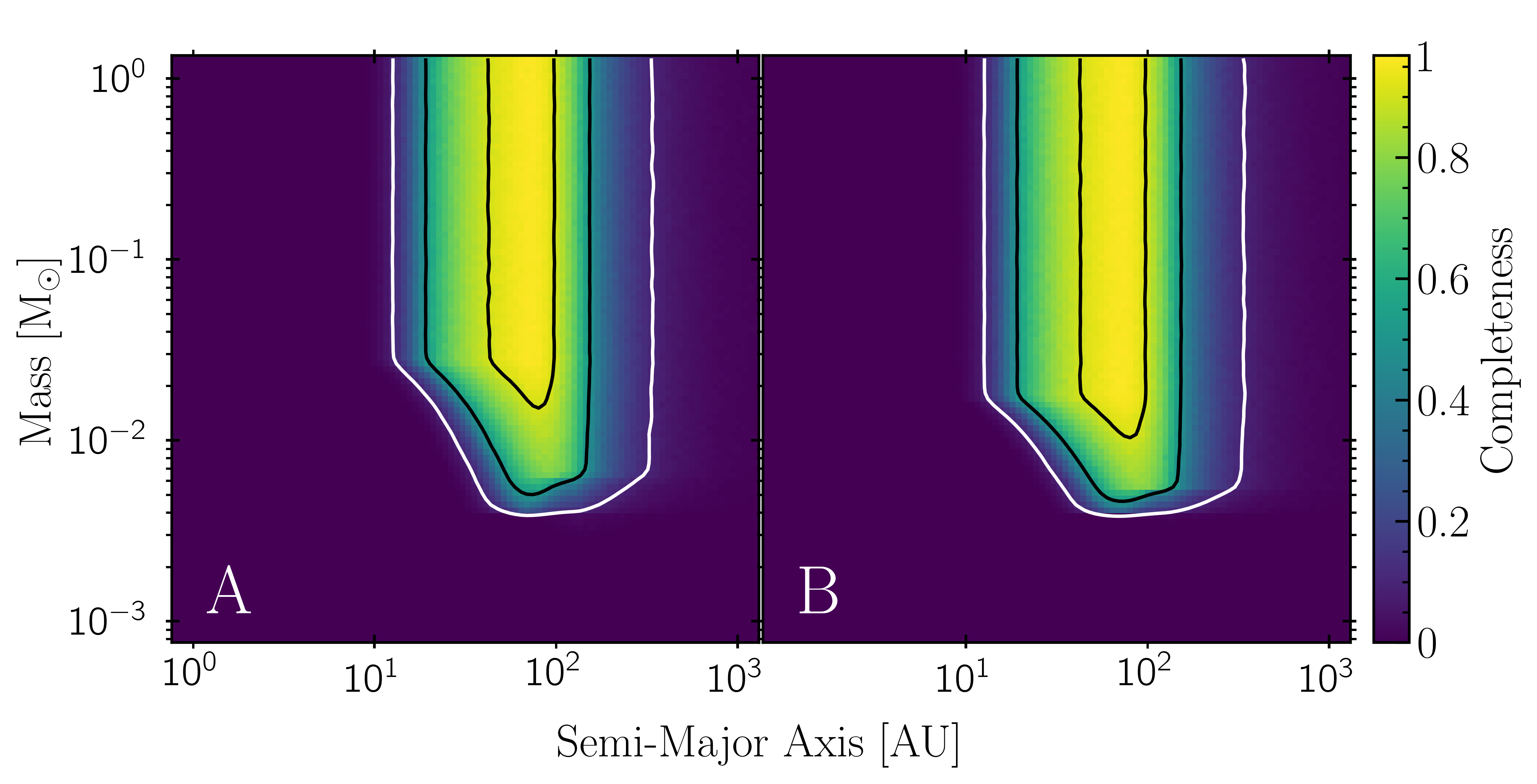}\\
\caption{Results of BDI-KLIP reduction of HD~37551. Top: reduced image of HD~37551~A reduced with HD~37551~B as reference (left) and vice versa (right) using 15 KLIP modes. North is up and East is to the left in both images. Middle: Contrast limits (left) and mass limits (right) as a function of separation for HD~37551~A (purple) and HD~37551~B (red).  These are the deepest contrast and mass limits in our survey, reaching as low as 5\Mjup\ in to 40 AU for both A and B.  Bottom: Survey completeness maps for both stars for a grid of (semi-major axis, mass) pairs.  Color map indicates fraction of simulated companions which would have been detected at each grid point.  \changed{Contours indicate 10\% (white), 50\%, and 90\% (inner-most contour) of simulated companions detected.}}
\label{fig:HD37551}
\end{figure*}

\movetabledown=0.7in
\begin{longrotatetable}
\begin{deluxetable*}{cccccccccccrcccl}
\tablecaption{Summary of Observations and Results by Binary System\label{tab:obs-summary}}
\tablewidth{0pt}
\tablehead{
\colhead{System}  & \colhead{Obs.} & \colhead{Obs.} & \colhead{N$_{\rm{images}}$} & \colhead{Filter} &
\colhead{T$_{\rm{int}}$} & \colhead{N$_{\rm{coadds}}$} & \colhead{Obs.} & \colhead{Binary} & \colhead{Inner} &
\colhead{Outer} & \colhead{} & \colhead{N$_{\rm{KLIP}}$} & \colhead{Best} & \colhead{Mass} & \colhead{}\\
\colhead{} & \colhead{Date} &\colhead{Mode\tablenotemark{$*$}} & \colhead{} & \colhead{} & \colhead{(s)} & \colhead{} &
\colhead{$\Delta$Mag} & \colhead{Sep (\arcsec)} & \colhead{Sep\tablenotemark{$\dagger$}} & \colhead{Sep} & \colhead{} & \colhead{Modes} & \colhead{Contrast} & \colhead{Limit} & \colhead{}\\
\colhead{} & \colhead{} & \colhead{} & \colhead{} & \colhead{} & \colhead{} & \colhead{} &
\colhead{} & \colhead{PA (deg)} & \colhead{(AU)} & \colhead{} & 
\colhead{} & \colhead{} & \colhead{($\Delta$mag)} & \colhead{(M$_\odot$)}\\
}
\startdata
HD 36705 & 2017-02-18 & ABBA & 28 & [3.95] & 5 & 6 & 2.0 & 8.881$\pm$0.003\arcsec & 4 & 1.5\arcsec & A & 20 & 5.3 & 0.07 & at 0.9\arcsec, 10 AU \\
\cline{12-16}
 & & & & & & & & 347.57$\pm$0.05$^{\circ}$ & & 22 AU & B & 20 & 6.8 & 0.01 & at 1.5\arcsec, 20 AU\\
\hline
HD 37551 & 2014-11-26 & Sky & 167 & [3.95] & 5 & 2 & 0.4 & 4.0165$\pm$0.0006\arcsec & 17 & 1.5\arcsec & A & 15 & 7.8 & \changed{0.02} & at 1.3\arcsec, 110 AU  \\
\cline{12-16}
 & & & & & & & & 115.167$\pm$0.007$^{\circ}$ & & 120 AU & B & 15 & 7.5 & \changed{0.02} & at 1.3\arcsec, 110 AU\\
\hline
HD 47787 & 2017-02-11 & ABBA & 10 & [3.95] & 3 & 20 & 0.1 & 2.178$\pm$0.003\arcsec& 10 & 1.6\arcsec & A & 9 & 6.0 & 0.01 & at 1.0\arcsec, 50 AU\\
\cline{12-16}
 & & & & & & & & 201.46$\pm$0.07$^{\circ}$ & & 72 AU & B & 9 & 5.75 & 0.01 & at 1.1\arcsec, 50 AU\\
 \hline
HD 76534 & 2017-02-18 & ABBA & 63 & [3.95] & 5 & 6 & 2.1 & 2.0794$\pm$0.0009\arcsec & 190 & 1.2\arcsec & A & 10 & 5.0 & 0.39\tablenotemark{$**$} & at 1.0\arcsec, 860 AU \\
\cline{12-16}
 & & & & & & & & 303.94$\pm$0.04$^{\circ}$ & & 970 AU & B & 10 & 3.0 & 0.39\tablenotemark{$**$} & at 1.0\arcsec, 860 AU \\
 \hline
HD 82984 & 2015-05-29 & ABBA & 16 & [3.95] & 4 & 5 & 0.7 & 2.018$\pm$0.006\arcsec & 60 & 1.2\arcsec & A & 15 & 6.5 & 0.47 & at 0.5\arcsec, 140 AU\\
\cline{12-16}
 & &  &  & &  &  & &  220.69$\pm$0.03$^{\circ}$ &  & 300 AU & B & 15 & 6.5 & 0.31 & at 0.8\arcsec, 200 AU\\
 \hline
HD 104231 & 2017-02-18 & Sky & 49 & [3.95] & 5 & 6 & 1.7 & 4.479$\pm$0.001\arcsec & 22 & 2.4\arcsec & A & 30 & 5.5 & 0.04 & at 1.9\arcsec, 1.9 AU \\
\cline{12-16}
 & & & & & & & &161.45$\pm$0.01$^{\circ}$ & & 240 AU & B & 30 & 6.0 & 0.009 & at 1.1\arcsec, 110AU \\
 \hline
HD 118072 & 2015-05-29 & Sky & 10 & [3.95] & 3 & 5 & 0.06 & 2.28$\pm$0.03\arcsec & 18 & 1.6\arcsec & A & 7 & 5.0 & 0.06 & at 1.0\arcsec, 80 AU\\
\cline{12-16}
 & & & & & & & & 80.7$\pm$0.9$^{\circ}$& & 120 AU & B & 15 & 6.3 & 0.02 & at 1.3\arcsec, 100 AU \\
 \cline{2-15}
 & 2017-02-19  &  Sky & 35 & [3.95] & 5 & 6 & 0.06 & 2.290$\pm$0.002\arcsec& 18 & 1.6\arcsec & A &  30 & 4.0 & 0.12 & at 1.1\arcsec, 90 AU\\
\cline{12-16}
 & &  & & & & & &79.49$\pm$0.05$^{\circ}$& & 120 AU & B & 10 & 4.0 & 0.13 & at 0.4\arcsec, 30 AU \\
 \hline
HD 118991 & 2015-05-30 & ABBA & 16 & [3.95] & 4 & 5 & 0.8 & 2.29$\pm$0.03\arcsec & 20 & 1.2\arcsec & A & 15 & 5.5 & 0.40 & at 0.7\arcsec, 60 AU\\
\cline{12-16}
 & & & & & & & &162.91$\pm$0.01$^{\circ}$ & & 100 AU & B & 15 & 5.5 & 0.27 & at 0.6\arcsec, 55 AU\\
 \hline
HD 137727 & 2017-02-20 & ABBA & 18 & [3.95] & 3 & 10 & 1.0 & 2.2108$\pm$0.0006\arcsec\ & 25 & 1.2\arcsec & A & 17 & 4.5 & 0.09 & at 1.1\arcsec, 120 AU \\
\cline{12-16}
 & & & & & & & & 185.3$\pm$0.1$^{\circ}$ & & 150 AU & B & 17 & 4.5 & 0.04 & at 0.6\arcsec, 70 AU\\
\hline
HD 147553 & 2015-05-24 & Sky & 52 & [3.95] & 4 & 3 & 0.2 & 6.274$\pm$0.004\arcsec & 30 & 0.8\arcsec & A & 15 & 5.8 & 0.04 & at 0.5\arcsec, 70 AU  \\
\cline{12-16}
 & & & & & & & & 152.48$\pm$0.04$^{\circ}$ & & 110 AU & B & 3 & 6.3 & 0.02 & at 0.5\arcsec, 70 AU\\
\cline{2-15}
& 2015-06-02 & ABBA & 31 & [3.95] & 4 & 10 & 0.2 & 6.259$\pm$0.002\arcsec & 30 & 0.8\arcsec & A & 30 & 5.8 & 0.04 & at 0.8\arcsec, 100 AU \\
\cline{12-16}
&  & & & & & & & 152.478$\pm$0.008$^{\circ}$ & & 110 AU & B & 30 & 5.5 & 0.04 & at 0.7\arcsec, 100 AU\\
\hline
HD 151771 & 2017-09-05 & ABBA & 25 & [3.95] & 5 & 1 & 1.3 & 6.79$\pm$0.01\arcsec & 55 & 1.2\arcsec & A & 20 & 6.3 & 0.66 & at 0.6\arcsec, 170 AU\\
\cline{12-16}
&  & & & & & & & 4.8$\pm$0.2$^{\circ}$ & & 300 AU & B & 20 & 6.0 & 0.43 & at 0.6\arcsec, 170 AU \\
 \hline
HD 164249 & 2017-09-04 & ABBA & 54 & [3.95] & 3.5 & 1 & 2.3 & 6.547$\pm$0.007\arcsec & 10 & 1.2\arcsec & A & 30 & 4.5 & 0.08 & at 0.7\arcsec, 40 AU \\
\cline{12-16}
&  & & & & & & & 89.52$\pm$0.03$^{\circ}$ & & 75 AU & B & 30 & 3.0 & 0.04 & at 0.6\arcsec, 30 AU \\
 \hline
HD 201247 & 2015-05-24 & ABBA & 34 & [3.95] & 4 & 3 & 0.1 & 4.190$\pm$0.003\arcsec & 6 & 1.2\arcsec & A & 5 & 7.5 & 0.03 & at 0.9\arcsec, 30 AU\\
\cline{12-16}
&  & & & & & & & 132.31$\pm$0.02$^{\circ}$ & & 37 AU & B & 30 & 7.3 & 0.03 & at 0.9\arcsec, 30 AU\\ 
 \cline{2-15}
 & 2017-09-02 & ABBA & 30 & [3.95] & 5 & 6 & 0.1 & 4.216$\pm$0.003\arcsec & 6 & 1.6\arcsec & A & 3 & 5.5 & 0.09 & at 0.5\arcsec, 20 AU\\
\cline{12-16}
&  & & & & & & & 132.61$\pm$0.03$^{\circ}$ & & 50 AU & B & 5 & 6.0 & 0.06 & at 1.0\arcsec, 35 AU\\ 
 \hline
HD 222259 & 2015-06-03 & Sky & 25 & [3.95] & 4 & 10 & 0.4 & 5.388$\pm$0.002\arcsec & 10 & 1.8\arcsec & A & 24 & 6.5 & 0.01 & at 1.1\arcsec, 50 AU\\
\cline{12-16}
&  &  & & & & & & 347.76$\pm$0.01$^{\circ}$ & & 82 AU & B & 24 & 6.5 & 0.01 & at 1.1\arcsec, 50 AU \\
 \cline{2-15}
 & 2017-09-05 & ABBA & 32 & [3.95] & 5 & 1 & 0.4 & 5.391$\pm$0.005\arcsec & 10 & 0.8\arcsec & A &  10 & 5.0 & 0.04 & at 0.5\arcsec, 20 AU \\
\cline{12-16}
&  &  & & & & & & 347.82$\pm$0.04$^{\circ}$ & & 33 AU & B & 10 & 5.0 & 0.03 & at 0.5\arcsec, 20 AU \\
\hline
HIP 67506 & 2015-05-31 & Sky & 44 & MKO L$^\prime$ & 0.6 & 20 & 0.4 & 9.424$\pm$0.004\arcsec & 50 & 1.43\arcsec & & 30 & 6.5 & 0.03 & at 0.8\arcsec, 90 AU \\
\cline{12-16}
\changed{TYC 7797-34-2} & & & & & & & & 326.92$\pm$0.02$^{\circ}$ & & 140 AU & & 30 & 6.5 & 0.02 & at 0.8\arcsec, \changed{1500 AU}\\
& & & & & & & & & & \changed{ / 2600 AU} \\
 \hline
TWA 13 & 2015-05-23 & ABBA & 27 & MKO L$^\prime$ & 0.5 & 200 & 0.05 & 5.080$\pm$0.007\arcsec & 15 & 1.6\arcsec & A & 10 & 4.8 & 0.02 & at 0.5\arcsec, 30 AU\\
\cline{12-16}
& &  & & & & & & 327.27$\pm$0.09$^{\circ}$ & & 90 AU & B & 25 & 4.8 & 0.02 & at 0.5\arcsec, 30 AU\\ 
 \hline
2MASS J01535076- & 2017-09-05 & ABBA & 132 & [3.95] & 4 & 1 & 0.1 &  2.875$\pm$0.006\arcsec & 8 & 1.6\arcsec & A & 10 & 6.0 & 0.01  & at 1.2\arcsec, 40 AU\\
\cline{12-16}
 1459503 & & & & & & & &291.1$\pm$0.1$^{\circ}$ & & 50 AU & B & 3 & 5.8 & 0.01 & at 0.6\arcsec, 20 AU \\
 \hline
\enddata
\tablenotetext{*}{Two observing modes: ABBA - two nods, A and B, 10 frames each, alternating in an ABBA pattern throughout observation; Sky: one nod position and a set of ``sky" frames without stars in the frame.}
\tablenotetext{**}{The age of HD 76534 is $<$ 1Myr below the age range of most isochrone models.}
\tablenotetext{\dagger}{All systems used inner working angle of 1.7 $\lambda$D$^{-1}$ = 0.2\arcsec}
\tablecomments{\changed{N$_{\rm{images}}$ is the number of images used in the KLIP reduction; N$_{\rm{coadds}}$ is the number of coadded frames per image; Observed $\Delta$ Mag is the contrast in magnitudes measured in our survey between components A and B; Binary separation (sep) and position angle (PA) are the mean and standard deviation of position measurements of images in each dataset; Inner/Outer sep are the inner and outer radius of working mask for BDI reduction; N$_{\rm{KLIP}}$ Modes records the number of basis modes used in the KLIP reduction of the star.}
}
\end{deluxetable*}
\end{longrotatetable}

\section{Results}\label{sec:results}

We report in Table \ref{tab:obs-summary} a summary of the deepest contrast achieved for each binary system as a function of number of KLIP modes (N$_{\rm{KLIP}}$) and separation in arcseconds and AU.  Contrast is reported in units of $\Delta$log$_{10}$(flux) between injected companion signal and host star at the 5-$\sigma$ level, with corresponding mass in \Msun.  Figure \ref{fig:HD37551} displays the results of our pipeline for HD~37551.  Figure \ref{fig:HD37551} (top) shows the reduced images of HD~37551~A (left) and B (right), with the inner 1~$\lambda$/D and outer ring masked. Figure \ref{fig:HD37551} (bottom) shows the 5-$\sigma$ flux contrast limits (left) and mass limits (right) for A (purple) and B (red) as a function of separation in AU and arcseconds.  Similar plots for all stars in our survey are included in the supplementary material and are available online.

\subsection{Factors affecting contrast \changed{and mass} limits}

\textit{Variable conditions.} We found that variable conditions during the observations dramatically affected achievable contrast.  Similarly, bad pixels, poor pixel correction, a high background level relative to star peak also decreased achievable contrast.  We found that limiting the datasets to only the very best quality images achieved deeper contrast limits compared to having more lower quality images in the basis set.  For each dataset we inspected by eye and retained only the sharpest images.  In Table \ref{tab:obs-summary} we report the number of images used in the final reduction for each dataset \changed{(N$_{\rm{images}}$)}.  The varying levels of contrast achieved from dataset to dataset is mostly a function of the image quality of that particular observation; i.e. the highest Strehl images achieved the deepest contrast limits.  

\begin{figure}
\centering
\includegraphics[width=0.48\textwidth]{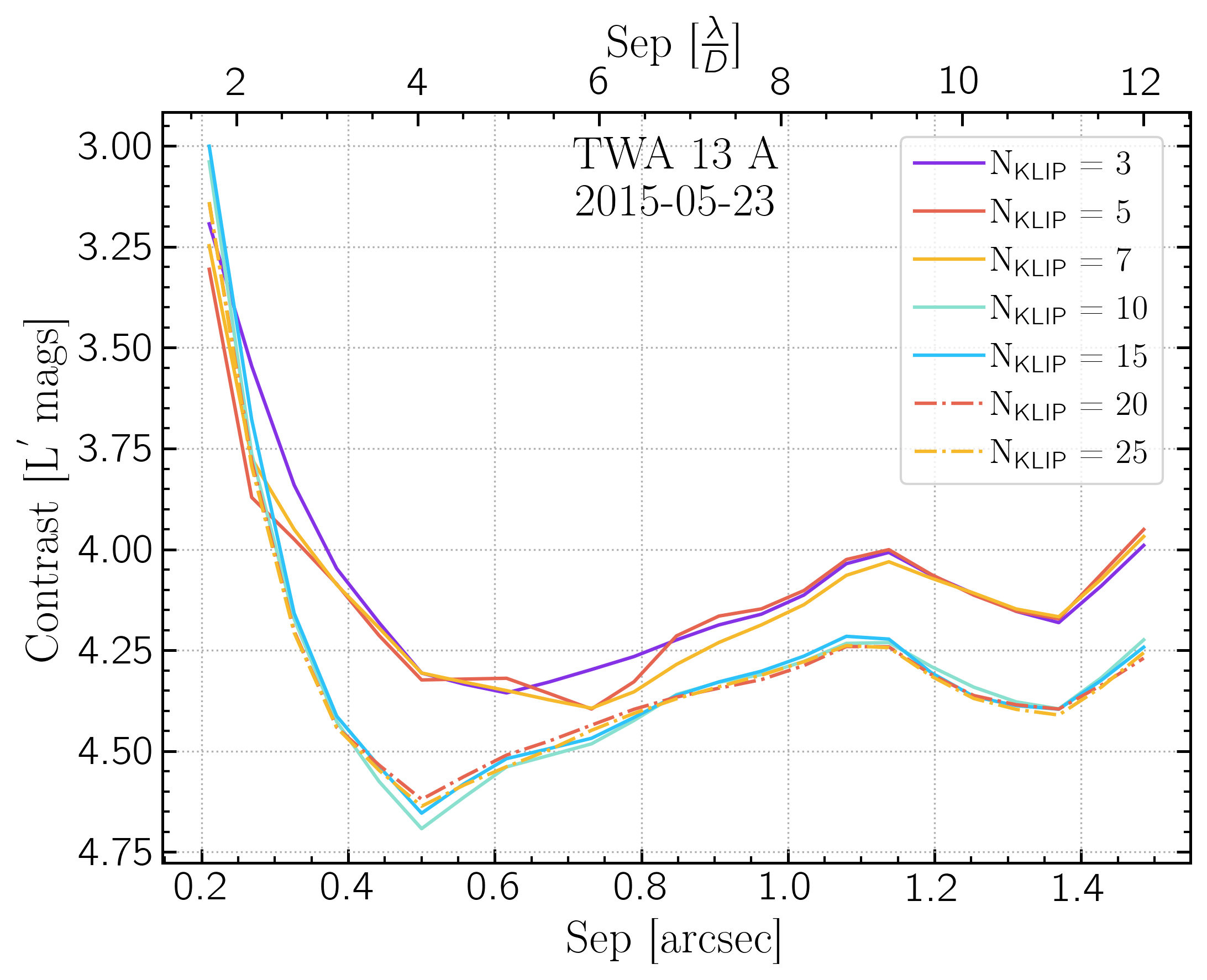}\\
\caption{Contrast for TWA~13~A as a function of number of basis modes (N$_{\rm{KLIP}}$). The deepest contrast is achieved at N$_{\rm{KLIP}}$ = 10 for this system.  Optimal number of basis modes varies between datasets and is reported in Table \ref{tab:obs-summary}.}
\label{fig:example KLIP}
\end{figure}

\textit{Number of KLIP basis modes.} We also found that the optimal number of KLIP modes to obtain the deepest contrast varied between datasets. Figure \ref{fig:example KLIP} displays an example of contrast limits as a function of KLIP modes for TWA~13~A.  In this example, there is dramatic improvement in contrast for N$_{\rm{KLIP}}$~$>$~7, and the deepest contrast is achieved at N$_{\rm{KLIP}}$~=~10 at 0.5\arcsec\ (4~$\lambda$/D).  The optimal N$_{\rm{KLIP}}$ for each system is reported in Table \ref{tab:obs-summary}.

\textit{Binary contrast.} The binary stars' contrast, reported as $\Delta$Mag in Table \ref{tab:obs-summary}, also affected the depth of the companion search. The strength of BDI relies on achieving identical PSF signal-to-noise between reference and target star, which will vary inversely with flux ratio between the two stars. 
In our survey, achieved contrast was generally poorer for systems with higher binary contrast.  

\changed{\textit{Age.} Finally, the assumed age for the system affects the final mass limits we derived from our measured contrast limits.  As discussed in Sections \ref{sec:systems} and \ref{sec:cont limits}, we made use of literature ages to derive mass limits corresponding to our contrast limits for each system.  For limits in the substellar regime, this introduces some uncertainty that is not captured in the reported  mass limits, as luminosity in the infrared depends on age for substellar objects.  For some systems in our survey there were several discrepant ages in the literature; for others, there was no independent age for the system, and we assumed the average age of the associated moving group, which has a range of possible ages of members.  For systems with literature age estimates, they were typically derived from model fitting, which can vary with the assumptions underlying the model.  Details of the age we used for each system are described in Appendix \ref{appendixA}.  Substellar objects cool with age, so for two hypothetical objects with the same properties but different ages, the younger one will appear brighter than the older one in observations.  So if the actual age of our system were younger than the age we assumed, our contrast limits would correspond to lower mass limits, and vice versa.  In most cases, the effect on limits would be minimal.  For example, 2MASS J01535076-1459503 is a Beta Pictoris Moving Group member \citep{messina__2017}, and we adopted the moving group age of 25$\pm$3 Myr \citep{messina_rotation-lithium_2016}.  Computing corresponding mass limits for ages 2-$\sigma$ younger (19 Myr) and 2-$\sigma$ older (31 Myr) results in a difference of $\sim$0.05~\Mjup\ at the highest contrast.  However in some cases there are widely discrepant ages in literature, such as for AB Dor AB, which has age estimates spanning 5-240~Myr.  This results in a $\sim$15~\Mjup\ difference in the mass limits at the highest contrast between the youngest and oldest ages estimates.  Our reported mass limits and completeness estimates assume the age given in Table \ref{tab:system-summary} for each system, and the variation induced by differing ages in not captured in those limits.
}

\subsection{Notable system results}

\begin{figure}
\centering
\includegraphics[width=0.48\textwidth]{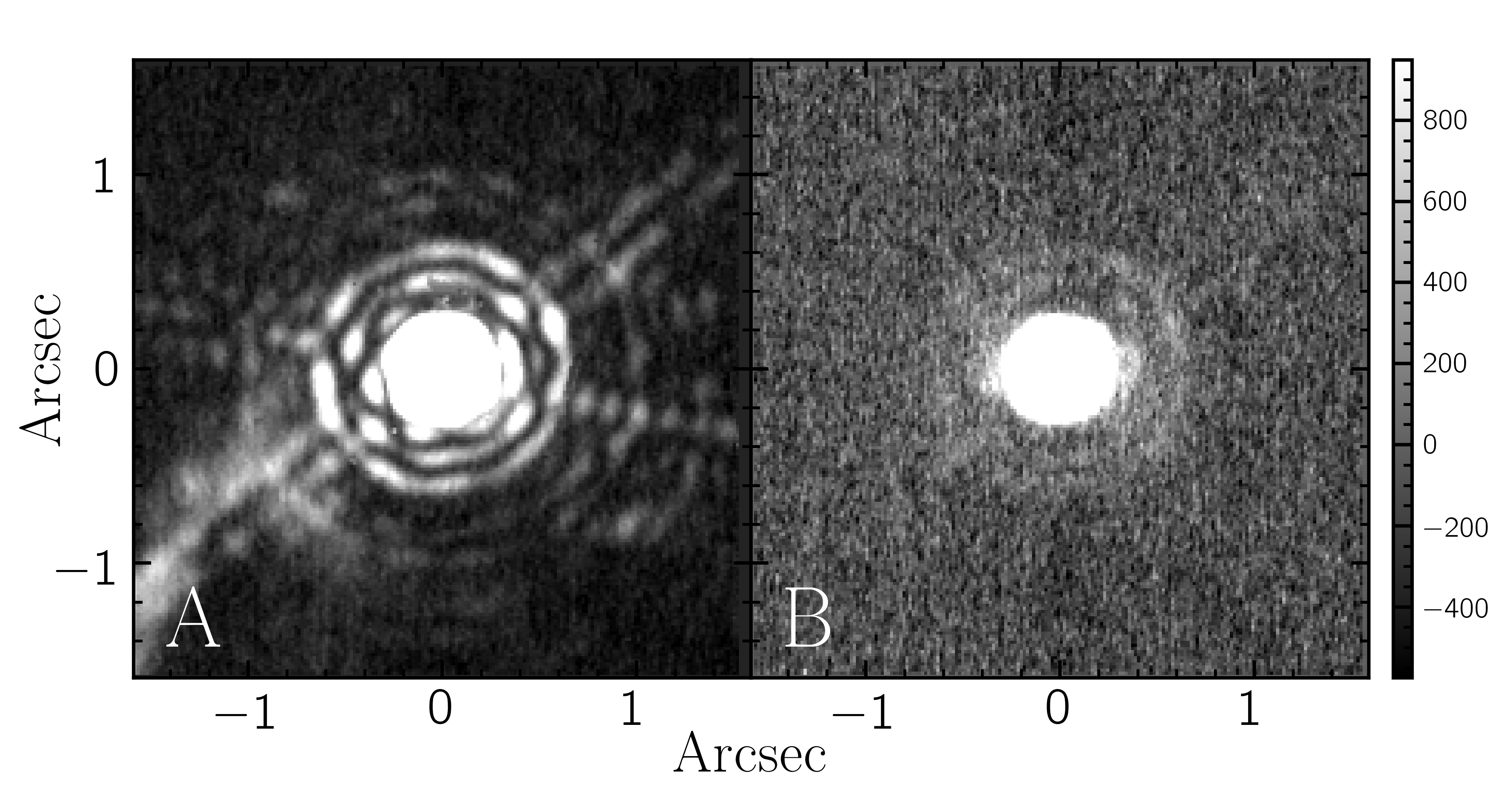}\\
\includegraphics[width=0.48\textwidth]{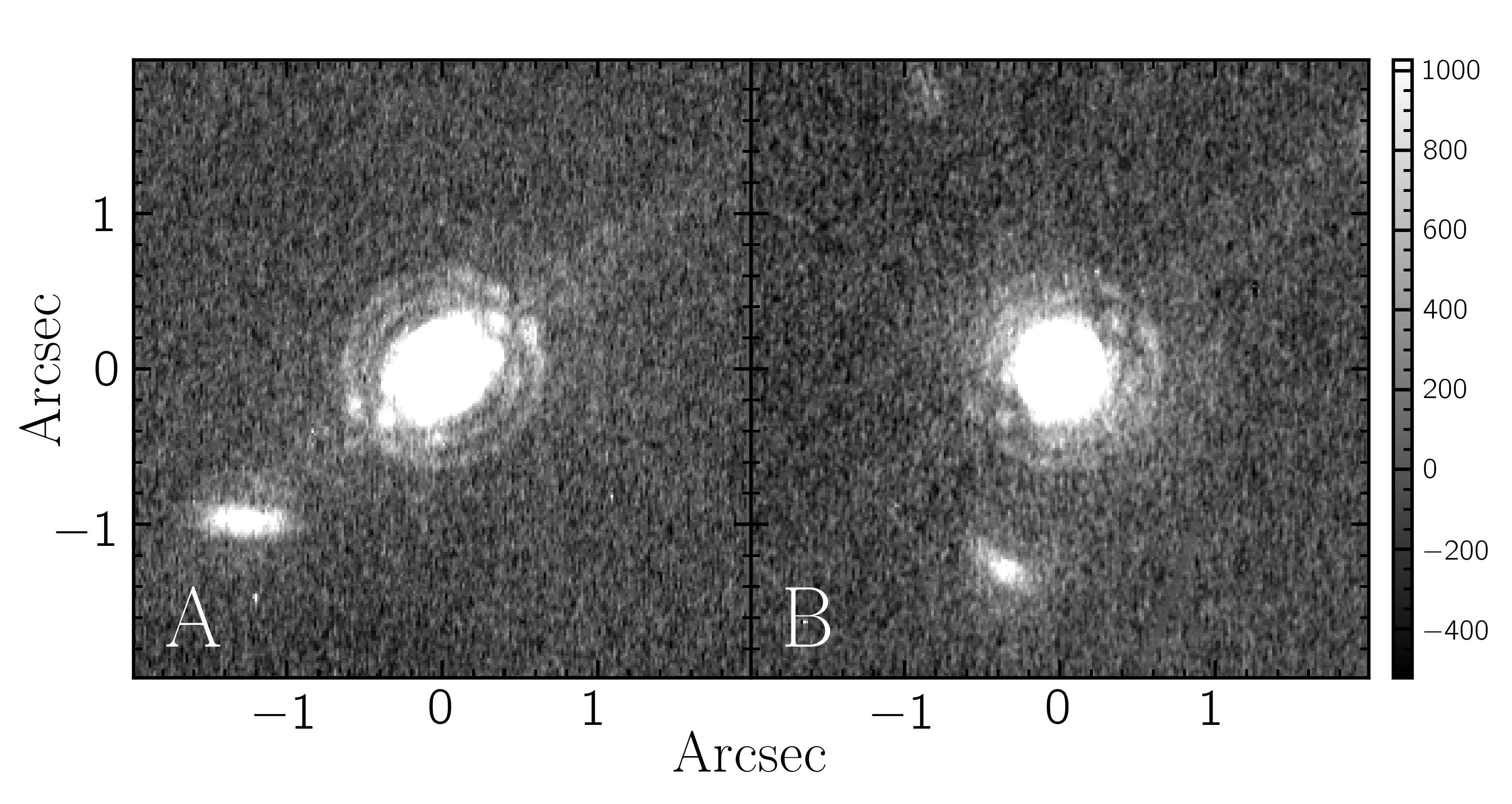}\\
\caption{Top: A selected image of HD~36705~A (left) and HD~36705~B (right) from our 2017 dataset, shown with a ZScale stretch to emphasize the faint PSF features, with gray scale showing pixel counts.  The image of the two stars appear significantly different due to their large relative contrast ($\Delta$mag~$\approx$~2) and the brightness of HD~36705~A in the MagAO [3.95] filter, with many features visible on A lost in the noise for B.  This resulted in contrast limits when used to perform BDI due to insufficient starlight subtraction.  Bottom: A selected image from the 2015 epoch HD~222259 observation, shown with a ZScale stretch.  The HD~222259~A PSF (left) contains a glint (bottom left corner), HD~222259~B (right) contains a different glint (bottom of frame); both PSF cores are elongated.  These features show up prominently in the BDI reduction.
}
\label{fig:ABDorPSF}
\end{figure}

Here we discuss some notable results of select binary systems in our sample.  The results for the remaining objects in our survey are available digitally\footnote{\url{https://github.com/logan-pearce/Pearce2022-BDI-Public-Data-Release}}.

\subsubsection{HD 37551 -- the deepest contrast}
HD~37551 achieved the deepest contrast limits in our sample \changed{ ($\Delta$[3.95]~=~7.8 and 7.6 magnitudes)} at the deepest points for A and B respectively).  This dataset also retained the highest number of high-quality images in the final BDI reduction, due to the stable seeing conditions and AO correction throughout the observation.  We did not identify any candidate companion signals in the reduced images.  Figure \ref{fig:HD37551} displays the reduced images and corresponding contrast and mass limits for HD~37551.

\begin{figure}
\centering
\includegraphics[width=0.49\textwidth]{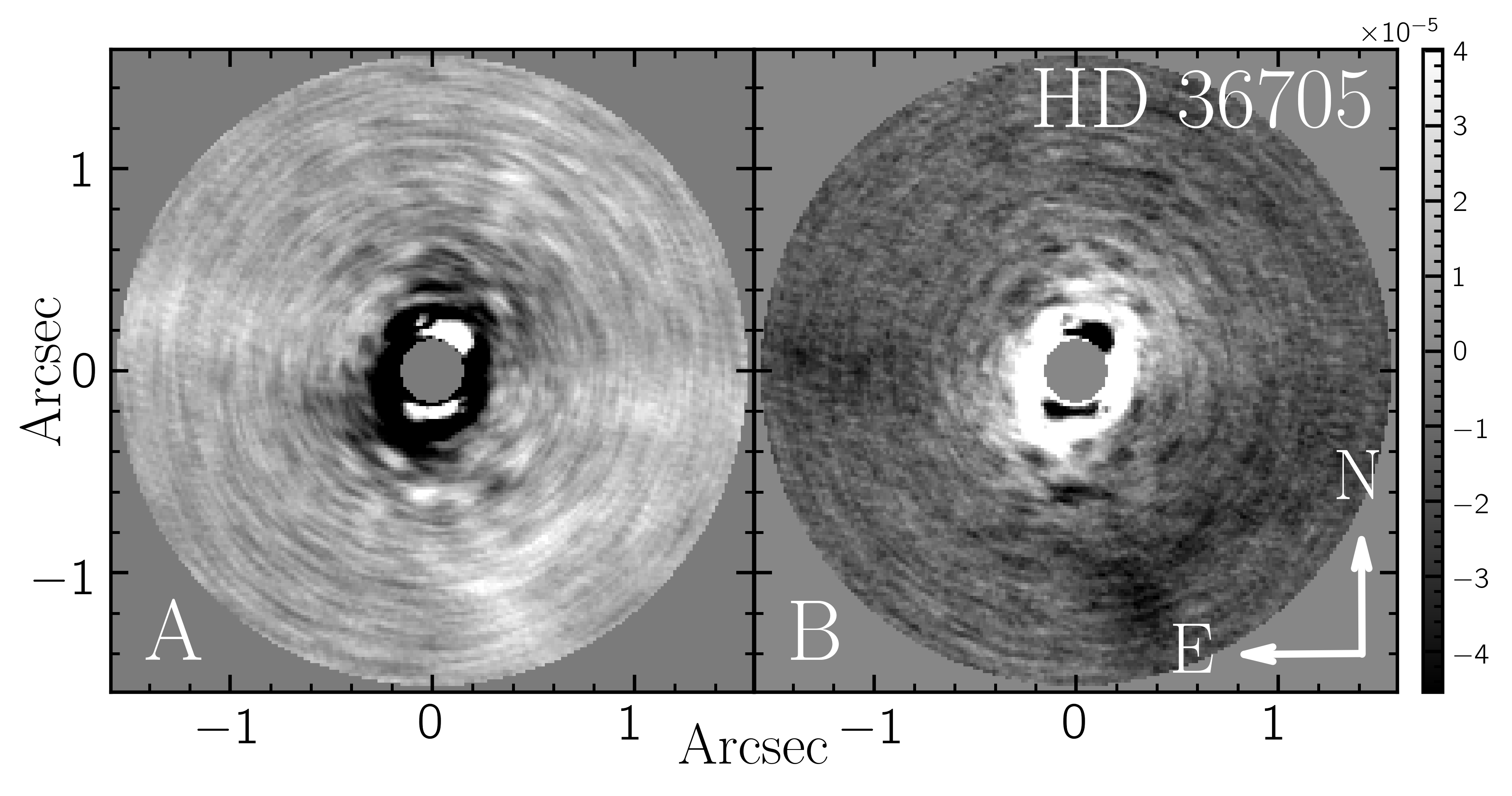}\\
\includegraphics[width=0.49\textwidth]{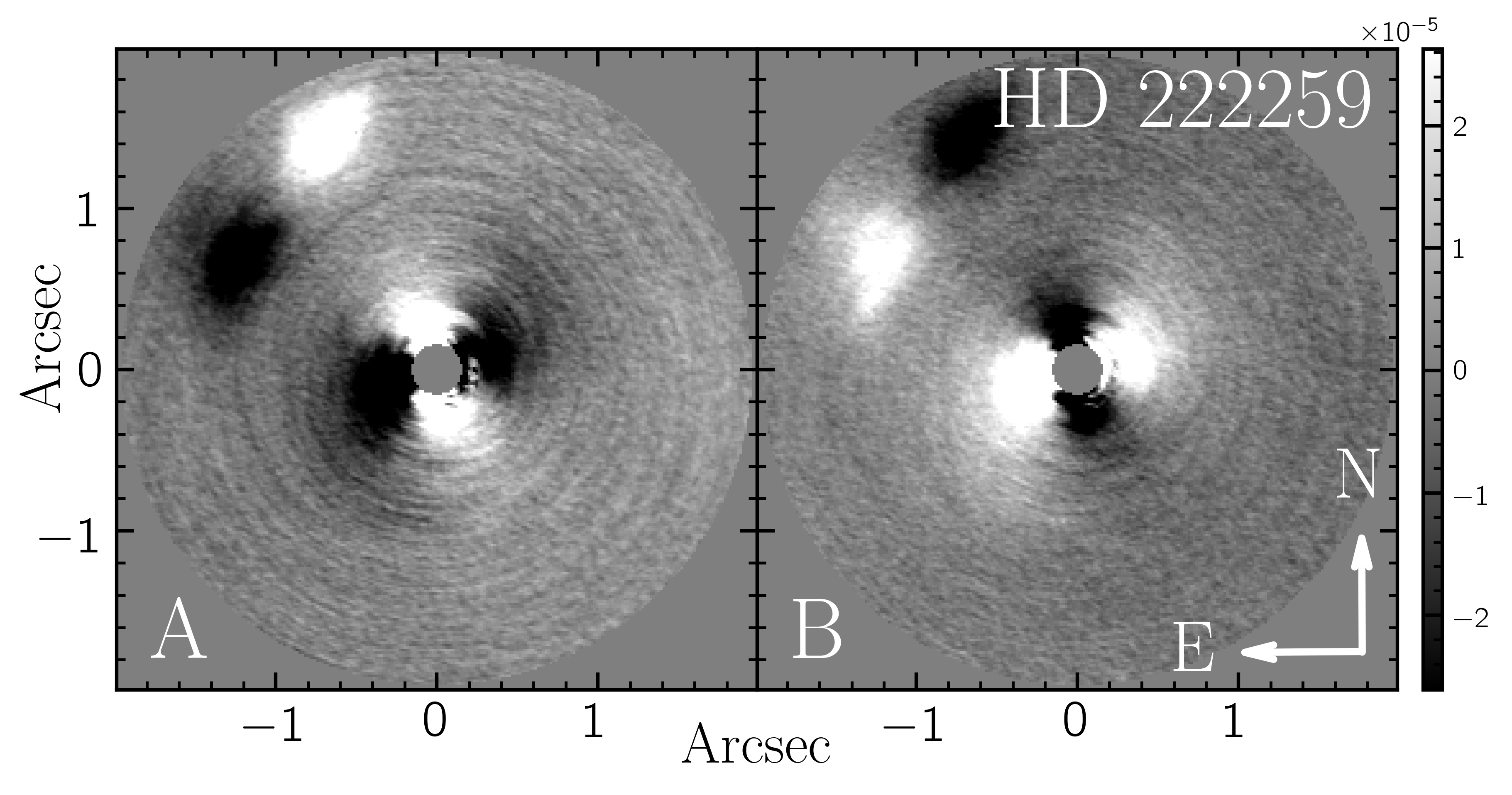}
\caption{Top: BDI reduction of HD~36705 using 20 KLIP modes.  North is up and East is to the left in both images.  HD~36705~A is significantly brighter than HD~36705~B, and PSF features visible in A are lost to noise in B, resulting in poor starlight subtraction and contrast limits, particularly for A.  Bottom: BDI reduction of HD~222259 using 24 KLIP modes.  North is up and East is to the left in both images. Regions of bright and dark pixels in the upper northeast corner are due to the ghosts visible in Figure \ref{fig:ABDorPSF}, as are bright and dark areas in the central regions.}
\label{ABDorResults}
\end{figure}

\subsubsection{HD 36705 -- the effect of binary contrast}\label{sec:ABDor contrast discussion}
HD~36705 is the most extreme case of the effect of binary contrast on the reduction in our sample.  HD~36705~A is a nearby (15 pc) bright (Gaia G mag = 6.7) K0V type star with an M5-6 binary companion.  We observed $\Delta$[3.95]$\approx$2.  Figure \ref{fig:ABDorPSF} displays a single image from the HD~36705 dataset with a ZScale stretch to emphasize faint PSF features.  

The image for HD~36705~A (left) is bright with many features apparent with strong signal-to-noise.  Several rings of the Airy pattern and a second set of diffraction spikes (oriented left-right) visible for A which are lost in the noise for B.  Figure \ref{ABDorResults} (top) shows the reduced image for HD~36705~A (left), in which both sets of diffraction spikes are visible in the residuals, showing incomplete starlight subtraction.  The resulting contrast limits are poor, especially for A, due to the residual starlight.  

As there was no infrared excess observed for this system (see Appendix \ref{appendixA}), we interpret the apparent ``fuzziness" of the residuals near the core of HD~36705~B to be due to incomplete starlight subtraction and not physical features.  We do not expect either of the known stellar companions to be visible in our reduction as they both have separations $<$0.2\arcsec.

\subsubsection{HD 222259 -- the effect of instrument ghosts}
The 2015 observation of HD~222259 contained an elongated PSF core shape due to residual vibrations from suboptimal tip/tilt gain setting, as well as different optical ghosts to the lower left of each star, shown in Figure \ref{fig:ABDorPSF} (bottom).  These features show up clearly in the BDI reduction in Figure \ref{ABDorResults} (bottom) as positive and negative valued areas to the northeast and around the center masked region in both reduced images, and degraded the achieved contrast.  Neither of these features are present in the 2017 epoch observation of HD~222259.

\begin{figure*}
\centering
\includegraphics[width=0.9\textwidth]{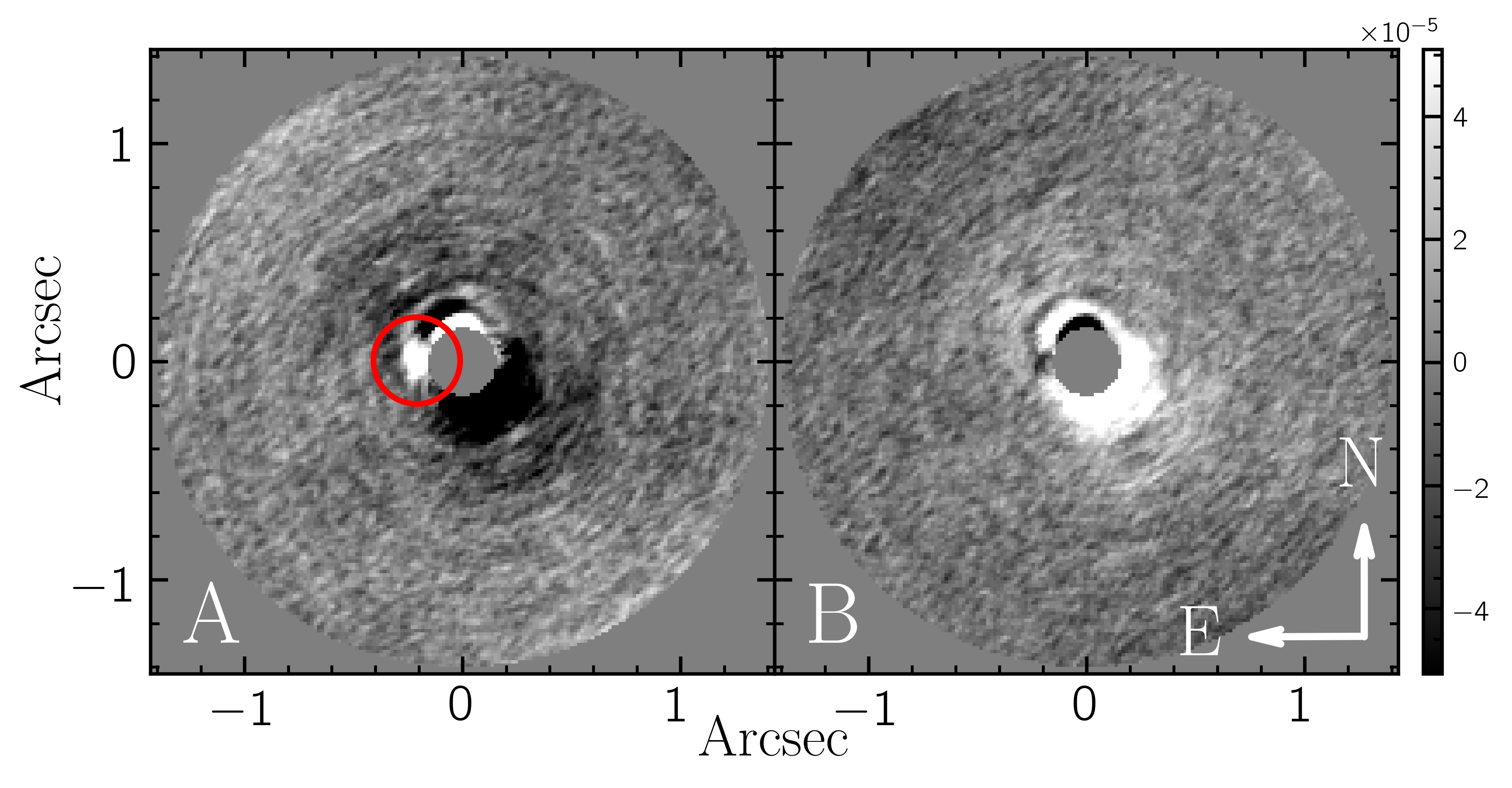}\\
\includegraphics[width=0.85\textwidth]{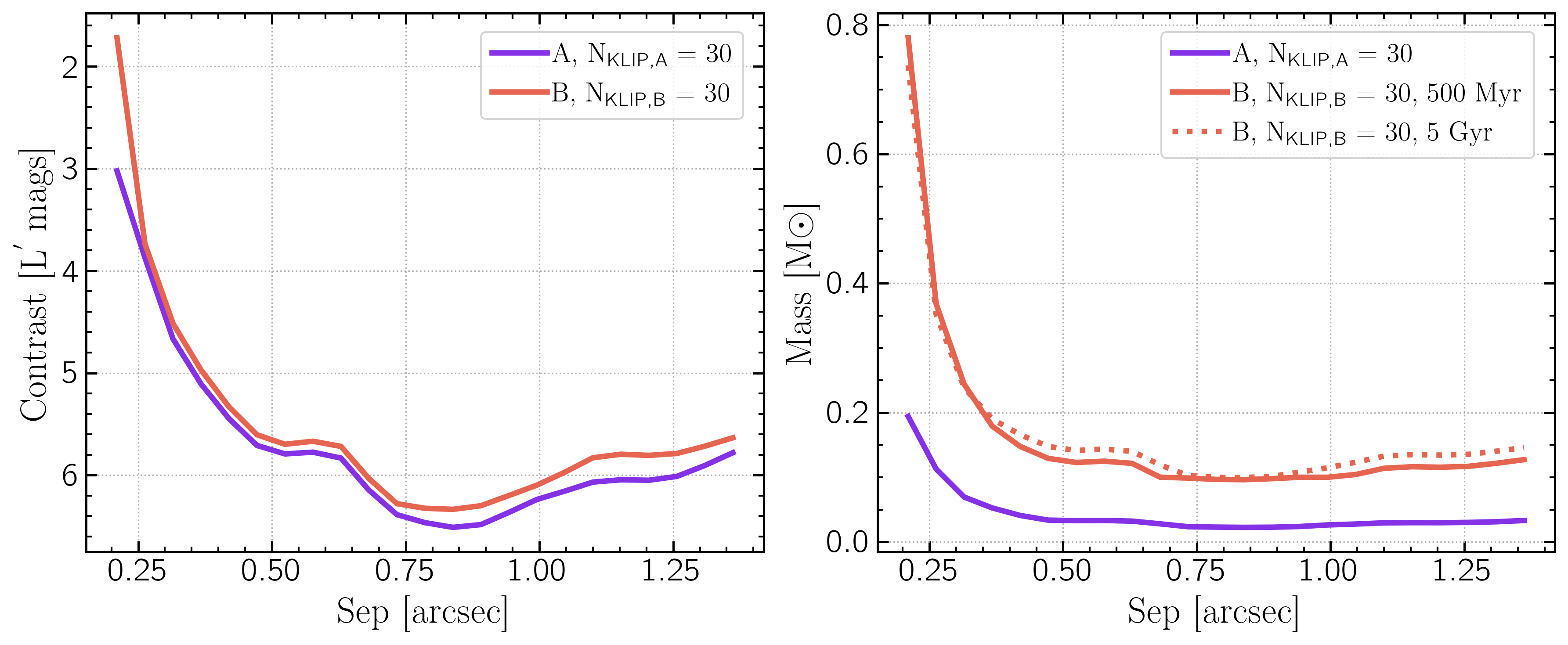}\\
\includegraphics[width=0.85\textwidth]{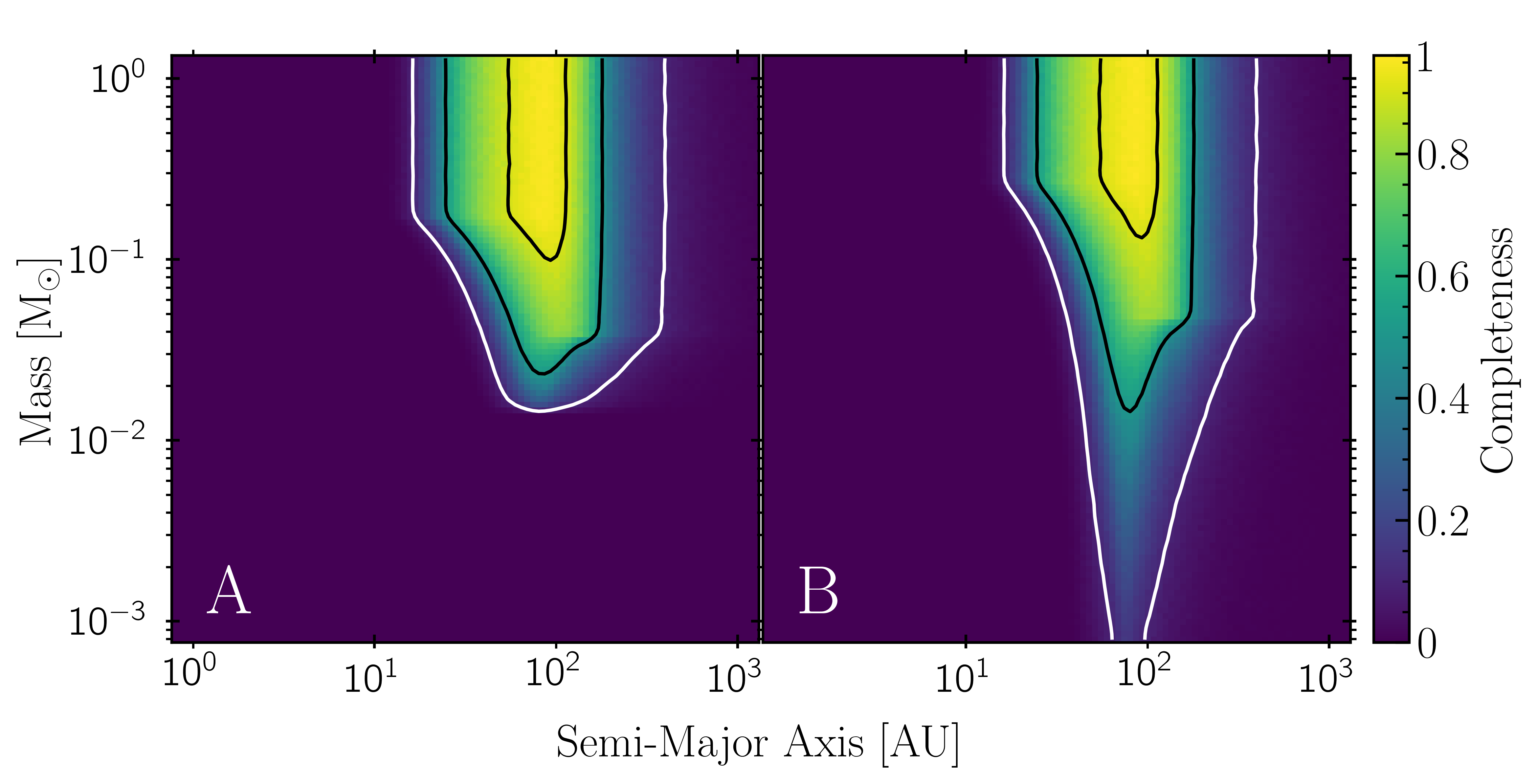}\\
\caption{Top: BDI reduction of HIP~67506 \changed{(labeled A) and TYC 7797-34-2 (labeled B)} using 30 KLIP modes.  North is up and East is to the left in both images.  The candidate companion signal is located $\sim$0.2\arcsec\ ($\sim$2$\lambda$/D) to the east of HIP~67506 (behind mask), indicated by the red circle.  The candidate signal rotated with the sky rotation, unlike the azimuthally broadened features at similar separation.  \changed{Middle: contrast curves for HIP 65706 and TYC 7797-34-2. \changed{We show mass limits for TYC 7797-34-2 using a young age (500 Myr) and a field age (5 Gyr)}.  Bottom: Completeness map for HIP 65706 and TYC 7797-34-2.  Contours show 10\%, 50\%, and 90\% completeness.}}
\label{fig:HIP67506candidatecompanion}
\end{figure*}

\begin{deluxetable}{cc}
\tablecaption{\changed{Gaia EDR3} Multiplicity Metrics for HIP 67506 \label{tab:BDI1350metrics}}
\tablewidth{0pt}
\tablehead{
\colhead{Metric} & \colhead{Value}
}
\startdata
RUWE & 2.02\\
\texttt{astrometric\_excess\_noise} & 0.22\\
\texttt{astrometric\_excess\_noise\_sig} & 75.2\\
\texttt{astrometric\_chi2\_al} & 2277.97\\
\texttt{ipd\_gof\_harmonic\_amplitude} & 0.0099\\
\texttt{ipd\_frac\_multi\_peak} & 0\\
\enddata
\end{deluxetable}

\subsubsection{HIP 67506 -- a candidate companion signal}

The BDI reduction of HIP~67506 contains a promising candidate companion signal, marked with a red circle in Figure \ref{fig:HIP67506candidatecompanion}, which shows the BDI reduction of HIP~67506 \changed{and TYC 7797-34-2 (labeled A and B)} in MKO L$^\prime$ reduced with 30 KLIP modes.  The candidate signal, located at separation $\approx$1.3 $\lambda$/D ($\approx$0.2\arcsec) and position angle~$\approx$~90$^\circ$, is more similar to a PSF shape than any other features in reduced images in our survey, although it is distorted due to its proximity to the star core.  The candidate signal rotated with the sky and did not smear azimuthally like the other features at similar separation.  

\textit{Other lines of evidence.} HIP~67506 has an elevated RUWE in EDR3 (RUWE~=~2.02; see Appendix \ref{appendixA}), indicating the possible presence of a companion unresolved in Gaia that caused it to deviate from the assumed single-star model \citep{lindegren_re-normalising_2018}.  RUWE has been shown to be highly sensitive to the presence of unresolved subsystems \citep{Stassun2021EclipsingBinaries,Penoyre2020BinaryDeviations,Belokurov2020UnresolvedBinariesDR2}.  Additionally, HIP~67506 has a statistically significant acceleration ($\chi^{2}$~=~24) in the Hipparcos-Gaia Catalog of Accelerating Stars (HGCA, \citealt{brandt_hipparcos-gaia_2018}).  \cite{kervella_stellar_2019} computed a statistically significant (S/N~=~5) proper motion anomaly in the Gaia DR2 epoch which could be caused by a $\sim$230~\Mjup\ object at the candidate signal separation of 18~AU (0.2\arcsec).  Similarly, the \cite{Kervella2022} PMa catalog for Gaia EDR3 astrometry measured a PMa which could be caused by a $\sim$200-300~\Mjup\ object at 18~AU.

While RUWE is the most complete and easy to interpret metric \citep{Lindegren2018RUWE}, other metrics in Gaia can probe multiplicity.  Perturbations of the source photocenter (caused by orbiting unresolved objects) compared to the center-of-mass motion (which moves as a single star) will cause the observations to be a poor match to the fitting model, which registers as excess noise via the \texttt{astrometric\_excess\_noise} parameter, and whose significance is captured in the \texttt{astrometric\_excess\_noise\_sig} parameter ($>$2 indicates significant excess noise).  The \texttt{astrometric\_chi2\_al} term reports the $\chi^2$ value of the observations to the fitting model. From the image parameter determination (IPD) phase, \texttt{ipd\_gof\_harmonic\_amplitude} is sensitive to elongated PSF shapes relative to the scan direction (larger values indicate more elongation), and \texttt{ipd\_frac\_multi\_peak} reports the percentage of observations which contained more than one peak in the windows\footnote{See \url{https://gea.esac.esa.int/archive/documentation/GEDR3/Gaia_archive/chap_datamodel/sec_dm_main_tables/ssec_dm_gaia_source.html} for complete description of Gaia catalog contents}.

Table \ref{tab:BDI1350metrics} shows values of these metrics for HIP~67506.  The IPD parameters are small, suggesting that there are no marginally resolved sources (separation larger than the resolution limit but smaller than the confusion limit, $\sim$0.1-1.2\arcsec, \citealt{gaiaEDR3}) present in the images, however the astrometric noise parameters are large and significant, affirming the presence of subsystems.  It appears possible that the subsystem(s) affecting the astrometry are closer than 0.1\arcsec, however the candidate signal's position of $\approx$0.2\arcsec\ is near the resolution limit so it is not ruled out as a genuine signal by these metrics.

\textit{Candidate signal properties.} Treating this candidate signal as a genuine companion, we estimated the mass by injecting a negative template PSF in the same manner as Section \ref{sec:cont limits}.  We varied the separation, position angle, and relative contrast of the negative signal to minimize the residual root-mean-square value of pixels within a diameter~=~1$\lambda$/D aperture centered on the injected signal.  We estimated the contrast between star and candidate companion to be $\Delta$L$^\prime$ $\approx$ 5~-~5.5 magnitudes.  We used the age of the system ($\approx$200 Myr) and L$^\prime$ magnitude to interpolate a mass estimate using \changed{\citetalias{Baraffe2015BHAC}} evolutionary atmosphere models, and estimated a corresponding mass of $\approx$60--90~M$_{\rm{Jup}}$, which spans the divide between high-mass brown dwarf and low-mass M-dwarf regimes.  \changed{This is however smaller than the mass estimates derived from the PMa.  Given the proximity to the star's core, at separation $\approx$1.3 $\lambda$/D, it is possible that some of the companion flux was subtracted in the reduction.
}\changed{However the smaller mass estimate} places the candidate companion in \changed{an (age, luminosity, mass) regime} with few other detected young high-mass brown dwarf companions \citep[see][Fig 34]{Faherty2016BDAnalogsToExoplanets}, \changed{so if the small-mass estimate is valid this will be an interesting benchmark object}.  This makes HIP~67506 a good target for follow-up observations to confirm the companion, obtain spectral type, T${_{\rm{eff}}}$, and log(g) estimates, and potentially a dynamical mass measurement.

\subsection{Completeness}

We determined the survey completeness to stellar and substellar companions using a Monte Carlo approach.  Over a grid that is uniform in log(mass) $\in$ [-3,0] \changed{$\Msun$} and log(semimajor axis) $\in$ [0,3] \changed{AU} we generated 5$\times$10$^{3}$ simulated companions for each grid point, randomly assigned orbital parameters from priors\footnote{eccentricity (e): P(e) = 2.1 - 2.2$\times$e, e $\in$ [0,0.95], following \citealt{nielsen_gemini_2019}; inclination (i): $\cos$(i) $\in$ Unif[-1,1]; argument of periastron ($\omega$): $\omega \in$ Unif[0,2$\pi$]; mean anomaly (M): M $\in$ Unif[0,2$\pi$]; since contrast curves are one-dimensional we did not simulate longitude of nodes}, and computed the projected separation.  A companion was considered detectable if it fell above the contrast curve and undetectable if below.  \changed{We determined completeness as the fraction of simulated companions at each grid point that would have been detected at at least SNR=5, with 1.0 corresponding to detecting every simulated companion, and 0.0 detecting none.  We computed survey completeness for each star in our survey, with contours at 10\%, 50\%, and 90\% of simulated companions detected.}

\begin{figure}
\centering
\includegraphics[width=0.49\textwidth]{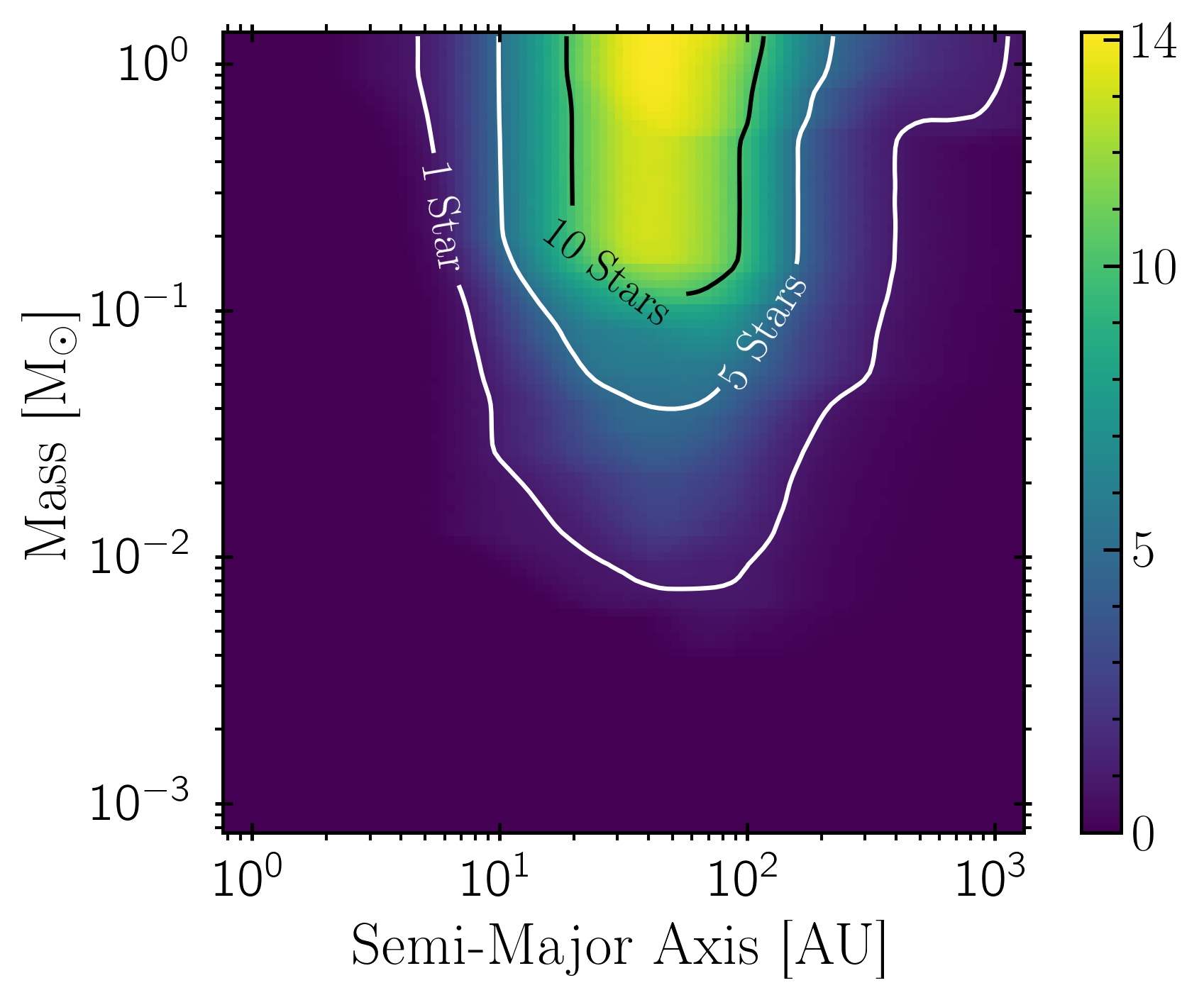}\\
\caption{Completeness map for every star in our survey as a function of mass and semi-major axis.  Colormap and contours give the number of stars for which a given (sma, mass) pair is complete.}
\label{fig:completeness}
\end{figure}

Figure \ref{fig:completeness} displays completeness for the entire survey, made by summing completeness maps for every star in the survey \citep{Lunine2008WorldsBeyond,nielsen_gemini_2019}.  Contours and colormap give number of stars for which the survey is complete for a given (sma, mass) pair.  Stars in our survey cover a variety of separation regimes, and so individual completeness plots do not line up; additionally
individual completeness plots never reach 100\% as some simulated planets fall outside the inner or outer working angles when projected, and become undetectable.  Thus the maximum value in composite completeness plot is $\sim$14 stars, even though all stars have some fractional sensitivity to companions.

\section{Discussion}\label{sec:discussion}

\subsection{\changed{BDI performance compared to other observing modes.}}

\citetalias{rodigas_direct_2015} described several advantages to BDI over ADI or ``classical" RDI: 1. PA rotation is not a consideration when planning and executing observations, 2. BDI allows reducing two stars with 1 observation (2$\times$ more efficient than ADI, 4$\times$ more efficient than RDI), and 3. It targets stars often excluded from large direct imaging surveys (wide binaries).  They used simulated companion injections into a single MagAO/Clio [3.95] dataset of HD~37551 to determine that BDI performed $\sim$0.5 mag better than ADI at small separations ($\sim$1\arcsec).

We did not explicitly test only-BDI vs only-ADI in our survey --- some amount of rotation was included with each BDI dataset but was not the source of diversity used to reconstruct the stellar PSF.  We found that the effectiveness of our reduction depends highly on the observing conditions and image/detector quality, but these are factors which would affect both ADI and BDI equally.  However, ADI is not susceptible to the contrast between the two binary components, as discussed in Section \ref{sec:results}.  \citetalias{rodigas_direct_2015} selected binaries with NIR $\Delta$m~$\lesssim$~2, but state this was not a strict requirement based on their analysis.  We found that for systems in the regime $\Delta$m$_{\rm{IR}}$~$\sim$~1-2, 5-$\sigma$ contrast limits were shallower for higher $\Delta$m$_{\rm{IR}}$, as PSF features visible in the bright star do not have sufficient signal-to-noise in the fainter star to be fully subtracted.  BDI is also susceptible to variation due to anisotropy, unlike ADI and SDI, and the separation between the stars should be designed to fall within the isoplanatic patch for the observing wavelength.

\subsection{\changed{The scientific context of our survey}}
\changed{The small number of stars in our survey and their diversity of characteristics does not allow us to make meaningful contributions to the occurrence rates discussed in Section \ref{sec:motivation}.  The 35 stars in our survey to date span a range of spectral types, ages, (lack of) group membership, and binary separations, and were chosen for their utility in the BDI technique.  This initial survey represents a contribution to probes of (sub)stellar companions in wide binaries
; further observations of wide binary systems are needed to continue to fill in the picture of brown dwarfs and giant planets in wide stellar binaries.}

\section{Conclusion}

We have presented the results of 17 binary star systems imaged in NIR with MagAO/Clio and reduced using Binary Differential Imaging and PCA techniques.  Our achieved contrast was limited by image quality, observing conditions, and binary star contrast.  We detected a candidate companion signal around HIP~67506~A which is near the stellar-substellar boundary, and merits follow-up to confirm companion status and characterize the companion.  

Targeting young wide multiple star systems with direct imaging surveys is advantageous from both a technical and astrophysical perspective.  Simultaneously imaging the science and reference star in the same filter within the same isoplanatic patch should provide superior PSF matching for starlight subtraction, particularly when combined with PCA for building a PSF model. This promises to be an even more powerful technique for space-based observations, including JWST, as the PSF is much more stable and is not subject to anisoplanatism.  Brown dwarf and giant planet formation and dynamical evolution in binaries is a data-starved problem with many unanswered questions, and is an important piece of the star and planet formation picture.

\vspace{1.5cm}
\nolinenumbers
The authors thank the referee for helpful and constructive comments that improved the quality of this manuscript.

\changed{The authors wish to thank T. J. Rodigas for designing the MagAO/Clio BDI survey and contributing to the collection of the data used in this work.}

L.A.P.~acknowledges research support from the NSF Graduate Research Fellowship.  This material is based upon work supported by the National Science Foundation Graduate Research Fellowship Program under Grant No. DGE-1746060. J.D.L.~thanks the Heising-Simons Foundation (Grant \#2020-1824) and NSF AST (\#1625441, MagAO-X).

MagAO was developed with support from the NSF (\#0321312, \#1206422, \#1506818, \#9618852). We thank the LCO and Magellan staffs for their outstanding assistance throughout our commissioning runs. We also thank the teams at the Steward Observatory Mirror Lab/CAAO (University of Arizona), Microgate (Italy), and ADS (Italy) for their contributions to the ASM.

This work used High Performance Computing (HPC) resources supported by the University of Arizona TRIF, UITS, and the Office for Research, Innovation, and Impact (RII) and maintained by the UArizona Research Technologies department.  

This work has made use of data from the European Space Agency (ESA) mission
{\it Gaia} (\url{https://www.cosmos.esa.int/gaia}), processed by the {\it Gaia}
Data Processing and Analysis Consortium (DPAC,
\url{https://www.cosmos.esa.int/web/gaia/dpac/consortium}). Funding for the DPAC
has been provided by national institutions, in particular the institutions
participating in the {\it Gaia} Multilateral Agreement.

This publication makes use of data products from the Wide-field Infrared Survey Explorer, which is a joint project of the University of California, Los Angeles, and the Jet Propulsion Laboratory/California Institute of Technology, funded by the National Aeronautics and Space Administration.

This research has made use of the NASA Exoplanet Archive, which is operated by the California Institute of Technology, under contract with the National Aeronautics and Space Administration under the Exoplanet Exploration Program.

This research has made use of ATRAN sky transmission files generated at the international Gemini Observatory, a program of NSF’s NOIRLab, which is managed by the Association of Universities for Research in Astronomy (AURA) under a cooperative agreement with the National Science Foundation on behalf of the Gemini Observatory partnership: the National Science Foundation (United States), National Research Council (Canada), Agencia Nacional de Investigación y Desarrollo (Chile), Ministerio de Ciencia, Tecnología e Innovación (Argentina), Ministério da Ciência, Tecnologia, Inovações e Comunicações (Brazil), and Korea Astronomy and Space Science Institute (Republic of Korea).


\vspace{5mm}
\facilities{Las Campanas Observatory, Magellan:Clay (MagAO+Clio)}

\software{Numpy \citep{Harris2020Numpy}, Astropy \citep{astropy:2018}, Matplotlib \citep{Hunter:2007}, Scipy \citep{2020SciPy-NMeth}, OpenCV \citep{opencv_library}, Photutils \citep{photutils_citation}}

\textit{Data Availability} 
The data underlying this article are available in at \url{https://github.com/logan-pearce/Pearce2022-BDI-Public-Data-Release} and at DOI: 10.5281/zenodo.6111597. Additional figures for each star in the survey are available online in the supplementary material.

\bibliography{zotero-references}{}

\begin{thebibliography}{}
\expandafter\ifx\csname natexlab\endcsname\relax\def\natexlab#1{#1}\fi
\providecommand{\url}[1]{\href{#1}{#1}}
\providecommand{\dodoi}[1]{doi:~\href{http://doi.org/#1}{\nolinkurl{#1}}}
\providecommand{\doeprint}[1]{\href{http://ascl.net/#1}{\nolinkurl{http://ascl.net/#1}}}
\providecommand{\doarXiv}[1]{\href{https://arxiv.org/abs/#1}{\nolinkurl{https://arxiv.org/abs/#1}}}

\bibitem[{Hip(1997)}]{HipAndTychoDoublesCat}
 1997, ESA Special Publication, Vol. 1200, {The HIPPARCOS and TYCHO catalogues.
  Astrometric and photometric star catalogues derived from the ESA HIPPARCO
  Space Astrometry Mission}

\bibitem[{{Anders} {et~al.}(2019){Anders}, {Khalatyan}, {Chiappini}, {Queiroz},
  {Santiago}, {Jordi}, {Girardi}, {Brown}, {Matijevi{\v{c}}}, {Monari},
  {Cantat-Gaudin}, {Weiler}, {Khan}, {Miglio}, {Carrillo}, {Romero-G{\'o}mez},
  {Minchev}, {de Jong}, {Antoja}, {Ramos}, {Steinmetz}, \&
  {Enke}}]{Anders2019AstrophysicalParametersGaiaDR2}
{Anders}, F., {Khalatyan}, A., {Chiappini}, C., {et~al.} 2019, \aap, 628, A94,
  \dodoi{10.1051/0004-6361/201935765}

\bibitem[{{Arun} {et~al.}(2019){Arun}, {Mathew}, {Manoj}, {Ujjwal}, {Kartha},
  {Viswanath}, {Narang}, \& {Paul}}]{Arun2019MassIRExcessHerbigABStars}
{Arun}, R., {Mathew}, B., {Manoj}, P., {et~al.} 2019, \aj, 157, 159,
  \dodoi{10.3847/1538-3881/ab0ca1}

\bibitem[{{Asensio-Torres} {et~al.}(2018){Asensio-Torres}, {Janson},
  {Bonavita}, {Desidera}, {Thalmann}, {Kuzuhara}, {Henning}, {Marzari},
  {Meyer}, {Calissendorff}, \& {Uyama}}]{Asensio-Torres2018SPOTS}
{Asensio-Torres}, R., {Janson}, M., {Bonavita}, M., {et~al.} 2018, \aap, 619,
  A43, \dodoi{10.1051/0004-6361/201833349}

\bibitem[{{Azulay} {et~al.}(2017){Azulay}, {Guirado}, {Marcaide},
  {Mart{\'\i}-Vidal}, {Ros}, {Tognelli}, {Jauncey}, {Lestrade}, \&
  {Reynolds}}]{Azulay2017ABDorACDynMass}
{Azulay}, R., {Guirado}, J.~C., {Marcaide}, J.~M., {et~al.} 2017, \aap, 607,
  A10, \dodoi{10.1051/0004-6361/201730641}

\bibitem[{Bailer-Jones {et~al.}(2018)Bailer-Jones, Rybizki, Fouesneau,
  Mantelet, \& Andrae}]{bailer-jones_estimating_2018}
Bailer-Jones, C. A.~L., Rybizki, J., Fouesneau, M., Mantelet, G., \& Andrae, R.
  2018, The Astronomical Journal, 156, 58, \dodoi{10.3847/1538-3881/aacb21}

\bibitem[{{Baraffe} {et~al.}(2015){Baraffe}, {Homeier}, {Allard}, \&
  {Chabrier}}]{Baraffe2015BHAC}
{Baraffe}, I., {Homeier}, D., {Allard}, F., \& {Chabrier}, G. 2015, \aap, 577,
  A42, \dodoi{10.1051/0004-6361/201425481}

\bibitem[{{Barrado y Navascu{\'e}s} {et~al.}(2004){Barrado y Navascu{\'e}s},
  {Stauffer}, \& {Jayawardhana}}]{Barrado2004LiDepl}
{Barrado y Navascu{\'e}s}, D., {Stauffer}, J.~R., \& {Jayawardhana}, R. 2004,
  \apj, 614, 386, \dodoi{10.1086/423485}

\bibitem[{Barrado Y~Navascués(2006)}]{barrado_y_navascues_age_2006}
Barrado Y~Navascués, D. 2006, Astronomy and Astrophysics, 459, 511,
  \dodoi{10.1051/0004-6361:20065717}

\bibitem[{Bazs\'{o} \& Pilat-Lohinger(2020)}]{bazso_fear_2020}
Bazs\'{o}, A., \& Pilat-Lohinger, E. 2020, AJ, 160, 2,
  \dodoi{10.3847/1538-3881/ab9104}

\bibitem[{Bell {et~al.}(2015)Bell, Mamajek, \&
  Naylor}]{bell_self-consistent_2015}
Bell, C. P.~M., Mamajek, E.~E., \& Naylor, T. 2015, Monthly Notices of the
  Royal Astronomical Society, 454, 593, \dodoi{10.1093/mnras/stv1981}

\bibitem[{{Belokurov} {et~al.}(2020){Belokurov}, {Penoyre}, {Oh}, {Iorio},
  {Hodgkin}, {Evans}, {Everall}, {Koposov}, {Tout}, {Izzard}, {Clarke}, \&
  {Brown}}]{Belokurov2020UnresolvedBinariesDR2}
{Belokurov}, V., {Penoyre}, Z., {Oh}, S., {et~al.} 2020, \mnras, 496, 1922,
  \dodoi{10.1093/mnras/staa1522}

\bibitem[{Belokurov {et~al.}(2020)Belokurov, Penoyre, Oh, Iorio, Hodgkin,
  Evans, Everall, Koposov, Tout, Izzard, Clarke, \&
  Brown}]{belokurov_unresolved_2020}
Belokurov, V., Penoyre, Z., Oh, S., {et~al.} 2020, Monthly Notices of the Royal
  Astronomical Society, 496, 1922, \dodoi{10.1093/mnras/staa1522}

\bibitem[{{Berrilli} {et~al.}(1992){Berrilli}, {Corciulo}, {Ingrosso},
  {Lorenzetti}, {Nisini}, \& {Strafella}}]{Berrilli1992IRDustHerbigAB}
{Berrilli}, F., {Corciulo}, G., {Ingrosso}, G., {et~al.} 1992, \apj, 398, 254,
  \dodoi{10.1086/171853}

\bibitem[{{Best} \& {Petrovich}(2022)}]{Best2022RetrogradeMultiplanetChaos}
{Best}, S., \& {Petrovich}, C. 2022, \apjl, 925, L5,
  \dodoi{10.3847/2041-8213/ac49e9}

\bibitem[{{Beuzit} {et~al.}(2008){Beuzit}, {Feldt}, {Dohlen}, {Mouillet},
  {Puget}, {Wildi}, {Abe}, {Antichi}, {Baruffolo}, {Baudoz}, {Boccaletti},
  {Carbillet}, {Charton}, {Claudi}, {Downing}, {Fabron}, {Feautrier},
  {Fedrigo}, {Fusco}, {Gach}, {Gratton}, {Henning}, {Hubin}, {Joos}, {Kasper},
  {Langlois}, {Lenzen}, {Moutou}, {Pavlov}, {Petit}, {Pragt}, {Rabou}, {Rigal},
  {Roelfsema}, {Rousset}, {Saisse}, {Schmid}, {Stadler}, {Thalmann}, {Turatto},
  {Udry}, {Vakili}, \& {Waters}}]{Beuzit2008SPHERE}
{Beuzit}, J.-L., {Feldt}, M., {Dohlen}, K., {et~al.} 2008, in Society of
  Photo-Optical Instrumentation Engineers (SPIE) Conference Series, Vol. 7014,
  Ground-based and Airborne Instrumentation for Astronomy II, ed. I.~S.
  {McLean} \& M.~M. {Casali}, 701418, \dodoi{10.1117/12.790120}

\bibitem[{{Binks} {et~al.}(2020){Binks}, {Jeffries}, \&
  {Wright}}]{Binks2020YoungStars}
{Binks}, A.~S., {Jeffries}, R.~D., \& {Wright}, N.~J. 2020, \mnras, 494, 2429,
  \dodoi{10.1093/mnras/staa909}

\bibitem[{{Bohlin} {et~al.}(2014){Bohlin}, {Gordon}, \&
  {Tremblay}}]{Bohlin2014CALSPECRef}
{Bohlin}, R.~C., {Gordon}, K.~D., \& {Tremblay}, P.~E. 2014, \pasp, 126, 711,
  \dodoi{10.1086/677655}

\bibitem[{{Bonavita} \&
  {Desidera}(2007)}]{Bonavita2007PlanetsInMultipleSystems}
{Bonavita}, M., \& {Desidera}, S. 2007, \aap, 468, 721,
  \dodoi{10.1051/0004-6361:20066671}

\bibitem[{{Bonavita} {et~al.}(2016){Bonavita}, {Desidera}, {Thalmann},
  {Janson}, {Vigan}, {Chauvin}, \& {Lannier}}]{Bonavita2016SPOTS}
{Bonavita}, M., {Desidera}, S., {Thalmann}, C., {et~al.} 2016, \aap, 593, A38,
  \dodoi{10.1051/0004-6361/201628231}

\bibitem[{{Borucki} {et~al.}(2010){Borucki}, {Koch}, {Basri}, {Batalha},
  {Brown}, {Caldwell}, {Caldwell}, {Christensen-Dalsgaard}, {Cochran},
  {DeVore}, {Dunham}, {Dupree}, {Gautier}, {Geary}, {Gilliland}, {Gould},
  {Howell}, {Jenkins}, {Kondo}, {Latham}, {Marcy}, {Meibom}, {Kjeldsen},
  {Lissauer}, {Monet}, {Morrison}, {Sasselov}, {Tarter}, {Boss}, {Brownlee},
  {Owen}, {Buzasi}, {Charbonneau}, {Doyle}, {Fortney}, {Ford}, {Holman},
  {Seager}, {Steffen}, {Welsh}, {Rowe}, {Anderson}, {Buchhave}, {Ciardi},
  {Walkowicz}, {Sherry}, {Horch}, {Isaacson}, {Everett}, {Fischer}, {Torres},
  {Johnson}, {Endl}, {MacQueen}, {Bryson}, {Dotson}, {Haas}, {Kolodziejczak},
  {Van Cleve}, {Chandrasekaran}, {Twicken}, {Quintana}, {Clarke}, {Allen},
  {Li}, {Wu}, {Tenenbaum}, {Verner}, {Bruhweiler}, {Barnes}, \&
  {Prsa}}]{Borucki2010Kepler}
{Borucki}, W.~J., {Koch}, D., {Basri}, G., {et~al.} 2010, Science, 327, 977,
  \dodoi{10.1126/science.1185402}

\bibitem[{Bradley {et~al.}(2020)Bradley, Sip{\H o}cz, Robitaille, Tollerud,
  Vin{\'{\i}}cius, Deil, Barbary, Wilson, Busko, G{\"u}nther, Cara, Conseil,
  Bostroem, Droettboom, Bray, Bratholm, Lim, Barentsen, Craig, Pascual, Perren,
  Greco, Donath, de~Val-Borro, Kerzendorf, Bach, Weaver, D'Eugenio, Souchereau,
  \& Ferreira}]{photutils_citation}
Bradley, L., Sip{\H o}cz, B., Robitaille, T., {et~al.} 2020, astropy/photutils:
  1.0.0, 1.0.0,  Zenodo, \dodoi{10.5281/zenodo.4044744}

\bibitem[{Bradski(2000)}]{opencv_library}
Bradski, G. 2000, Dr. Dobb's Journal of Software Tools

\bibitem[{{Brandt} {et~al.}(2021){Brandt}, {Brandt}, {Dupuy}, {Li}, \&
  {Michalik}}]{Brandt2021DynamicalMassBetaPicBBetaPicC}
{Brandt}, G.~M., {Brandt}, T.~D., {Dupuy}, T.~J., {Li}, Y., \& {Michalik}, D.
  2021, \aj, 161, 179, \dodoi{10.3847/1538-3881/abdc2e}

\bibitem[{Brandt(2018)}]{brandt_hipparcos-gaia_2018}
Brandt, T.~D. 2018, The Astrophysical Journal Supplement Series, 239, 31,
  \dodoi{10.3847/1538-4365/aaec06}

\bibitem[{{Brandt}(2021)}]{Brandt2021_HGCA_ERD3}
{Brandt}, T.~D. 2021, \apjs, 254, 42, \dodoi{10.3847/1538-4365/abf93c}

\bibitem[{Brandt {et~al.}(2019)Brandt, Dupuy, \& Bowler}]{brandt_precise_2019}
Brandt, T.~D., Dupuy, T.~J., \& Bowler, B.~P. 2019, {\textbackslash}aj, 158,
  140, \dodoi{10.3847/1538-3881/ab04a8}

\bibitem[{{Bryan} {et~al.}(2019){Bryan}, {Knutson}, {Lee}, {Fulton}, {Batygin},
  {Ngo}, \& {Meshkat}}]{Bryan2019JupiterAnalogs}
{Bryan}, M.~L., {Knutson}, H.~A., {Lee}, E.~J., {et~al.} 2019, \aj, 157, 52,
  \dodoi{10.3847/1538-3881/aaf57f}

\bibitem[{Bryan {et~al.}(2020)Bryan, Chiang, Bowler, Morley, Millholland,
  Blunt, Ashok, Nielsen, Ngo, Mawet, \& Knutson}]{bryan_obliquity_2020}
Bryan, M.~L., Chiang, E., Bowler, B.~P., {et~al.} 2020, The Astronomical
  Journal, 159, 181, \dodoi{10.3847/1538-3881/ab76c6}

\bibitem[{{Cadman} {et~al.}(2022){Cadman}, {Hall}, {Fontanive}, \&
  {Rice}}]{Cadman2022BinariesGravitationalInstability}
{Cadman}, J., {Hall}, C., {Fontanive}, C., \& {Rice}, K. 2022, \mnras,
  \dodoi{10.1093/mnras/stac033}

\bibitem[{{Chandler} {et~al.}(2016){Chandler}, {McDonald}, \&
  {Kane}}]{Chandler2016HabitableZones}
{Chandler}, C.~O., {McDonald}, I., \& {Kane}, S.~R. 2016, \aj, 151, 59,
  \dodoi{10.3847/0004-6256/151/3/59}

\bibitem[{{Chauvin} {et~al.}(2010){Chauvin}, {Lagrange}, {Bonavita},
  {Zuckerman}, {Dumas}, {Bessell}, {Beuzit}, {Bonnefoy}, {Desidera}, {Farihi},
  {Lowrance}, {Mouillet}, \& {Song}}]{Chauvin2010DeepImagingSurvey}
{Chauvin}, G., {Lagrange}, A.~M., {Bonavita}, M., {et~al.} 2010, \aap, 509,
  A52, \dodoi{10.1051/0004-6361/200911716}

\bibitem[{{Choi} {et~al.}(2016){Choi}, {Dotter}, {Conroy}, {Cantiello},
  {Paxton}, \& {Johnson}}]{Choi2016MIST}
{Choi}, J., {Dotter}, A., {Conroy}, C., {et~al.} 2016, \apj, 823, 102,
  \dodoi{10.3847/0004-637X/823/2/102}

\bibitem[{{Christian} {et~al.}(2022){Christian}, {Vanderburg}, {Becker},
  {Yahalomi}, {Pearce}, {Zhou}, {Collins}, {Kraus}, {Stassun}, {de Beurs},
  {Ricker}, {Vanderspek}, {Latham}, {Winn}, {Seager}, {Jenkins}, {Abe},
  {Agabi}, {Amado}, {Baker}, {Barkaoui}, {Benkhaldoun}, {Benni}, {Berberian},
  {Berlind}, {Bieryla}, {Esparza-Borges}, {Bowen}, {Brown}, {Buchhave},
  {Burke}, {Buttu}, {Cadieux}, {Caldwell}, {Charbonneau}, {Chazov},
  {Chimaladinne}, {Collins}, {Combs}, {Conti}, {Crouzet}, {de Leon},
  {Deljookorani}, {Diamond}, {Doyon}, {Dragomir}, {Dransfield}, {Essack},
  {Evans}, {Fukui}, {Gan}, {Esquerdo}, {Gillon}, {Girardin}, {Guerra},
  {Guillot}, {Habich}, {Henriksen}, {Hoch}, {Isogai}, {Jehin}, {Jensen},
  {Johnson}, {Livingston}, {Kielkopf}, {Kim}, {Kawauchi}, {Krushinsky},
  {Kunzle}, {Laloum}, {Leger}, {Lewin}, {Mallia}, {Massey}, {Mori}, {McLeod},
  {M{\'e}karnia}, {Mireles}, {Mishevskiy}, {Tamura}, {Murgas}, {Narita},
  {Naves}, {Nelson}, {Osborn}, {Palle}, {Parviainen}, {Plavchan}, {Pozuelos},
  {Rabus}, {Relles}, {Rodr{\'\i}guez L{\'o}pez}, {Quinn}, {Schmider},
  {Schlieder}, {Schwarz}, {Shporer}, {Sibbald}, {Srdoc}, {Stibbards},
  {Stickler}, {Suarez}, {Stockdale}, {Tan}, {Terada}, {Triaud}, {Tronsgaard},
  {Waalkes}, {Wang}, {Watanabe}, {Wenceslas}, {Wingham}, {Wittrock}, \&
  {Ziegler}}]{Christian2022OrbitalAlignments}
{Christian}, S., {Vanderburg}, A., {Becker}, J., {et~al.} 2022, arXiv e-prints,
  arXiv:2202.00042.
\newblock \doarXiv{2202.00042}

\bibitem[{{Climent} {et~al.}(2019){Climent}, {Berger}, {Guirado}, {Marcaide},
  {Mart{\'\i}-Vidal}, {M{\'e}rand}, {Tognelli}, \&
  {Wittkowski}}]{Climent2019ABDorC}
{Climent}, J.~B., {Berger}, J.~P., {Guirado}, J.~C., {et~al.} 2019, \apjl, 886,
  L9, \dodoi{10.3847/2041-8213/ab5065}

\bibitem[{{Close} {et~al.}(2005){Close}, {Lenzen}, {Guirado}, {Nielsen},
  {Mamajek}, {Brandner}, {Hartung}, {Lidman}, \& {Biller}}]{Close2005ABDorC}
{Close}, L.~M., {Lenzen}, R., {Guirado}, J.~C., {et~al.} 2005, \nat, 433, 286,
  \dodoi{10.1038/nature03225}

\bibitem[{Close {et~al.}(2013)Close, Males, Morzinski, Kopon, Follette,
  Rodigas, Hinz, Wu, Puglisi, Esposito, Riccardi, Pinna, Xompero, Briguglio,
  Uomoto, \& Hare}]{close_diffraction-limited_2013}
Close, L.~M., Males, J.~R., Morzinski, K., {et~al.} 2013, The Astrophysical
  Journal, 774, 94, \dodoi{10.1088/0004-637X/774/2/94}

\bibitem[{{Corbally}(1984)}]{Corbally1984BinarySpT}
{Corbally}, C.~J. 1984, \apjs, 55, 657, \dodoi{10.1086/190973}

\bibitem[{Correa-Otto \& Gil-Hutton(2017)}]{correa-otto_galactic_2017}
Correa-Otto, J.~A., \& Gil-Hutton, R.~A. 2017, Astronomy and Astrophysics, 608,
  A116, \dodoi{10.1051/0004-6361/201731229}

\bibitem[{{Currie} {et~al.}(2021){Currie}, {Brandt}, {Kuzuhara}, {Chilcote},
  {Cashman}, {Liu}, {Lawson}, {Tobin}, {Brandt}, {Guyon}, {Lozi}, {Deo},
  {Vievard}, {Ahn}, \& {Skaf}}]{Currie2021SurveyofAcceleratingStars}
{Currie}, T., {Brandt}, T., {Kuzuhara}, M., {et~al.} 2021, arXiv e-prints,
  arXiv:2109.09745.
\newblock \doarXiv{2109.09745}

\bibitem[{{Cutri} {et~al.}(2012){Cutri}, {Wright}, {Conrow}, {Bauer},
  {Benford}, {Brandenburg}, {Dailey}, {Eisenhardt}, {Evans}, {Fajardo-Acosta},
  {Fowler}, {Gelino}, {Grillmair}, {Harbut}, {Hoffman}, {Jarrett},
  {Kirkpatrick}, {Leisawitz}, {Liu}, {Mainzer}, {Marsh}, {Masci}, {McCallon},
  {Padgett}, {Ressler}, {Royer}, {Skrutskie}, {Stanford}, {Wyatt}, {Tholen},
  {Tsai}, {Wachter}, {Wheelock}, {Yan}, {Alles}, {Beck}, {Grav}, {Masiero},
  {McCollum}, {McGehee}, {Papin}, \& {Wittman}}]{Cutri2012WISE}
{Cutri}, R.~M., {Wright}, E.~L., {Conrow}, T., {et~al.} 2012, {Explanatory
  Supplement to the WISE All-Sky Data Release Products}, Explanatory Supplement
  to the WISE All-Sky Data Release Products

\bibitem[{David \& Hillenbrand(2015)}]{david_ages_2015}
David, T.~J., \& Hillenbrand, L.~A. 2015, The Astrophysical Journal, 804, 146,
  \dodoi{10.1088/0004-637X/804/2/146}

\bibitem[{Deacon \& Kraus(2020)}]{deacon_wide_2020}
Deacon, N.~R., \& Kraus, A.~L. 2020, arXiv:2006.06679 [astro-ph].
\newblock \url{http://arxiv.org/abs/2006.06679}

\bibitem[{Deacon {et~al.}(2016)Deacon, Kraus, Mann, Magnier, Chambers,
  Wainscoat, Tonry, Kaiser, Waters, Flewelling, Hodapp, \&
  Burgett}]{deacon_pan-starrs_2016}
Deacon, N.~R., Kraus, A.~L., Mann, A.~W., {et~al.} 2016, Monthly Notices of the
  Royal Astronomical Society, 455, 4212, \dodoi{10.1093/mnras/stv2132}

\bibitem[{{Dommanget} \& {Nys}(2000)}]{Dommanget2000VisualDoubleStarsCatalog}
{Dommanget}, J., \& {Nys}, O. 2000, \aap, 363, 991

\bibitem[{{Dupuy} {et~al.}(2022){Dupuy}, {Kraus}, {Kratter}, {Rizzuto}, {Mann},
  {Huber}, \& {Ireland}}]{Dupuy2022OrbitalArchII}
{Dupuy}, T.~J., {Kraus}, A.~L., {Kratter}, K.~M., {et~al.} 2022, arXiv
  e-prints, arXiv:2202.00013.
\newblock \doarXiv{2202.00013}

\bibitem[{{Ekstr{\"o}m} {et~al.}(2012){Ekstr{\"o}m}, {Georgy}, {Eggenberger},
  {Meynet}, {Mowlavi}, {Wyttenbach}, {Granada}, {Decressin}, {Hirschi},
  {Frischknecht}, {Charbonnel}, \& {Maeder}}]{Ekstrom2021GenevaStelEvCode}
{Ekstr{\"o}m}, S., {Georgy}, C., {Eggenberger}, P., {et~al.} 2012, \aap, 537,
  A146, \dodoi{10.1051/0004-6361/201117751}

\bibitem[{El-Badry {et~al.}(2021)El-Badry, Rix, \&
  Heintz}]{el-badry_million_2021}
El-Badry, K., Rix, H.-W., \& Heintz, T.~M. 2021, arXiv e-prints, 2101,
  arXiv:2101.05282.
\newblock \url{http://adsabs.harvard.edu/abs/2021arXiv210105282E}

\bibitem[{{Elliott} {et~al.}(2014){Elliott}, {Bayo}, {Melo}, {Torres},
  {Sterzik}, \& {Quast}}]{Elliott2014SACYV}
{Elliott}, P., {Bayo}, A., {Melo}, C.~H.~F., {et~al.} 2014, \aap, 568, A26,
  \dodoi{10.1051/0004-6361/201423856}

\bibitem[{{Fabricius} {et~al.}(2002){Fabricius}, {H{\o}g}, {Makarov}, {Mason},
  {Wycoff}, \& {Urban}}]{Fabricius2002TychoDoubleStarCat}
{Fabricius}, C., {H{\o}g}, E., {Makarov}, V.~V., {et~al.} 2002, \aap, 384, 180,
  \dodoi{10.1051/0004-6361:20011822}

\bibitem[{{Faherty} {et~al.}(2016){Faherty}, {Riedel}, {Cruz}, {Gagne},
  {Filippazzo}, {Lambrides}, {Fica}, {Weinberger}, {Thorstensen}, {Tinney},
  {Baldassare}, {Lemonier}, \& {Rice}}]{Faherty2016BDAnalogsToExoplanets}
{Faherty}, J.~K., {Riedel}, A.~R., {Cruz}, K.~L., {et~al.} 2016, \apjs, 225,
  10, \dodoi{10.3847/0067-0049/225/1/10}

\bibitem[{{Finkenzeller} \&
  {Mundt}(1984)}]{Finkenzeller1984HerbigABwithNebulosity}
{Finkenzeller}, U., \& {Mundt}, R. 1984, \aaps, 55, 109

\bibitem[{{Fitton} {et~al.}(2022){Fitton}, {Tofflemire}, \&
  {Kraus}}]{Fitton2022RUWEDisks}
{Fitton}, S., {Tofflemire}, B.~M., \& {Kraus}, A.~L. 2022, Research Notes of
  the American Astronomical Society, 6, 18, \dodoi{10.3847/2515-5172/ac4bb7}

\bibitem[{{Fontanive} {et~al.}(2019){Fontanive}, {Rice}, {Bonavita}, {Lopez},
  {Mu{\v{z}}i{\'c}}, {}, \& {Biller}}]{Fontanive2019HighBinarity}
{Fontanive}, C., {Rice}, K., {Bonavita}, M., {et~al.} 2019, \mnras, 485, 4967,
  \dodoi{10.1093/mnras/stz671}

\bibitem[{Gagné {et~al.}(2018)Gagné, Mamajek, Malo, Riedel, Rodriguez,
  Lafrenière, Faherty, Roy-Loubier, Pueyo, Robin, \&
  Doyon}]{gagne_banyan_2018}
Gagné, J., Mamajek, E.~E., Malo, L., {et~al.} 2018, The Astrophysical Journal,
  856, 23, \dodoi{10.3847/1538-4357/aaae09}

\bibitem[{{Gaia Collaboration} {et~al.}(2016){Gaia Collaboration}, Prusti,
  de~Bruijne, Brown, Vallenari, Babusiaux, Bailer-Jones, Bastian, Biermann,
  Evans, Eyer, Jansen, Jordi, Klioner, Lammers, Lindegren, Luri, Mignard,
  Milligan, Panem, Poinsignon, Pourbaix, Randich, Sarri, Sartoretti, Siddiqui,
  Soubiran, Valette, van Leeuwen, Walton, Aerts, Arenou, Cropper, Drimmel,
  Høg, Katz, Lattanzi, O'Mullane, Grebel, Holland, Huc, Passot, Bramante,
  Cacciari, Castañeda, Chaoul, Cheek, De~Angeli, Fabricius, Guerra,
  Hernández, Jean-Antoine-Piccolo, Masana, Messineo, Mowlavi, Nienartowicz,
  Ordóñez-Blanco, Panuzzo, Portell, Richards, Riello, Seabroke, Tanga,
  Thévenin, Torra, Els, Gracia-Abril, Comoretto, Garcia-Reinaldos, Lock,
  Mercier, Altmann, Andrae, Astraatmadja, Bellas-Velidis, Benson, Berthier,
  Blomme, Busso, Carry, Cellino, Clementini, Cowell, Creevey, Cuypers,
  Davidson, De~Ridder, de~Torres, Delchambre, Dell'Oro, Ducourant, Frémat,
  García-Torres, Gosset, Halbwachs, Hambly, Harrison, Hauser, Hestroffer,
  Hodgkin, Huckle, Hutton, Jasniewicz, Jordan, Kontizas, Korn, Lanzafame,
  Manteiga, Moitinho, Muinonen, Osinde, Pancino, Pauwels, Petit, Recio-Blanco,
  Robin, Sarro, Siopis, Smith, Smith, Sozzetti, Thuillot, van Reeven, Viala,
  Abbas, Abreu~Aramburu, Accart, Aguado, Allan, Allasia, Altavilla, Álvarez,
  Alves, Anderson, Andrei, Anglada~Varela, Antiche, Antoja, Antón, Arcay,
  Atzei, Ayache, Bach, Baker, Balaguer-Núñez, Barache, Barata, Barbier,
  Barblan, Baroni, Barrado~y Navascués, Barros, Barstow, Becciani, Bellazzini,
  Bellei, Bello~García, Belokurov, Bendjoya, Berihuete, Bianchi, Bienaymé,
  Billebaud, Blagorodnova, Blanco-Cuaresma, Boch, Bombrun, Borrachero,
  Bouquillon, Bourda, Bouy, Bragaglia, Breddels, Brouillet, Brüsemeister,
  Bucciarelli, Budnik, Burgess, Burgon, Burlacu, Busonero, Buzzi, Caffau,
  Cambras, Campbell, Cancelliere, Cantat-Gaudin, Carlucci, Carrasco,
  Castellani, Charlot, Charnas, Charvet, Chassat, Chiavassa, Clotet, Cocozza,
  Collins, Collins, Costigan, Crifo, Cross, Crosta, Crowley, Dafonte, Damerdji,
  Dapergolas, David, David, De~Cat, de~Felice, de~Laverny, De~Luise, De~March,
  de~Martino, de~Souza, Debosscher, del Pozo, Delbo, Delgado, Delgado,
  di~Marco, Di~Matteo, Diakite, Distefano, Dolding, Dos~Anjos, Drazinos,
  Durán, Dzigan, Ecale, Edvardsson, Enke, Erdmann, Escolar, Espina, Evans,
  Eynard~Bontemps, Fabre, Fabrizio, Faigler, Falcão, Farràs~Casas, Faye,
  Federici, Fedorets, Fernández-Hernández, Fernique, Fienga, Figueras,
  Filippi, Findeisen, Fonti, Fouesneau, Fraile, Fraser, Fuchs, Furnell, Gai,
  Galleti, Galluccio, Garabato, García-Sedano, Garé, Garofalo, Garralda,
  Gavras, Gerssen, Geyer, Gilmore, Girona, Giuffrida, Gomes, González-Marcos,
  González-Núñez, González-Vidal, Granvik, Guerrier, Guillout, Guiraud,
  Gúrpide, Gutiérrez-Sánchez, Guy, Haigron, Hatzidimitriou, Haywood, Heiter,
  Helmi, Hobbs, Hofmann, Holl, Holland, Hunt, Hypki, Icardi, Irwin, Jevardat~de
  Fombelle, Jofré, Jonker, Jorissen, Julbe, Karampelas, Kochoska, Kohley,
  Kolenberg, Kontizas, Koposov, Kordopatis, Koubsky, Kowalczyk, Krone-Martins,
  Kudryashova, Kull, Bachchan, Lacoste-Seris, Lanza, Lavigne,
  Le~Poncin-Lafitte, Lebreton, Lebzelter, Leccia, Leclerc, Lecoeur-Taibi,
  Lemaitre, Lenhardt, Leroux, Liao, Licata, Lindstrøm, Lister, Livanou, Lobel,
  Löffler, López, Lopez-Lozano, Lorenz, Loureiro, MacDonald,
  Magalhães~Fernandes, Managau, Mann, Mantelet, Marchal, Marchant, Marconi,
  Marie, Marinoni, Marrese, Marschalkó, Marshall, Martín-Fleitas, Martino,
  Mary, Matijevič, Mazeh, McMillan, Messina, Mestre, Michalik, Millar,
  Miranda, Molina, Molinaro, Molinaro, Molnár, Moniez, Montegriffo, Monteiro,
  Mor, Mora, Morbidelli, Morel, Morgenthaler, Morley, Morris, Mulone, Muraveva,
  Musella, Narbonne, Nelemans, Nicastro, Noval, Ordénovic, Ordieres-Meré,
  Osborne, Pagani, Pagano, Pailler, Palacin, Palaversa, Parsons, Paulsen,
  Pecoraro, Pedrosa, Pentikäinen, Pereira, Pichon, Piersimoni, Pineau, Plachy,
  Plum, Poujoulet, Prša, Pulone, Ragaini, Rago, Rambaux, Ramos-Lerate,
  Ranalli, Rauw, Read, Regibo, Renk, Reylé, Ribeiro, Rimoldini, Ripepi, Riva,
  Rixon, Roelens, Romero-Gómez, Rowell, Royer, Rudolph, Ruiz-Dern, Sadowski,
  Sagristà~Sellés, Sahlmann, Salgado, Salguero, Sarasso, Savietto, Schnorhk,
  Schultheis, Sciacca, Segol, Segovia, Segransan, Serpell, Shih, Smareglia,
  Smart, Smith, Solano, Solitro, Sordo, Soria~Nieto, Souchay, Spagna, Spoto,
  Stampa, Steele, Steidelmüller, Stephenson, Stoev, Suess, Süveges, Surdej,
  Szabados, Szegedi-Elek, Tapiador, Taris, Tauran, Taylor, Teixeira, Terrett,
  Tingley, Trager, Turon, Ulla, Utrilla, Valentini, van Elteren, Van~Hemelryck,
  van Leeuwen, Varadi, Vecchiato, Veljanoski, Via, Vicente, Vogt, Voss,
  Votruba, Voutsinas, Walmsley, Weiler, Weingrill, Werner, Wevers, Whitehead,
  Wyrzykowski, Yoldas, Žerjal, Zucker, Zurbach, Zwitter, Alecu, Allen,
  Allende~Prieto, Amorim, Anglada-Escudé, Arsenijevic, Azaz, Balm, Beck,
  Bernstein, Bigot, Bijaoui, Blasco, Bonfigli, Bono, Boudreault, Bressan,
  Brown, Brunet, Bunclark, Buonanno, Butkevich, Carret, Carrion, Chemin,
  Chéreau, Corcione, Darmigny, de~Boer, de~Teodoro, de~Zeeuw, Delle~Luche,
  Domingues, Dubath, Fodor, Frézouls, Fries, Fustes, Fyfe, Gallardo, Gallegos,
  Gardiol, Gebran, Gomboc, Gómez, Grux, Gueguen, Heyrovsky, Hoar, Iannicola,
  Isasi~Parache, Janotto, Joliet, Jonckheere, Keil, Kim, Klagyivik, Klar,
  Knude, Kochukhov, Kolka, Kos, Kutka, Lainey, LeBouquin, Liu, Loreggia,
  Makarov, Marseille, Martayan, Martinez-Rubi, Massart, Meynadier, Mignot,
  Munari, Nguyen, Nordlander, Ocvirk, O'Flaherty, Olias~Sanz, Ortiz, Osorio,
  Oszkiewicz, Ouzounis, Palmer, Park, Pasquato, Peltzer, Peralta, Péturaud,
  Pieniluoma, Pigozzi, Poels, Prat, Prod'homme, Raison, Rebordao, Risquez,
  Rocca-Volmerange, Rosen, Ruiz-Fuertes, Russo, Sembay, Serraller~Vizcaino,
  Short, Siebert, Silva, Sinachopoulos, Slezak, Soffel, Sosnowska, Straižys,
  ter Linden, Terrell, Theil, Tiede, Troisi, Tsalmantza, Tur, Vaccari, Vachier,
  Valles, Van~Hamme, Veltz, Virtanen, Wallut, Wichmann, Wilkinson, Ziaeepour,
  \& Zschocke}]{gaia_collaboration_gaia_2016}
{Gaia Collaboration}, Prusti, T., de~Bruijne, J. H.~J., {et~al.} 2016,
  Astronomy and Astrophysics, 595, A1, \dodoi{10.1051/0004-6361/201629272}

\bibitem[{{Gaia Collaboration} {et~al.}(2021){Gaia Collaboration}, {Brown},
  {Vallenari}, {Prusti}, {de Bruijne}, {Babusiaux}, {Biermann}, {Creevey},
  {Evans}, {Eyer}, {Hutton}, {Jansen}, {Jordi}, {Klioner}, {Lammers},
  {Lindegren}, {Luri}, {Mignard}, {Panem}, {Pourbaix}, {Randich}, {Sartoretti},
  {Soubiran}, {Walton}, {Arenou}, {Bailer-Jones}, {Bastian}, {Cropper},
  {Drimmel}, {Katz}, {Lattanzi}, {van Leeuwen}, {Bakker}, {Cacciari},
  {Casta{\~n}eda}, {De Angeli}, {Ducourant}, {Fabricius}, {Fouesneau},
  {Fr{\'e}mat}, {Guerra}, {Guerrier}, {Guiraud}, {Jean-Antoine Piccolo},
  {Masana}, {Messineo}, {Mowlavi}, {Nicolas}, {Nienartowicz}, {Pailler},
  {Panuzzo}, {Riclet}, {Roux}, {Seabroke}, {Sordo}, {Tanga}, {Th{\'e}venin},
  {Gracia-Abril}, {Portell}, {Teyssier}, {Altmann}, {Andrae}, {Bellas-Velidis},
  {Benson}, {Berthier}, {Blomme}, {Brugaletta}, {Burgess}, {Busso}, {Carry},
  {Cellino}, {Cheek}, {Clementini}, {Damerdji}, {Davidson}, {Delchambre},
  {Dell'Oro}, {Fern{\'a}ndez-Hern{\'a}ndez}, {Galluccio}, {Garc{\'\i}a-Lario},
  {Garcia-Reinaldos}, {Gonz{\'a}lez-N{\'u}{\~n}ez}, {Gosset}, {Haigron},
  {Halbwachs}, {Hambly}, {Harrison}, {Hatzidimitriou}, {Heiter},
  {Hern{\'a}ndez}, {Hestroffer}, {Hodgkin}, {Holl}, {Jan{\ss}en}, {Jevardat de
  Fombelle}, {Jordan}, {Krone-Martins}, {Lanzafame}, {L{\"o}ffler}, {Lorca},
  {Manteiga}, {Marchal}, {Marrese}, {Moitinho}, {Mora}, {Muinonen}, {Osborne},
  {Pancino}, {Pauwels}, {Petit}, {Recio-Blanco}, {Richards}, {Riello},
  {Rimoldini}, {Robin}, {Roegiers}, {Rybizki}, {Sarro}, {Siopis}, {Smith},
  {Sozzetti}, {Ulla}, {Utrilla}, {van Leeuwen}, {van Reeven}, {Abbas}, {Abreu
  Aramburu}, {Accart}, {Aerts}, {Aguado}, {Ajaj}, {Altavilla}, {{\'A}lvarez},
  {{\'A}lvarez Cid-Fuentes}, {Alves}, {Anderson}, {Anglada Varela}, {Antoja},
  {Audard}, {Baines}, {Baker}, {Balaguer-N{\'u}{\~n}ez}, {Balbinot}, {Balog},
  {Barache}, {Barbato}, {Barros}, {Barstow}, {Bartolom{\'e}}, {Bassilana},
  {Bauchet}, {Baudesson-Stella}, {Becciani}, {Bellazzini}, {Bernet}, {Bertone},
  {Bianchi}, {Blanco-Cuaresma}, {Boch}, {Bombrun}, {Bossini}, {Bouquillon},
  {Bragaglia}, {Bramante}, {Breedt}, {Bressan}, {Brouillet}, {Bucciarelli},
  {Burlacu}, {Busonero}, {Butkevich}, {Buzzi}, {Caffau}, {Cancelliere},
  {C{\'a}novas}, {Cantat-Gaudin}, {Carballo}, {Carlucci}, {Carnerero},
  {Carrasco}, {Casamiquela}, {Castellani}, {Castro-Ginard}, {Castro Sampol},
  {Chaoul}, {Charlot}, {Chemin}, {Chiavassa}, {Cioni}, {Comoretto}, {Cooper},
  {Cornez}, {Cowell}, {Crifo}, {Crosta}, {Crowley}, {Dafonte}, {Dapergolas},
  {David}, {David}, {de Laverny}, {De Luise}, {De March}, {De Ridder}, {de
  Souza}, {de Teodoro}, {de Torres}, {del Peloso}, {del Pozo}, {Delbo},
  {Delgado}, {Delgado}, {Delisle}, {Di Matteo}, {Diakite}, {Diener},
  {Distefano}, {Dolding}, {Eappachen}, {Edvardsson}, {Enke}, {Esquej}, {Fabre},
  {Fabrizio}, {Faigler}, {Fedorets}, {Fernique}, {Fienga}, {Figueras},
  {Fouron}, {Fragkoudi}, {Fraile}, {Franke}, {Gai}, {Garabato},
  {Garcia-Gutierrez}, {Garc{\'\i}a-Torres}, {Garofalo}, {Gavras}, {Gerlach},
  {Geyer}, {Giacobbe}, {Gilmore}, {Girona}, {Giuffrida}, {Gomel}, {Gomez},
  {Gonzalez-Santamaria}, {Gonz{\'a}lez-Vidal}, {Granvik},
  {Guti{\'e}rrez-S{\'a}nchez}, {Guy}, {Hauser}, {Haywood}, {Helmi}, {Hidalgo},
  {Hilger}, {H{\l}adczuk}, {Hobbs}, {Holland}, {Huckle}, {Jasniewicz},
  {Jonker}, {Juaristi Campillo}, {Julbe}, {Karbevska}, {Kervella}, {Khanna},
  {Kochoska}, {Kontizas}, {Kordopatis}, {Korn}, {Kostrzewa-Rutkowska},
  {Kruszy{\'n}ska}, {Lambert}, {Lanza}, {Lasne}, {Le Campion}, {Le Fustec},
  {Lebreton}, {Lebzelter}, {Leccia}, {Leclerc}, {Lecoeur-Taibi}, {Liao},
  {Licata}, {Lindstr{\o}m}, {Lister}, {Livanou}, {Lobel}, {Madrero Pardo},
  {Managau}, {Mann}, {Marchant}, {Marconi}, {Marcos Santos}, {Marinoni},
  {Marocco}, {Marshall}, {Martin Polo}, {Mart{\'\i}n-Fleitas}, {Masip},
  {Massari}, {Mastrobuono-Battisti}, {Mazeh}, {McMillan}, {Messina},
  {Michalik}, {Millar}, {Mints}, {Molina}, {Molinaro}, {Moln{\'a}r},
  {Montegriffo}, {Mor}, {Morbidelli}, {Morel}, {Morris}, {Mulone}, {Munoz},
  {Muraveva}, {Murphy}, {Musella}, {Noval}, {Ord{\'e}novic}, {Orr{\`u}},
  {Osinde}, {Pagani}, {Pagano}, {Palaversa}, {Palicio}, {Panahi}, {Pawlak},
  {Pe{\~n}alosa Esteller}, {Penttil{\"a}}, {Piersimoni}, {Pineau}, {Plachy},
  {Plum}, {Poggio}, {Poretti}, {Poujoulet}, {Pr{\v{s}}a}, {Pulone}, {Racero},
  {Ragaini}, {Rainer}, {Raiteri}, {Rambaux}, {Ramos}, {Ramos-Lerate}, {Re
  Fiorentin}, {Regibo}, {Reyl{\'e}}, {Ripepi}, {Riva}, {Rixon}, {Robichon},
  {Robin}, {Roelens}, {Rohrbasser}, {Romero-G{\'o}mez}, {Rowell}, {Royer},
  {Rybicki}, {Sadowski}, {Sagrist{\`a} Sell{\'e}s}, {Sahlmann}, {Salgado},
  {Salguero}, {Samaras}, {Sanchez Gimenez}, {Sanna}, {Santove{\~n}a},
  {Sarasso}, {Schultheis}, {Sciacca}, {Segol}, {Segovia}, {S{\'e}gransan},
  {Semeux}, {Shahaf}, {Siddiqui}, {Siebert}, {Siltala}, {Slezak}, {Smart},
  {Solano}, {Solitro}, {Souami}, {Souchay}, {Spagna}, {Spoto}, {Steele},
  {Steidelm{\"u}ller}, {Stephenson}, {S{\"u}veges}, {Szabados}, {Szegedi-Elek},
  {Taris}, {Tauran}, {Taylor}, {Teixeira}, {Thuillot}, {Tonello}, {Torra},
  {Torra}, {Turon}, {Unger}, {Vaillant}, {van Dillen}, {Vanel}, {Vecchiato},
  {Viala}, {Vicente}, {Voutsinas}, {Weiler}, {Wevers}, {Wyrzykowski}, {Yoldas},
  {Yvard}, {Zhao}, {Zorec}, {Zucker}, {Zurbach}, \& {Zwitter}}]{gaiaEDR3}
{Gaia Collaboration}, {Brown}, A.~G.~A., {Vallenari}, A., {et~al.} 2021, \aap,
  649, A1, \dodoi{10.1051/0004-6361/202039657}

\bibitem[{{G{\'a}sp{\'a}r} {et~al.}(2013){G{\'a}sp{\'a}r}, {Rieke}, \&
  {Balog}}]{Gaspar2013DebrisDisks}
{G{\'a}sp{\'a}r}, A., {Rieke}, G.~H., \& {Balog}, Z. 2013, \apj, 768, 25,
  \dodoi{10.1088/0004-637X/768/1/25}

\bibitem[{{Georgy} {et~al.}(2013){Georgy}, {Ekstr{\"o}m}, {Granada}, {Meynet},
  {Mowlavi}, {Eggenberger}, \& {Maeder}}]{Georgy2013BStarIsochrones}
{Georgy}, C., {Ekstr{\"o}m}, S., {Granada}, A., {et~al.} 2013, \aap, 553, A24,
  \dodoi{10.1051/0004-6361/201220558}

\bibitem[{Gray {et~al.}(2006)Gray, Corbally, Garrison, McFadden, Bubar,
  McGahee, O'Donoghue, \& Knox}]{gray_contributions_2006}
Gray, R.~O., Corbally, C.~J., Garrison, R.~F., {et~al.} 2006, The Astronomical
  Journal, 132, 161, \dodoi{10.1086/504637}

\bibitem[{{Gray} \& {Garrison}(1987)}]{GrayGarrison1987EarlyATypeStars}
{Gray}, R.~O., \& {Garrison}, R.~F. 1987, \apjs, 65, 581,
  \dodoi{10.1086/191237}

\bibitem[{{Hagelberg} {et~al.}(2020){Hagelberg}, {Engler}, {Fontanive},
  {Daemgen}, {Quanz}, {K{\"u}hn}, {Reggiani}, {Meyer}, {Jayawardhana}, \&
  {Kostov}}]{Hagelberg2020VIBES}
{Hagelberg}, J., {Engler}, N., {Fontanive}, C., {et~al.} 2020, \aap, 643, A98,
  \dodoi{10.1051/0004-6361/202039173}

\bibitem[{{Hamers}(2017)}]{Hamers2017HintsofHiddenCompanions}
{Hamers}, A.~S. 2017, \apjl, 835, L24, \dodoi{10.3847/2041-8213/835/2/L24}

\bibitem[{{Hamers} \& {Tremaine}(2017)}]{Hamers2017HJinGC}
{Hamers}, A.~S., \& {Tremaine}, S. 2017, \aj, 154, 272,
  \dodoi{10.3847/1538-3881/aa9926}

\bibitem[{Harris {et~al.}(2020)Harris, Millman, van~der Walt, Gommers,
  Virtanen, Cournapeau, Wieser, Taylor, Berg, Smith, Kern, Picus, Hoyer, van
  Kerkwijk, Brett, Haldane, del R{'{\i}}o, Wiebe, Peterson,
  G{'{e}}rard-Marchant, Sheppard, Reddy, Weckesser, Abbasi, Gohlke, \&
  Oliphant}]{Harris2020Numpy}
Harris, C.~R., Millman, K.~J., van~der Walt, S.~J., {et~al.} 2020, Nature, 585,
  357, \dodoi{10.1038/s41586-020-2649-2}

\bibitem[{{Harris} {et~al.}(2012){Harris}, {Andrews}, {Wilner}, \&
  {Kraus}}]{Harris2012TaurusMultipleStarSys}
{Harris}, R.~J., {Andrews}, S.~M., {Wilner}, D.~J., \& {Kraus}, A.~L. 2012,
  \apj, 751, 115, \dodoi{10.1088/0004-637X/751/2/115}

\bibitem[{{Heinze} {et~al.}(2010){Heinze}, {Hinz}, {Sivanandam}, {Kenworthy},
  {Meyer}, \& {Miller}}]{Heinze2010LongPeriodPlanets}
{Heinze}, A.~N., {Hinz}, P.~M., {Sivanandam}, S., {et~al.} 2010, \apj, 714,
  1551, \dodoi{10.1088/0004-637X/714/2/1551}

\bibitem[{{Herczeg} \& {Hillenbrand}(2014)}]{Herczeg2014TTauriPhotospheres}
{Herczeg}, G.~J., \& {Hillenbrand}, L.~A. 2014, \apj, 786, 97,
  \dodoi{10.1088/0004-637X/786/2/97}

\bibitem[{{Herman} {et~al.}(2019){Herman}, {Zhu}, \&
  {Wu}}]{Herman2019LongPeriodTransit}
{Herman}, M.~K., {Zhu}, W., \& {Wu}, Y. 2019, \aj, 157, 248,
  \dodoi{10.3847/1538-3881/ab1f70}

\bibitem[{{Hern{\'a}ndez} {et~al.}(2005){Hern{\'a}ndez}, {Calvet}, {Hartmann},
  {Brice{\~n}o}, {Sicilia-Aguilar}, \& {Berlind}}]{Hernandez2005HerbigOBAssoc}
{Hern{\'a}ndez}, J., {Calvet}, N., {Hartmann}, L., {et~al.} 2005, \aj, 129,
  856, \dodoi{10.1086/426918}

\bibitem[{{Hjorth} {et~al.}(2021){Hjorth}, {Albrecht}, {Hirano}, {Winn},
  {Dawson}, {Zanazzi}, {Knudstrup}, \& {Sato}}]{Hjorth2021BackwardSpinning}
{Hjorth}, M., {Albrecht}, S., {Hirano}, T., {et~al.} 2021, Proceedings of the
  National Academy of Science, 118, 2017418118, \dodoi{10.1073/pnas.2017418118}

\bibitem[{{Holman} \& {Wiegert}(1999)}]{Holman1999PlanetsInBinaries}
{Holman}, M.~J., \& {Wiegert}, P.~A. 1999, \aj, 117, 621,
  \dodoi{10.1086/300695}

\bibitem[{{Hoogerwerf}(2000)}]{Hoogerwerf2000OBAssocMembers}
{Hoogerwerf}, R. 2000, \mnras, 313, 43,
  \dodoi{10.1046/j.1365-8711.2000.03192.x}

\bibitem[{{Houk}(1978)}]{Houk1978HDSpectralTypes}
{Houk}, N. 1978, {Michigan catalogue of two-dimensional spectral types for the
  HD stars}

\bibitem[{{Houk} \& {Cowley}(1975)}]{Houk1975SpTofHDStarsSouthernDecl}
{Houk}, N., \& {Cowley}, A.~P. 1975, {University of Michigan Catalogue of
  two-dimensional spectral types for the HD stars. Volume I. Declinations -90\_
  to -53\_{\textflorin}0.}

\bibitem[{Hunter(2007)}]{Hunter:2007}
Hunter, J.~D. 2007, Computing In Science \& Engineering, 9, 90,
  \dodoi{10.1109/MCSE.2007.55}

\bibitem[{Kaib {et~al.}(2013)Kaib, Raymond, \& Duncan}]{kaib_planetary_2013}
Kaib, N.~A., Raymond, S.~N., \& Duncan, M. 2013, Nature, 493, 381,
  \dodoi{10.1038/nature11780}

\bibitem[{{Kasper} {et~al.}(2007){Kasper}, {Apai}, {Janson}, \&
  {Brandner}}]{Kasper2007LBandImaging}
{Kasper}, M., {Apai}, D., {Janson}, M., \& {Brandner}, W. 2007, \aap, 472, 321,
  \dodoi{10.1051/0004-6361:20077646}

\bibitem[{{Kenworthy} {et~al.}(2007){Kenworthy}, {Codona}, {Hinz}, {Angel},
  {Heinze}, \& {Sivanandam}}]{Kenworthy2007APP}
{Kenworthy}, M.~A., {Codona}, J.~L., {Hinz}, P.~M., {et~al.} 2007, \apj, 660,
  762, \dodoi{10.1086/513596}

\bibitem[{Kervella {et~al.}(2019)Kervella, Arenou, Mignard, \&
  Thévenin}]{kervella_stellar_2019}
Kervella, P., Arenou, F., Mignard, F., \& Thévenin, F. 2019, A\&A, 623, A72,
  \dodoi{10.1051/0004-6361/201834371}

\bibitem[{{Kervella} {et~al.}(2022){Kervella}, {Arenou}, \&
  {Th{\'e}venin}}]{Kervella2022}
{Kervella}, P., {Arenou}, F., \& {Th{\'e}venin}, F. 2022, \aap, 657, A7,
  \dodoi{10.1051/0004-6361/202142146}

\bibitem[{{Knutson} {et~al.}(2014){Knutson}, {Fulton}, {Montet}, {Kao}, {Ngo},
  {Howard}, {Crepp}, {Hinkley}, {Bakos}, {Batygin}, {Johnson}, {Morton}, \&
  {Muirhead}}]{Knutson2014FOHJ1}
{Knutson}, H.~A., {Fulton}, B.~J., {Montet}, B.~T., {et~al.} 2014, \apj, 785,
  126, \dodoi{10.1088/0004-637X/785/2/126}

\bibitem[{Kozai(1962)}]{kozai_secular_1962}
Kozai, Y. 1962, The Astronomical Journal, 67, 591, \dodoi{10.1086/108790}

\bibitem[{{Kratter} \& {Perets}(2012)}]{Kratter2012StarHoppers}
{Kratter}, K.~M., \& {Perets}, H.~B. 2012, \apj, 753, 91,
  \dodoi{10.1088/0004-637X/753/1/91}

\bibitem[{Kraus {et~al.}(2016)Kraus, Ireland, Huber, Mann, \&
  Dupuy}]{kraus_impact_2016}
Kraus, A.~L., Ireland, M.~J., Huber, D., Mann, A.~W., \& Dupuy, T.~J. 2016, The
  Astronomical Journal, 152, 8, \dodoi{10.3847/0004-6256/152/1/8}

\bibitem[{Kraus {et~al.}(2014)Kraus, Shkolnik, Allers, \&
  Liu}]{kraus_stellar_2014}
Kraus, A.~L., Shkolnik, E.~L., Allers, K.~N., \& Liu, M.~C. 2014, The
  Astronomical Journal, 147, 146, \dodoi{10.1088/0004-6256/147/6/146}

\bibitem[{{Lafreni{\`e}re} {et~al.}(2007){Lafreni{\`e}re}, {Marois}, {Doyon},
  {Nadeau}, \& {Artigau}}]{lafreniere_new_2007}
{Lafreni{\`e}re}, D., {Marois}, C., {Doyon}, R., {Nadeau}, D., \& {Artigau},
  {\'E}. 2007, \apj, 660, 770, \dodoi{10.1086/513180}

\bibitem[{{Lalitha} {et~al.}(2013){Lalitha}, {Fuhrmeister}, {Wolter},
  {Schmitt}, {Engels}, \& {Wieringa}}]{Lalitha2013ABDorAtmosphere}
{Lalitha}, S., {Fuhrmeister}, B., {Wolter}, U., {et~al.} 2013, \aap, 560, A69,
  \dodoi{10.1051/0004-6361/201321419}

\bibitem[{Lidov(1962)}]{lidov_evolution_1962}
Lidov, M.~L. 1962, Planetary and Space Science, 9, 719,
  \dodoi{10.1016/0032-0633(62)90129-0}

\bibitem[{Lindegren(2018{\natexlab{a}})}]{lindegren_re-normalising_2018}
Lindegren, L. 2018{\natexlab{a}}, Re-normalising the astrometric chi-square in
  {Gaia} DR2.
\newblock \url{http://www.rssd.esa.int/doc_fetch.php?id=3757412}

\bibitem[{Lindegren(2018{\natexlab{b}})}]{Lindegren2018RUWE}
---. 2018{\natexlab{b}}.
\newblock \url{http://www.rssd.esa.int/doc_fetch.php?id=3757412}

\bibitem[{Lindegren {et~al.}(2018)Lindegren, Hernández, Bombrun, Klioner,
  Bastian, Ramos-Lerate, de~Torres, Steidelmüller, Stephenson, Hobbs, Lammers,
  Biermann, Geyer, Hilger, Michalik, Stampa, McMillan, Castañeda, Clotet,
  Comoretto, Davidson, Fabricius, Gracia, Hambly, Hutton, Mora, Portell, van
  Leeuwen, Abbas, Abreu, Altmann, Andrei, Anglada, Balaguer-Núñez, Barache,
  Becciani, Bertone, Bianchi, Bouquillon, Bourda, Brüsemeister, Bucciarelli,
  Busonero, Buzzi, Cancelliere, Carlucci, Charlot, Cheek, Crosta, Crowley,
  de~Bruijne, de~Felice, Drimmel, Esquej, Fienga, Fraile, Gai, Garralda,
  González-Vidal, Guerra, Hauser, Hofmann, Holl, Jordan, Lattanzi, Lenhardt,
  Liao, Licata, Lister, Löffler, Marchant, Martin-Fleitas, Messineo, Mignard,
  Morbidelli, Poggio, Riva, Rowell, Salguero, Sarasso, Sciacca, Siddiqui,
  Smart, Spagna, Steele, Taris, Torra, van Elteren, van Reeven, \&
  Vecchiato}]{lindegren_gaia_2018}
Lindegren, L., Hernández, J., Bombrun, A., {et~al.} 2018, Astronomy and
  Astrophysics, 616, A2, \dodoi{10.1051/0004-6361/201832727}

\bibitem[{{Lord}(1992)}]{atran}
{Lord}, S.~D. 1992, NASA Technical Memorandum 103957: A new software tool for
  computing Earth's atmospheric transmission of near- and far-infrared
  radiation, Tech. rep.
\newblock \url{https://ntrs.nasa.gov/citations/19930010877}

\bibitem[{{Luhman} {et~al.}(2005){Luhman}, {Stauffer}, \&
  {Mamajek}}]{Luhman2005ABDorAge}
{Luhman}, K.~L., {Stauffer}, J.~R., \& {Mamajek}, E.~E. 2005, \apjl, 628, L69,
  \dodoi{10.1086/432617}

\bibitem[{{Lunine} {et~al.}(2008){Lunine}, {Fischer}, {Hammel}, {Henning},
  {Hillenbrand}, {Kasting}, {Laughlin}, {Macintosh}, {Marley}, {Melnick},
  {Monet}, {Noecker}, {Peale}, {Quirrenbach}, {Seager}, \&
  {Winn}}]{Lunine2008WorldsBeyond}
{Lunine}, J.~I., {Fischer}, D., {Hammel}, H., {et~al.} 2008, arXiv e-prints,
  arXiv:0808.2754.
\newblock \doarXiv{0808.2754}

\bibitem[{{Macintosh} {et~al.}(2014){Macintosh}, {Graham}, {Ingraham},
  {Konopacky}, {Marois}, {Perrin}, {Poyneer}, {Bauman}, {Barman}, {Burrows},
  {Cardwell}, {Chilcote}, {De Rosa}, {Dillon}, {Doyon}, {Dunn}, {Erikson},
  {Fitzgerald}, {Gavel}, {Goodsell}, {Hartung}, {Hibon}, {Kalas}, {Larkin},
  {Maire}, {Marchis}, {Marley}, {McBride}, {Millar-Blanchaer}, {Morzinski},
  {Norton}, {Oppenheimer}, {Palmer}, {Patience}, {Pueyo}, {Rantakyro},
  {Sadakuni}, {Saddlemyer}, {Savransky}, {Serio}, {Soummer},
  {Sivaramakrishnan}, {Song}, {Thomas}, {Wallace}, {Wiktorowicz}, \&
  {Wolff}}]{Macintosh2014GPI}
{Macintosh}, B., {Graham}, J.~R., {Ingraham}, P., {et~al.} 2014, Proceedings of
  the National Academy of Science, 111, 12661, \dodoi{10.1073/pnas.1304215111}

\bibitem[{{Mamajek} \& {Hillenbrand}(2008)}]{MamajekAges2008}
{Mamajek}, E.~E., \& {Hillenbrand}, L.~A. 2008, \apj, 687, 1264,
  \dodoi{10.1086/591785}

\bibitem[{{Marley} {et~al.}(2021){Marley}, {Saumon}, {Visscher}, {Lupu},
  {Freedman}, {Morley}, {Fortney}, {Seay}, {Smith}, {Teal}, \&
  {Wang}}]{Marley2021SonoraBobcat}
{Marley}, M.~S., {Saumon}, D., {Visscher}, C., {et~al.} 2021, \apj, 920, 85,
  \dodoi{10.3847/1538-4357/ac141d}

\bibitem[{{Marois} {et~al.}(2000){Marois}, {Doyon}, {Racine}, \&
  {Nadeau}}]{Marois2000SDI}
{Marois}, C., {Doyon}, R., {Racine}, R., \& {Nadeau}, D. 2000, \pasp, 112, 91,
  \dodoi{10.1086/316492}

\bibitem[{Marois {et~al.}(2006)Marois, Lafrenière, Doyon, Macintosh, \&
  Nadeau}]{marois_angular_2006}
Marois, C., Lafrenière, D., Doyon, R., Macintosh, B., \& Nadeau, D. 2006, The
  Astrophysical Journal, 641, 556, \dodoi{10.1086/500401}

\bibitem[{{Mason} {et~al.}(2001){Mason}, {Wycoff}, {Hartkopf}, {Douglass}, \&
  {Worley}}]{Mason2001WDS}
{Mason}, B.~D., {Wycoff}, G.~L., {Hartkopf}, W.~I., {Douglass}, G.~G., \&
  {Worley}, C.~E. 2001, \aj, 122, 3466, \dodoi{10.1086/323920}

\bibitem[{Mawet {et~al.}(2014)Mawet, Milli, Wahhaj, Pelat, Absil, Delacroix,
  Boccaletti, Kasper, Kenworthy, Marois, Mennesson, \&
  Pueyo}]{mawet_fundamental_2014}
Mawet, D., Milli, J., Wahhaj, Z., {et~al.} 2014, The Astrophysical Journal,
  792, 97, \dodoi{10.1088/0004-637X/792/2/97}

\bibitem[{Maíz~Apellániz {et~al.}(2021)Maíz~Apellániz,
  Pantaleoni~González, \& Barbá}]{maiz_apellaniz_validation_2021}
Maíz~Apellániz, J., Pantaleoni~González, M., \& Barbá, R.~H. 2021, A\&A,
  Volume 649, id.A13, 10 pp., 649, A13, \dodoi{10.1051/0004-6361/202140418}

\bibitem[{{McCarthy} \& {White}(2012)}]{McCarthy2012SizesOfNearestYoungStars}
{McCarthy}, K., \& {White}, R.~J. 2012, \aj, 143, 134,
  \dodoi{10.1088/0004-6256/143/6/134}

\bibitem[{McDonald {et~al.}(2012)McDonald, Zijlstra, \&
  Boyer}]{mcdonald_fundamental_2012}
McDonald, I., Zijlstra, A.~A., \& Boyer, M.~L. 2012, Monthly Notices of the
  Royal Astronomical Society, 427, 343,
  \dodoi{10.1111/j.1365-2966.2012.21873.x}

\bibitem[{Messina {et~al.}(2016)Messina, Lanzafame, Feiden, Millward, Desidera,
  Buccino, Curtis, Jofré, Kehusmaa, Medhi, Monard, \&
  Petrucci}]{messina_rotation-lithium_2016}
Messina, S., Lanzafame, A.~C., Feiden, G.~A., {et~al.} 2016, Astronomy and
  Astrophysics, 596, A29, \dodoi{10.1051/0004-6361/201628524}

\bibitem[{Messina {et~al.}(2017)Messina, Lanzafame, Malo, Desidera, Buccino,
  Zhang, Artemenko, Millward, \& Hambsch}]{messina__2017}
Messina, S., Lanzafame, A.~C., Malo, L., {et~al.} 2017, Astronomy and
  Astrophysics, 607, A3, \dodoi{10.1051/0004-6361/201730444}

\bibitem[{{Mewe} {et~al.}(1996){Mewe}, {Kaastra}, {White}, \&
  {Pallavicini}}]{Mewe1996ABDoradus}
{Mewe}, R., {Kaastra}, J.~S., {White}, S.~M., \& {Pallavicini}, R. 1996, \aap,
  315, 170

\bibitem[{{Mittal} {et~al.}(2015){Mittal}, {Chen}, {Jang-Condell}, {Manoj},
  {Sargent}, {Watson}, \& {Lisse}}]{Mittal2015DebrisDisk}
{Mittal}, T., {Chen}, C.~H., {Jang-Condell}, H., {et~al.} 2015, \apj, 798, 87,
  \dodoi{10.1088/0004-637X/798/2/87}

\bibitem[{Moe \& Kratter(2019)}]{moe_impact_2019}
Moe, M., \& Kratter, K.~M. 2019, arXiv e-prints, 1912, arXiv:1912.01699.
\newblock \url{http://adsabs.harvard.edu/abs/2019arXiv191201699M}

\bibitem[{Morzinski {et~al.}(2015)Morzinski, Males, Skemer, Close, Hinz,
  Rodigas, Puglisi, Esposito, Riccardi, Pinna, Xompero, Briguglio, Bailey,
  Follette, Kopon, Weinberger, \& Wu}]{morzinski_magellan_2015}
Morzinski, K.~M., Males, J.~R., Skemer, A.~J., {et~al.} 2015, The Astrophysical
  Journal, 815, 108, \dodoi{10.1088/0004-637X/815/2/108}

\bibitem[{{Mudryk} \& {Wu}(2006)}]{Mudryk2006ResonanceOverlap}
{Mudryk}, L.~R., \& {Wu}, Y. 2006, \apj, 639, 423, \dodoi{10.1086/499347}

\bibitem[{{Mustill} {et~al.}(2021){Mustill}, {Davies}, {Blunt}, \&
  {Howard}}]{Mustill2021HR5183bKozaiLidoz}
{Mustill}, A.~J., {Davies}, M.~B., {Blunt}, S., \& {Howard}, A. 2021, arXiv
  e-prints, arXiv:2102.06031.
\newblock \doarXiv{2102.06031}

\bibitem[{Newton {et~al.}(2019)Newton, Mann, Tofflemire, Pearce, Rizzuto,
  Vanderburg, Martinez, Wang, Ruffio, Kraus, Johnson, Thao, Wood, Rampalli,
  Nielsen, Collins, Dragomir, Hellier, Anderson, Barclay, Brown, Feiden, Hart,
  Isopi, Kielkopf, Mallia, Nelson, Rodriguez, Stockdale, Waite, Wright,
  Lissauer, Ricker, Vanderspek, Latham, Seager, Winn, Jenkins, Bouma, Burke,
  Davies, Fausnaugh, Li, Morris, Mukai, Villaseñor, Villeneuva, De~Rosa,
  Macintosh, Mengel, Okumura, \& Wittenmyer}]{newton_tess_2019}
Newton, E.~R., Mann, A.~W., Tofflemire, B.~M., {et~al.} 2019, The Astrophysical
  Journal Letters, 880, L17, \dodoi{10.3847/2041-8213/ab2988}

\bibitem[{{Newton} {et~al.}(2021){Newton}, {Mann}, {Kraus}, {Livingston},
  {Vanderburg}, {Curtis}, {Thao}, {Hawkins}, {Wood}, {Rizzuto}, {Soubkiou},
  {Tofflemire}, {Zhou}, {Crossfield}, {Pearce}, {Collins}, {Conti}, {Tan},
  {Villeneuva}, {Spencer}, {Dragomir}, {Quinn}, {Jensen}, {Collins},
  {Stockdale}, {Cloutier}, {Hellier}, {Benkhaldoun}, {Ziegler}, {Brice{\~n}o},
  {Law}, {Benneke}, {Christiansen}, {Gorjian}, {Kane}, {Kreidberg}, {Morales},
  {Werner}, {Twicken}, {Levine}, {Ciardi}, {Guerrero}, {Hesse}, {Quintana},
  {Shiao}, {Smith}, {Torres}, {Ricker}, {Vanderspek}, {Seager}, {Winn},
  {Jenkins}, \& {Latham}}]{Newton2021ThreeSmallTESSPlanets}
{Newton}, E.~R., {Mann}, A.~W., {Kraus}, A.~L., {et~al.} 2021, \aj, 161, 65,
  \dodoi{10.3847/1538-3881/abccc6}

\bibitem[{{Ngo} {et~al.}(2015){Ngo}, {Knutson}, {Hinkley}, {Crepp}, {Bechter},
  {Batygin}, {Howard}, {Johnson}, {Morton}, \& {Muirhead}}]{Ngo2015FOHJ2}
{Ngo}, H., {Knutson}, H.~A., {Hinkley}, S., {et~al.} 2015, \apj, 800, 138,
  \dodoi{10.1088/0004-637X/800/2/138}

\bibitem[{{Ngo} {et~al.}(2016){Ngo}, {Knutson}, {Hinkley}, {Bryan}, {Crepp},
  {Batygin}, {Crossfield}, {Hansen}, {Howard}, {Johnson}, {Mawet}, {Morton},
  {Muirhead}, \& {Wang}}]{Ngo2016FOHJ4}
---. 2016, \apj, 827, 8, \dodoi{10.3847/0004-637X/827/1/8}

\bibitem[{Nielsen {et~al.}(2019)Nielsen, De~Rosa, Macintosh, Wang, Ruffio,
  Chiang, Marley, Saumon, Savransky, Ammons, Bailey, Barman, Blain, Bulger,
  Burrows, Chilcote, Cotten, Czekala, Doyon, Duchêne, Esposito, Fabrycky,
  Fitzgerald, Follette, Fortney, Gerard, Goodsell, Graham, Greenbaum, Hibon,
  Hinkley, Hirsch, Hom, Hung, Dawson, Ingraham, Kalas, Konopacky, Larkin, Lee,
  Lin, Maire, Marchis, Marois, Metchev, Millar-Blanchaer, Morzinski,
  Oppenheimer, Palmer, Patience, Perrin, Poyneer, Pueyo, Rafikov, Rajan,
  Rameau, Rantakyrö, Ren, Schneider, Sivaramakrishnan, Song, Soummer, Tallis,
  Thomas, Ward-Duong, \& Wolff}]{nielsen_gemini_2019}
Nielsen, E.~L., De~Rosa, R.~J., Macintosh, B., {et~al.} 2019, The Astronomical
  Journal, 158, 13, \dodoi{10.3847/1538-3881/ab16e9}

\bibitem[{{Osborn} {et~al.}(2020){Osborn}, {Ansdell}, {Ioannou}, {Sasdelli},
  {Angerhausen}, {Caldwell}, {Jenkins}, {R{\"a}issi}, \&
  {Smith}}]{Osborn2020ClassificationOfTESS}
{Osborn}, H.~P., {Ansdell}, M., {Ioannou}, Y., {et~al.} 2020, \aap, 633, A53,
  \dodoi{10.1051/0004-6361/201935345}

\bibitem[{{Otten} {et~al.}(2014){Otten}, {Snik}, {Kenworthy}, {Miskiewicz}, \&
  {Escuti}}]{Otten2014APP}
{Otten}, G. P.~P.~L., {Snik}, F., {Kenworthy}, M.~A., {Miskiewicz}, M.~N., \&
  {Escuti}, M.~J. 2014, Optics Express, 22, 30287, \dodoi{10.1364/OE.22.030287}

\bibitem[{Pecaut {et~al.}(2012)Pecaut, Mamajek, \& Bubar}]{pecaut_revised_2012}
Pecaut, M.~J., Mamajek, E.~E., \& Bubar, E.~J. 2012, The Astrophysical Journal,
  746, 154, \dodoi{10.1088/0004-637X/746/2/154}

\bibitem[{{Penoyre} {et~al.}(2021){Penoyre}, {Belokurov}, \&
  {Evans}}]{Penoyre2021Astrometric1}
{Penoyre}, Z., {Belokurov}, V., \& {Evans}, N.~W. 2021, arXiv e-prints,
  arXiv:2111.10380.
\newblock \doarXiv{2111.10380}

\bibitem[{{Penoyre} {et~al.}(2020){Penoyre}, {Belokurov}, {Wyn Evans},
  {Everall}, \& {Koposov}}]{Penoyre2020BinaryDeviations}
{Penoyre}, Z., {Belokurov}, V., {Wyn Evans}, N., {Everall}, A., \& {Koposov},
  S.~E. 2020, \mnras, 495, 321, \dodoi{10.1093/mnras/staa1148}

\bibitem[{{Piskorz} {et~al.}(2015){Piskorz}, {Knutson}, {Ngo}, {Muirhead},
  {Batygin}, {Crepp}, {Hinkley}, \& {Morton}}]{Piskorz2015FOHJ3}
{Piskorz}, D., {Knutson}, H.~A., {Ngo}, H., {et~al.} 2015, \apj, 814, 148,
  \dodoi{10.1088/0004-637X/814/2/148}

\bibitem[{{Poleski} {et~al.}(2021){Poleski}, {Skowron}, {Mr{\'o}z}, {Udalski},
  {Szyma{\'n}ski}, {Pietrukowicz}, {Ulaczyk}, {Rybicki}, {Iwanek}, {Wrona}, \&
  {Gromadzki}}]{Poleski2021MicrolensingWidePlanetsCommon}
{Poleski}, R., {Skowron}, J., {Mr{\'o}z}, P., {et~al.} 2021, \actaa, 71, 1,
  \dodoi{10.32023/0001-5237/71.1.1}

\bibitem[{{Price-Whelan} {et~al.}(2018){Price-Whelan}, {Sip{'{o}}cz},
  {G{"u}nther}, {Lim}, {Crawford}, {Conseil}, {Shupe}, {Craig}, {Dencheva},
  {Ginsburg}, {VanderPlas}, {Bradley}, {P{'e}rez-Su{'a}rez}, {de Val-Borro},
  {Paper Contributors}, {Aldcroft}, {Cruz}, {Robitaille}, {Tollerud},
  {Coordination Committee}, {Ardelean}, {Babej}, {Bach}, {Bachetti}, {Bakanov},
  {Bamford}, {Barentsen}, {Barmby}, {Baumbach}, {Berry}, {Biscani}, {Boquien},
  {Bostroem}, {Bouma}, {Brammer}, {Bray}, {Breytenbach}, {Buddelmeijer},
  {Burke}, {Calderone}, {Cano Rodr{'i}guez}, {Cara}, {Cardoso}, {Cheedella},
  {Copin}, {Corrales}, {Crichton}, {D{ extquoteright}Avella}, {Deil},
  {Depagne}, {Dietrich}, {Donath}, {Droettboom}, {Earl}, {Erben}, {Fabbro},
  {Ferreira}, {Finethy}, {Fox}, {Garrison}, {Gibbons}, {Goldstein}, {Gommers},
  {Greco}, {Greenfield}, {Groener}, {Grollier}, {Hagen}, {Hirst}, {Homeier},
  {Horton}, {Hosseinzadeh}, {Hu}, {Hunkeler}, {Ivezi{'c}}, {Jain}, {Jenness},
  {Kanarek}, {Kendrew}, {Kern}, {Kerzendorf}, {Khvalko}, {King}, {Kirkby},
  {Kulkarni}, {Kumar}, {Lee}, {Lenz}, {Littlefair}, {Ma}, {Macleod},
  {Mastropietro}, {McCully}, {Montagnac}, {Morris}, {Mueller}, {Mumford},
  {Muna}, {Murphy}, {Nelson}, {Nguyen}, {Ninan}, {N{"o}the}, {Ogaz}, {Oh},
  {Parejko}, {Parley}, {Pascual}, {Patil}, {Patil}, {Plunkett}, {Prochaska},
  {Rastogi}, {Reddy Janga}, {Sabater}, {Sakurikar}, {Seifert}, {Sherbert},
  {Sherwood-Taylor}, {Shih}, {Sick}, {Silbiger}, {Singanamalla}, {Singer},
  {Sladen}, {Sooley}, {Sornarajah}, {Streicher}, {Teuben}, {Thomas},
  {Tremblay}, {Turner}, {Terr{'o}n}, {van Kerkwijk}, {de la Vega}, {Watkins},
  {Weaver}, {Whitmore}, {Woillez}, {Zabalza}, \& {Contributors}}]{astropy:2018}
{Price-Whelan}, A.~M., {Sip{'{o}}cz}, B.~M., {G{"u}nther}, H.~M., {et~al.}
  2018, aj, 156, 123, \dodoi{10.3847/1538-3881/aabc4f}

\bibitem[{{Racine} {et~al.}(1999){Racine}, {Walker}, {Nadeau}, {Doyon}, \&
  {Marois}}]{Racine1999SDI}
{Racine}, R., {Walker}, G. A.~H., {Nadeau}, D., {Doyon}, R., \& {Marois}, C.
  1999, \pasp, 111, 587, \dodoi{10.1086/316367}

\bibitem[{{Rameau} {et~al.}(2015){Rameau}, {Chauvin}, {Lagrange}, {Maire},
  {Boccaletti}, \& {Bonnefoy}}]{Rameau2015SDIDetLimits}
{Rameau}, J., {Chauvin}, G., {Lagrange}, A.~M., {et~al.} 2015, \aap, 581, A80,
  \dodoi{10.1051/0004-6361/201525879}

\bibitem[{{Riaz} {et~al.}(2006){Riaz}, {Gizis}, \&
  {Harvin}}]{Riaz2006NewMDwarfs}
{Riaz}, B., {Gizis}, J.~E., \& {Harvin}, J. 2006, \aj, 132, 866,
  \dodoi{10.1086/505632}

\bibitem[{{Richey-Yowell} {et~al.}(2019){Richey-Yowell}, {Shkolnik},
  {Schneider}, {Osby}, {Barman}, \& {Meadows}}]{RicheyYowell2019HAZMATKStars}
{Richey-Yowell}, T., {Shkolnik}, E.~L., {Schneider}, A.~C., {et~al.} 2019,
  \apj, 872, 17, \dodoi{10.3847/1538-4357/aafa74}

\bibitem[{{Ricker} {et~al.}(2015){Ricker}, {Winn}, {Vanderspek}, {Latham},
  {Bakos}, {Bean}, {Berta-Thompson}, {Brown}, {Buchhave}, {Butler}, {Butler},
  {Chaplin}, {Charbonneau}, {Christensen-Dalsgaard}, {Clampin}, {Deming},
  {Doty}, {De Lee}, {Dressing}, {Dunham}, {Endl}, {Fressin}, {Ge}, {Henning},
  {Holman}, {Howard}, {Ida}, {Jenkins}, {Jernigan}, {Johnson}, {Kaltenegger},
  {Kawai}, {Kjeldsen}, {Laughlin}, {Levine}, {Lin}, {Lissauer}, {MacQueen},
  {Marcy}, {McCullough}, {Morton}, {Narita}, {Paegert}, {Palle}, {Pepe},
  {Pepper}, {Quirrenbach}, {Rinehart}, {Sasselov}, {Sato}, {Seager},
  {Sozzetti}, {Stassun}, {Sullivan}, {Szentgyorgyi}, {Torres}, {Udry}, \&
  {Villasenor}}]{Ricker2015TESSMission}
{Ricker}, G.~R., {Winn}, J.~N., {Vanderspek}, R., {et~al.} 2015, Journal of
  Astronomical Telescopes, Instruments, and Systems, 1, 014003,
  \dodoi{10.1117/1.JATIS.1.1.014003}

\bibitem[{Rodigas {et~al.}(2015)Rodigas, Weinberger, Mamajek, Males, Close,
  Morzinski, Hinz, \& Kaib}]{rodigas_direct_2015}
Rodigas, T.~J., Weinberger, A., Mamajek, E.~E., {et~al.} 2015, The
  Astrophysical Journal, 811, 157, \dodoi{10.1088/0004-637X/811/2/157}

\bibitem[{{Samus'} {et~al.}(2003){Samus'}, {Goranskii}, {Durlevich}, {Zharova},
  {Kazarovets}, {Kireeva}, {Pastukhova}, {Williams}, \&
  {Hazen}}]{Samus2003GenCatVariableStars}
{Samus'}, N.~N., {Goranskii}, V.~P., {Durlevich}, O.~V., {et~al.} 2003,
  Astronomy Letters, 29, 468, \dodoi{10.1134/1.1589864}

\bibitem[{{Sanghi} {et~al.}(2021){Sanghi}, {Zhou}, \&
  {Bowler}}]{Sanghi2021RDIwithWFCArchive}
{Sanghi}, A., {Zhou}, Y., \& {Bowler}, B.~P. 2021, arXiv e-prints,
  arXiv:2112.10777.
\newblock \doarXiv{2112.10777}

\bibitem[{{Schneider} {et~al.}(2012){Schneider}, {Melis}, \&
  {Song}}]{Schneider2012TWAMembership}
{Schneider}, A., {Melis}, C., \& {Song}, I. 2012, \apj, 754, 39,
  \dodoi{10.1088/0004-637X/754/1/39}

\bibitem[{{Sebastian} {et~al.}(2021){Sebastian}, {Gillon}, {Ducrot},
  {Pozuelos}, {Garcia}, {G{\"u}nther}, {Delrez}, {Queloz}, {Demory}, {Triaud},
  {Burgasser}, {de Wit}, {Burdanov}, {Dransfield}, {Jehin}, {McCormac},
  {Murray}, {Niraula}, {Pedersen}, {Rackham}, {Sohy}, {Thompson}, \& {Van
  Grootel}}]{Sebastian2021Speculoos}
{Sebastian}, D., {Gillon}, M., {Ducrot}, E., {et~al.} 2021, \aap, 645, A100,
  \dodoi{10.1051/0004-6361/202038827}

\bibitem[{Soummer {et~al.}(2012)Soummer, Pueyo, \&
  Larkin}]{soummer_detection_2012}
Soummer, R., Pueyo, L., \& Larkin, J. 2012, The Astrophysical Journal Letters,
  755, L28, \dodoi{10.1088/2041-8205/755/2/L28}

\bibitem[{{Spencer Jones} \& {Jackson}(1939)}]{SpencerJones1939SpT}
{Spencer Jones}, H., \& {Jackson}, J. 1939, {Catalogue of 20554 faint stars in
  the Cape Astrographic Zone -40 deg. to -52 deg. For the equinox of 1900.0
  giving positions, precessions, proper motions and photographic magnitudes}

\bibitem[{{Stassun} \& {Torres}(2021)}]{Stassun2021EclipsingBinaries}
{Stassun}, K.~G., \& {Torres}, G. 2021, \apjl, 907, L33,
  \dodoi{10.3847/2041-8213/abdaad}

\bibitem[{Tetzlaff {et~al.}(2011)Tetzlaff, Neuhäuser, \&
  Hohle}]{tetzlaff_catalogue_2011}
Tetzlaff, N., Neuhäuser, R., \& Hohle, M.~M. 2011, Monthly Notices of the
  Royal Astronomical Society, 410, 190,
  \dodoi{10.1111/j.1365-2966.2010.17434.x}

\bibitem[{{Thalmann} {et~al.}(2014){Thalmann}, {Desidera}, {Bonavita},
  {Janson}, {Usuda}, {Henning}, {K{\"o}hler}, {Carson}, {Boccaletti},
  {Bergfors}, {Brandner}, {Feldt}, {Goto}, {Klahr}, {Marzari}, \&
  {Mordasini}}]{Thalmann2014SPOTS}
{Thalmann}, C., {Desidera}, S., {Bonavita}, M., {et~al.} 2014, \aap, 572, A91,
  \dodoi{10.1051/0004-6361/201424581}

\bibitem[{{Tobal}(2000)}]{Tobal2000VisualBinaries}
{Tobal}, T. 2000, Observations et Travaux, 52, 67

\bibitem[{Torres {et~al.}(2000)Torres, da~Silva, Quast, de~la Reza, \&
  Jilinski}]{torres_new_2000}
Torres, C. A.~O., da~Silva, L., Quast, G.~R., de~la Reza, R., \& Jilinski, E.
  2000, The Astronomical Journal, 120, 1410, \dodoi{10.1086/301539}

\bibitem[{Torres {et~al.}(2006)Torres, Quast, da~Silva, de~La~Reza, Melo, \&
  Sterzik}]{torres_search_2006}
Torres, C. A.~O., Quast, G.~R., da~Silva, L., {et~al.} 2006, Astronomy and
  Astrophysics, 460, 695, \dodoi{10.1051/0004-6361:20065602}

\bibitem[{{van Leeuwen}(2007)}]{vanLeeuwen2007HipReduction}
{van Leeuwen}, F. 2007, \aap, 474, 653, \dodoi{10.1051/0004-6361:20078357}

\bibitem[{Venner {et~al.}(2021)Venner, Vanderburg, \&
  Pearce}]{venner_true_2021}
Venner, A., Vanderburg, A., \& Pearce, L.~A. 2021, AJ, 162, 12,
  \dodoi{10.3847/1538-3881/abf932}

\bibitem[{{Vican}(2012)}]{Vican2012AgeDEBRISSurvey}
{Vican}, L. 2012, \aj, 143, 135, \dodoi{10.1088/0004-6256/143/6/135}

\bibitem[{{Vilhu} \& {Linsky}(1987)}]{Vilhu1987ABDorB}
{Vilhu}, O., \& {Linsky}, J.~L. 1987, \pasp, 99, 1071, \dodoi{10.1086/132079}

\bibitem[{Virtanen {et~al.}(2020)Virtanen, Gommers, Oliphant, Haberland, Reddy,
  Cournapeau, Burovski, Peterson, Weckesser, Bright, {van der Walt}, Brett,
  Wilson, Millman, Mayorov, Nelson, Jones, Kern, Larson, Carey, Polat, Feng,
  Moore, {VanderPlas}, Laxalde, Perktold, Cimrman, Henriksen, Quintero, Harris,
  Archibald, Ribeiro, Pedregosa, {van Mulbregt}, \& {SciPy 1.0
  Contributors}}]{2020SciPy-NMeth}
Virtanen, P., Gommers, R., Oliphant, T.~E., {et~al.} 2020, Nature Methods, 17,
  261, \dodoi{10.1038/s41592-019-0686-2}

\bibitem[{{von Zeipel}(1910)}]{VonZeipel1910}
{von Zeipel}, H. 1910, Astronomische Nachrichten, 183, 345,
  \dodoi{10.1002/asna.19091832202}

\bibitem[{{Xuan} {et~al.}(2020){Xuan}, {Kennedy}, {Wyatt}, \&
  {Yelverton}}]{Xuan2020MutualInclinations}
{Xuan}, J.~W., {Kennedy}, G.~M., {Wyatt}, M.~C., \& {Yelverton}, B. 2020,
  \mnras, 499, 5059, \dodoi{10.1093/mnras/staa3155}

\bibitem[{Ziegler {et~al.}(2020)Ziegler, Tokovinin, Briceño, Mang, Law, \&
  Mann}]{ziegler_soar_2020}
Ziegler, C., Tokovinin, A., Briceño, C., {et~al.} 2020, The Astronomical
  Journal, 159, 19, \dodoi{10.3847/1538-3881/ab55e9}

\bibitem[{{Z{\'u}{\~n}iga-Fern{\'a}ndez}
  {et~al.}(2021){Z{\'u}{\~n}iga-Fern{\'a}ndez}, {Bayo}, {Elliott}, {Zamora},
  {Corval{\'a}n}, {Haubois}, {Corral-Santana}, {Olofsson}, {Hu{\'e}lamo},
  {Sterzik}, {Torres}, {Quast}, \& {Melo}}]{Zuniga-Fernandez2021UpdatedSACY}
{Z{\'u}{\~n}iga-Fern{\'a}ndez}, S., {Bayo}, A., {Elliott}, P., {et~al.} 2021,
  \aap, 645, A30, \dodoi{10.1051/0004-6361/202037830}

\bibitem[{{Zuckerman}(2019)}]{Zuckerman2019ArgusAssoc}
{Zuckerman}, B. 2019, \apj, 870, 27, \dodoi{10.3847/1538-4357/aaee66}

\bibitem[{Zuckerman(2019)}]{zuckerman_nearby_2019}
Zuckerman, B. 2019, The Astrophysical Journal, 870, 27,
  \dodoi{10.3847/1538-4357/aaee66}

\bibitem[{Zuckerman {et~al.}(2013)Zuckerman, Vican, Song, \&
  Schneider}]{zuckerman_young_2013}
Zuckerman, B., Vican, L., Song, I., \& Schneider, A. 2013, The Astrophysical
  Journal, 778, 5, \dodoi{10.1088/0004-637X/778/1/5}

\end{thebibliography}
\bibliographystyle{aasjournal}

\appendix
\section{Binary System Details}\label{appendixA}
\changed{Here we present details of each binary system in our survey.} In the following discussion, we have made use of the Gaia EDR3 Renormalized Unit Weight Error (RUWE) metric as a signpost for the (non-)existence of unresolved companions.  RUWE encapsulates all sources of error in the fit to the assumed single star astrometric model, corrected for correlation with source color and magnitude. RUWE~$\approx$~1 is expected for a well-behaved solution \citep{lindegren_re-normalising_2018}\footnote{\url{https://www.cosmos.esa.int/web/gaia/dr2-known-issues\#AstrometryConsiderations}}.  \changed{RUWE has been shown to be sensitive to companions on separations from $\sim$0.2\arcsec-1.2\arcsec\ \citep{Kervella2022}, periods of months $\lesssim$ P $\lesssim$ 10 years, and mass and luminosity ratios $<$1, for which photocenter motion is perturbed from motion of a single star model \citep{Penoyre2021Astrometric1}}.
RUWE 1--1.4 has been shown to be \changed{very strongly correlated with photocenter perturbation from an unresolved companion} \citep{Stassun2021EclipsingBinaries, Belokurov2020UnresolvedBinariesDR2}; RUWE 1.4--2 indicates deviation from a single star model but the astrometry may still be reliable \citep{maiz_apellaniz_validation_2021}; RUWE $>$2 indicates signficant devation from a single star model.  \changed{Elevated RUWE in young sources ($\tau \lesssim$ 10 Myr) may also be attributed to the presence of a disk \citep{Fitton2022RUWEDisks}}. 

Additionally, we have made use of the Hipparcos--Gaia Catalog of Accelerations \citep{Brandt2021_HGCA_ERD3} as a signpost for unresolved companions on wider orbits for which RUWE is less sensitive.  Significant difference between the long-baseline proper motion vector and the instantaneous PM vectors in Hipparcos and Gaia observation epochs (proper motion anomaly, PMa) can indicate the presence of an unresolved companion causing acceleration.  We made use of the \citealt{kervella_stellar_2019} (for DR2) and \citealt{Kervella2022} (for EDR3) PMa catalogs to indicate the (non-)existence of significant PMa; S/N $>$ 3 is considered significant in \citealt{kervella_stellar_2019}.  \changed{We note that PMa sensitivity depends on mass, distance, and orbital period, and use it as a indicator only and not a tool for prediction of companion properties.}

\textbf{\textit{HD 36705}} --- HD 36705 (AB Dor) is a nearby (15 pc), K0V+M5-6 \citep{torres_search_2006}, 9\arcsec\ T-Tauri type binary in the AB Doradus moving group with masses 0.865~$\pm$~0.034~\Msun\ \citep{Close2005ABDorC} and  0.37\Msun\ \citep{Sebastian2021Speculoos} respectively.  AB~Dor~A is an ultra-fast rotator that is chromospherically active \citep{Lalitha2013ABDorAtmosphere}.  AB~Dor~B (RST~137B, HBC~434) was first detected by \cite{Vilhu1987ABDorB} in X-ray emission.  \cite{Close2005ABDorC} placed the age of AB Dor A at 50$^{+50}_{-20}$ Myr due to lithium \citep{Mewe1996ABDoradus}, X-ray activity, and rotation rate, younger than the average age of 149$^{+51}_{-19}$ for the AB Dor moving group \citep{bell_self-consistent_2015}.  A wide variety of ages have been estimated for AB Dor spanning 5 Myr to 240 Myr (75-150 Myr-- \citealt{Luhman2005ABDorAge}; 100 Myr-- \citealt{MamajekAges2008}, 70 Myr-- \citealt{Chauvin2010DeepImagingSurvey}, 240 Myr-- \citealt{Vican2012AgeDEBRISSurvey}, 10 Myr-- \citealt{ Gaspar2013DebrisDisks}, 150 Myr-- \citealt{RicheyYowell2019HAZMATKStars}, 5.6 Myr-- \citealt{Binks2020YoungStars}).  We adopted the average age of 100 Myr for our analysis.

Both stars have their own subsystems.  \cite{Close2005ABDorC} detected a significantly redder companion to AB~Dor~A, which they named AB~Dor~C 
, at 0.156~$\pm$~0.010\arcsec\ and position angle 127~$\pm$~1$^{\circ}$, with a dynamical mass of 0.090~$\pm$~0.008~\Msun \citep{Azulay2017ABDorACDynMass}. \cite{Climent2019ABDorC} inferred the presence of a companion to AB~Dor~C in VLTI/AMBER J,H,K band, with 38~$\pm$~1 mas separation and masses of 0.072~$\pm$~0.013 and 0.013~$\pm$~0.01~\Msun\ for AB~Dor~Ca,~Cb respectively.  \cite{Close2005ABDorC} also detected a 0.070\arcsec\ companion to AB~Dor~B (AB~Dor~Ba,~Bb) at position angle 238.6~$\pm$~0.38$^{\circ}$.  AB~Dor~A and B are resolved in Gaia EDR3, both with large RUWE values (A: RUWE~=~25.13; B: RUWE~=~3.52), \changed{nevertheless uncertainties of astrometric quantities are small, the parallaxes are consistent with Hipparcos \citep{vanLeeuwen2007HipReduction}, and separation/PA is consistent with the Washington Double Star catalog \citep[WDS;][]{Mason2001WDS}}.  No significant IR excess was detected by \cite{mcdonald_fundamental_2012} (\changed{average excess infrared (EIR)} = 1.108 for 4.2-25~$\mu$m, where EIR~=~1 indicates no excess).

\textbf{\textit{HD 37551}} --- HD 37551 (WX Col) is a young, 4\arcsec\ binary at a distance of 80 pc.  HD 37551 A and B have a mass of 0.93\Msun\ and 0.80\Msun\ respectively \citep{Anders2019AstrophysicalParametersGaiaDR2}, and spectral type G7V and K1V \citep{torres_search_2006}.  Both stars have RUWE values $\sim$1 (A: RUWE = 0.97; B: RUWE = 0.96), indicating that unresolved companions are unlikely \citep{lindegren_gaia_2018}.  

\cite{rodigas_direct_2015} used this system as a test case in their BDI paper.  They noted that it had previously been identified as an AB Dor member \citep{torres_search_2006, Elliott2014SACYV}, \changed{and} that low-mass AB Dor members have Li-depletion boundary ages indistinguishable from that of the Pleiades, i.e., 130~$\pm$~20~Myr \citep{Barrado2004LiDepl}, and adopted this age for the system.  \cite{Binks2020YoungStars} retained it as an AB Dor member, and BANYAN $\Sigma$ \citep{gagne_banyan_2018} gives 86.7\% AB Dor membership probability.  \changed{However} \cite{Binks2020YoungStars} determined ages of 18.3$^{+3.6}_{-4.1}$ and 11.6$^{+4.1}_{-5.1}$ via SED fitting for A and B respectfully.  \changed{We adopted an age of 130~$\pm$~20~Myr for our analysis.  The younger age of \cite{Binks2020YoungStars} would result in mass limits $\sim$15\Mjup\ smaller.}

\textbf{\textit{HD 47787}} --- HD~47787 is a young (16.5~$\pm$~6.5 Myr \changed{derived from evolutionary models}, \citealt{tetzlaff_catalogue_2011}) 2\arcsec\ binary of roughly equal brightness at 48 pc.  It is not a member of a known young moving group (99.9\% field in Banyan $\Sigma$).   A second possible companion is seen at 12.2\arcsec\ \citep{Dommanget2000VisualDoubleStarsCatalog,Fabricius2002TychoDoubleStarCat} however the Gaia EDR3 parallax is significantly different from HD~47787~A, suggesting they may not be associated.  A and B are both spectral type K1IV \citep{torres_search_2006}, and have mass 0.85\Msun\ and 0.89\Msun\ respectively \citep{Anders2019AstrophysicalParametersGaiaDR2}.  Both have RUWE~=~1.1, suggesting any unresolved companions are unlikely to be resolvable in imaging if present.

\textbf{\textit{HD~76534}} --- HD~76534 (OU~Vel) is a very young Herbig Be star with spectral type B2Vn \citep{Houk1978HDSpectralTypes} in a 2\arcsec\ binary at 869 pc with associated nebulosity.  The age is 0.27~$\pm$~0.01 Myr, \changed{derived from MESA Isochrones and Stellar Tracks \citep{Choi2016MIST} and Gaia DR2 colors}, in  \citealt{Arun2019MassIRExcessHerbigABStars}.   \cite{Finkenzeller1984HerbigABwithNebulosity} observed a double peaked H$\alpha$ emission line with an unshifted absorption line; \cite{Berrilli1992IRDustHerbigAB} confirmed dust structures around this and other Herbig Ae/Be stars with optical and Mid- to Far IR luminosities. HD~76534~A has a mass of 6.31~$\pm$0.05\Msun\ \citep{Arun2019MassIRExcessHerbigABStars}, HD~76534~B has a mass of $\sim$2\Msun\ \citep{Anders2019AstrophysicalParametersGaiaDR2}.  HD~76534~A has an elevated RUWE (RUWE~=~1.53), suggesting a possible unresolved companion, however RUWE can also be elevated for highly variable stars \citep{belokurov_unresolved_2020}.  
HD~76534~B has RUWE~=~0.88, making a companion unlikely.

\textbf{\textit{HD 82984}} --- HD~82984 is a 2\arcsec\ pre-main sequence field star binary at a distance of 274 pc, and an age of 53.4~$\pm$~15.1 \changed{Myr} (\changed{derived from evolutionary models}, \citealt{tetzlaff_catalogue_2011}).  Both stars are nearly equal magnitude, with mass 6.3~$\pm$~0.1 \changed{\Msun} \citep{tetzlaff_catalogue_2011} and spectral type B4-5III \citep{Houk1978HDSpectralTypes,tetzlaff_catalogue_2011}.  HD~82984~AB has RUWE = 1.04 and 1.25 respectively.  \cite{kervella_stellar_2019} identified a statistically significant PMa in both Hipparcos (S/N~=~9.24) and Gaia DR2 (S/N~=~10.04) astrometry for HD~82984~A, indicating the possible presence of a companion.  They computed that a normalized mass of m$_2^{\prime}$~=~513.26~\Mjup~AU$^{-1/2}$ would cause the observed acceleration.  Extending to the binary separation, this becomes
\begin{equation}
    m_2^\prime = 513.26\;\Mjup\;\mathrm{AU^{-1/2}} = \frac{m_2}{\sqrt{550\,\mathrm{AU}}} ; \; m_2 = 11\;\Msun
\end{equation}
so the influence of the secondary might contribute to the observed PMa.  PMa reported for Gaia EDR3 \citep{Kervella2022} is consistent with this estimate.  

\textbf{\textit{HD 104231}} --- HD~104231 is a 4.5\arcsec\ binary at 100 pc in Lower Centaurs Crux \citep{Hoogerwerf2000OBAssocMembers} with an age of 21 Myr \citep[derived from isochrone model fitting, ][]{pecaut_revised_2012}.  HD~104231~A has spectral type F5V \citep{Houk1975SpTofHDStarsSouthernDecl} and mass 1.33\Msun; HD~104231~B has mass 0.30\Msun\ \citep{Hagelberg2020VIBES}.  HD~104231~AB have RUWE = 0.82 and 2.29, suggesting a possible unresolved companion around B, although there is no significant PMa in \cite{kervella_stellar_2019}.  \cite{Mittal2015DebrisDisk} observed statistically significant infrared excess luminosity for HD~104231~A in \textsl{Spitzer} 10$\mu$m (S/N~=~6.85) and 20$\mu$m (S/N~=~12.82) bands, corresponding to silicate emission line features.  \cite{Tobal2000VisualBinaries} reported astrometry for a companion at 7.7\arcsec\ observed in 1997, labeled HD~104231~B in the Washington Double Star catalog (WDS), however no further observations of this companion are reported, nor is there a corresponding source in Gaia EDR3.  We conclude this is spurious and adopt the 4.5\arcsec\ companion (labeled HD~104231~C in WDS) to be HD~104231~B.

\textbf{\textit{HD 118072}} --- HD~118072 (V347 Hya) is a G-type 2.3\arcsec\ binary at 80 pc in the Argus Association (89.2\% Banyan $\Sigma$ probability).  We adopted the mean age of 40-50 Myr for the Argus Association \citep{zuckerman_nearby_2019}.   HD~118072~A has a spectral type G3V \citep{torres_new_2000} and mass 1.11\Msun\ \citep{Chandler2016HabitableZones}.  Both have RUWE close to one (A: RUWE~=~1.027, B: RUWE~=~0.962) \changed{and no significant PMa}.

\textbf{\textit{HD 118991}} --- HD~118991 (Q Cen) is a SpT B8.5 + A2.5 \citep{GrayGarrison1987EarlyATypeStars}, 5.6\arcsec binary at 88 pc in the Scorpius-Centaurus Association (76.7\% Lower Centaurus-Crux, 21.5\% Upper Centaurus-Lupus, 1.8\% Field Banyan $\Sigma$ probabilities), with an age of 130-140~Myr \citep[derived from isochrone fitting of Str\"{o}mgren photometry, ][]{david_ages_2015}.  HD~118991~A has a mass of 3.6-3.7~\Msun\ \citep{david_ages_2015}.  Both have RUWE close to one (A: RUWE~=~1.108, B: RUWE~=~1.065) \changed{and no significant PMa}.

\textbf{\textit{HD 137727}} --- HD~137727 contains a pair of G-type stars (G9III+G6IV, \citealt{torres_new_2000}) in a 2.2\arcsec\ binary at 112 pc, with an age of 8 Myr \citep{tetzlaff_catalogue_2011}.  HD~137727~A has a mass of 0.88\Msun\ \citep{Chandler2016HabitableZones}.  HD~137727~A has \changed{RUWE~=~1.416}, but \cite{kervella_stellar_2019} did not detect a significant PMa.  HD~137727~B has RUWE~=~0.88 and no significant PMa.

\textbf{\textit{HD 147553}} --- HD~147553 is a 6\arcsec\ B9.5V + A1V \citep{Corbally1984BinarySpT} binary in Upper Centaurus-Lupus \changed{(Banyan $\Sigma$: 90.7\% probability UCL, 9.0\% USco)} at a distance of 138 pc. \changed{We adopted the median age of UCL 16$\pm$1~Myr derived from isochrone fitting}  \citep{pecaut_revised_2012}.  HD~147553~A has a mass of 2.7~\Msun\ \citep{Hernandez2005HerbigOBAssoc}.  Both have RUWE close to one (A:~RUWE~=~0.927, B:~RUWE~=~1.017) and no significant PMa.  

\textbf{\textit{HD 151771}} --- HD~151771 contains a pair of late B-type (B8III + B9.5, \citealt{Corbally1984BinarySpT}) field stars with separation 7\arcsec\ at 270 pc. \cite{kervella_stellar_2019} found a slightly significant (S/N~=~3.79) proper motion anomaly on the Gaia DR2 epoch only for HD~151771~A, which at the current separation of the binary ($\sim$1890~AU) would result from an object of mass 3.5~\Msun, and so is likely explained by the influence of the secondary star.  HD~151771~A and B have RUWE = 1.22 and 0.797 respectively in Gaia EDR3.

We could not find an age in the literature for either star.  
We used SYCLIST isochrones for B-type stars computed in \citealt{Georgy2013BStarIsochrones}, which used the Geneva stellar evolution code \citep{Ekstrom2021GenevaStelEvCode} to compute grids from 1.7 to 15 M$_\odot$ for three metalicities (Z~=~0.014 (solar), Z~=~0.006, and Z~=~0.002) and a range of rotation rates ($\Omega$) from zero to critical velocity ($\Omega_{\rm{crit}}$).  We used the stellar luminosity estimate of \citealt{mcdonald_fundamental_2012} (L~=~311.76~L$_\odot$) and the 2MASS J-K color (J-K~=~0.08) to interpolate age from the \citealt{Georgy2013BStarIsochrones} isochrones.  We estimated ages for all three metalicities and for three rotation rate values ($\Omega$/$\Omega_{\rm{crit}}$~=~0.0, 0.5, 0.9), which returned age estimates spanning 200-300 Myr.  We performed our analysis for all nine estimated ages.

\textbf{\textit{HD 164249}} --- HD~164249 is a 6\arcsec\ binary at 50 pc in the $\beta$ Pic moving group \citep{messina__2017}.  We adopted the average age of the $\beta$~Pictoris moving group of 25$\pm$3 Myr, \changed{derived from lithium depletion boundary modeling} \citep{messina_rotation-lithium_2016}.  HD~164249~AB have masses 1.29 and 0.54~\Msun \citep{Zuniga-Fernandez2021UpdatedSACY} and SpT F6V + M2V \citep{torres_search_2006} respectively.  
\changed{HD~164249~AB have RUWE~=~1.09 and 1.22.}
HD~164249~A does not have a significant PMa value.

\textbf{\textit{HD 201247}} --- HD~201247 is a pair of G-type stars (G5V + G7V; \citealt{gray_contributions_2006}) at 33 pc with separation 4\arcsec\ and age of 200-300 Myr \citep[derived from chromospheric and coronal X-ray activity and Li EW,][]{zuckerman_young_2013}.  HD~201247~A has mass 0.94\Msun\ and HD~201247~B has mass 0.89\Msun\ \citep{Osborn2020ClassificationOfTESS}.  Both have RUWE close to one (A: RUWE~=~1.064, B: RUWE~=~1.002) and no significant PMa (S/N~$<$~3).

\textbf{\textit{HD 222259}} --- HD~222259 (DS Tuc) is a 5\arcsec\ binary of SpT G6V + K3V \citep{torres_new_2000} at 44 pc in the Tucana-Horologium Moving Group.   \changed{We adopted the average age of 45~Myr derived from isochrone fitting of moving group members} \citep{bell_self-consistent_2015}.  DS~Tuc~A and B have masses 1.01~$\pm$~0.06 and 0.84~$\pm$~0.06 \Msun\ \citep{newton_tess_2019}.  \changed{Both have RUWE close to one (A: RUWE~=~0.95, B: RUWE~=~0.91) and no significant PMa (S/N~$<$~3).}  \cite{newton_tess_2019} detected a transiting planet around DS~Tuc~A with the \textsl{Transiting Exoplanet Survey Satellite} \citep[\textsl{TESS};][]{Ricker2015TESSMission} with an 8 day period who's orbit is likely aligned with the binary orbit.  They did not find additional companions using the Gemini Planet Imager (GPI, \citealt{Macintosh2014GPI}) integral field spectroscopy in \textsl{H}-band or in the \textsl{TESS} photometry.

\textbf{\textit{HIP 67506 \changed{and TYC 7797-34-2}}} --- HIP~67506 \changed{is a} field star (99.9\% probability in Banyan $\Sigma$).  \changed{It was identified as a wide binary in the Hipparcos and Tycho Doubles and Multiples Catalog \citep{HipAndTychoDoublesCat} with another star (TYC 7797-34-2) with separation 9\arcsec\, and dubbed HIP~67506~A and B
}.  HIP~67506 has a distance 89.5~$\pm$~30~pc according to the Hipparcos catalog \citep{vanLeeuwen2007HipReduction}.  HIP~67506 is identified as type G5 \citep{SpencerJones1939SpT} and mass 1.2\Msun\ \citep{Chandler2016HabitableZones}. \changed{Both stars have Gaia ERD3 RUWE~$>$~1.4 (A: RUWE~=~2.02; B: RUWE~=~1.73).}

There is no age in the literature for these stars.  To estimate the age of HIP~67506, we used the luminosity and effective temperature estimates of \cite{mcdonald_fundamental_2012} for the primary (T$_{\rm{eff}}$~=~6077~$\pm$~150~K; L~=~0.37~$\pm$~0.07~L$_\odot$) to interpolate an age \changed{of $\approx$~200~Myr using \citetalias{Baraffe2015BHAC} isochrones.  There is no literature age nor stellar parameter estimates for TYC 7797-34-2, so we performed our analysis assuming a range of ages from 200 Myr to 10 Gyr.}

\textit{Status as gravitationally bound binary.}  Gaia EDR3 shows two sources corresponding to the expected separation and position angle given by Hipparcos, yet differing parallax solutions (A: source id~=~6109011780753115776, $\pi$~=~4.51 mas; B: source id~=~6109011742094383744, $\pi$~=~0.55 mas). The dramatically different Gaia parallaxes for A and B question if the two stars are actually a gravitationally bound pair versus a chance alignment of unassociated stars at different distances.  

\changed{We conducted a common proper motion analysis using Washington Double Star Catalog astrometry of the pair (WDS J13500-4303A and B) and determined the previously-identified HIP~67506~B (TYC 7797-34-2) is in fact an unassociated star that is much further distant.  This analysis will be presented in a forthcoming follow up paper on this system, Pearce et al. (in prep).  This has no impact on the utility of the pair for BDI, however it does impact the contrast and mass limits we are able to achieve for the much further distant star.
}

\textbf{\textit{TWA 13}} --- TWA~13 is a pair of M1Ve \citep{torres_search_2006} T-Tauri stars \citep{Samus2003GenCatVariableStars} at 60 pc and 5\arcsec\ separation in the TW Hydra Association \citep{Schneider2012TWAMembership}.  \changed{We adopted the mean age of TW Hydra (10$^{+10}_{-7}$ Myr, from isochrone fitting, lithium equivalent width, and H$\alpha$ emission in}  \citealt{barrado_y_navascues_age_2006}.  Both have mass of 0.57\Msun\ \citep{Herczeg2014TTauriPhotospheres}.  \changed{TWA~13~A has RUWE~=~1.085 while TWA~13~B has a slightly elevated RUWE~=~1.266.}

\textbf{\textit{2MASS J01535076-1459503}} --- 2MASS~J01535076-1459503 is a 3\arcsec\ young (25~$\pm$~3 Myr, \citealt{messina_rotation-lithium_2016}) binary in the $\beta$ Pictoris Moving Group \citep{messina__2017} at 33 pc.  We adopted the mean age of $\beta$ Pictoris Moving Group (25$\pm$3 Myr, \citealt{messina_rotation-lithium_2016}).  
2MASS~J01535076-1459503~A has a mass of 0.34\Msun \citep{Osborn2020ClassificationOfTESS} and SpT M3 \citep{Riaz2006NewMDwarfs}.  2MASS~J01535076-1459503~A has a slightly elevated RUWE while 2MASS~J01535076-1459503~B has \changed{an RUWE close to one} (A:~RUWE~=~1.220, B:~RUWE~=~1.089).

\end{document}